# Nanoparticles and quantum dots as emerging optical sensing platforms for Ni(II) detection: Recent approaches and perspectives

Sudhanshu Naithani[a,†], Heena[a,†], Pooja Sharma[a], Samar Layek[b,*], Franck Thetiot[c,*], Tapas Goswami [a,*], Sushil Kumar [a,*]

[†] *These authors contributed equally to this work*
*Department of Chemistry, School of Engineering, UPES, Dehradun 248007, Uttarakhand, India. Email: sushil.k@ddn.upes.ac.in; tgoswami@ddn.upes.ac.in*

[b] *Department of Physics, School of Engineering, UPES, Dehradun 248007, Uttarakhand, India. Email: samar.layek@ddn.upes.ac.in*

[c] *CEMCA, CNRS, UMR 6521, Université de Bretagne Occidentale, Brest 29238, France.*

*Email: franck.thetiot@univ-brest.fr*

**Abstract**

Over the preceding years, nickel (Ni) and its compounds have been increasingly employed in various aspects of human social life, metallurgical/industrial manufactures, healthcare and chemical processes. Although Ni is considered as an essential trace element in biological system, excessive intake or metabolic deficiency of $Ni^{2+}$ ions may cause detrimental health effects to the living organisms. Therefore, a facile and accurate detection of $Ni^{2+}$, especially in environment and biological settings, is of huge significance. As an efficient detection method, assaying $Ni^{2+}$ using optical (colorimetric and/or fluorogenic) sensors has experienced quite a vigorous growth period with large number of excellent researches. Nanomaterial-based optical sensors including metal nanoparticles (MNPs), quantum dots (QDs), and carbon dots (CDs) offer distinct advantages over conventional small-molecule organic and inorganic sensors. This study mainly provides an overview of the recent advancements and challenges related to the design strategies of various optical nanosensors to selectively detect $Ni^{2+}$ ion. Emphasis has also been placed on comparing the sensing performance of various nanosensors along with exploring future perspectives.

**Content**

## 1. Introduction

Metals ions play indispensable role in maintaining the balance of entire ecological system, serving as vital components in different biological, environmental and chemical processes [1–5]. Among various metal ions, nickel(II) is essential for numerous biological processes, and a wide range of organisms including plants, bacteria, unicellular eukaryotes, and archaea rely on this metal ion for crucial metabolic processes [6–8]. Indeed, it is widely employed in variety of industrial and metallurgical applications (such as alloys formation, electroplating, batteries, dentistry, and catalysis) due to its unique features of high melting point, corrosion resistance, and tensile strength [9–11]. Despite its many benefits, both excessive and deficient levels of $Ni^{2+}$ may lead to severe health effects on humans and environment [12,13]. Nickel pollution has increasingly become a global environmental concern due to its rapid industrialization, with an accentuation over the years [14,15]. Consequently, the World Health Organization (WHO) has established a concentration limit of 0.02 mg/L for $Ni^{2+}$ ions in the drinking water [16]. This underscores an urgent need to develop efficient methods for selective recognition of $Ni^{2+}$, particularly in environmental and biological contexts.

Typically, the conventional $Ni^{2+}$ detection methods rely on the use of laboratory based sophisticated analytical techniques such as voltammetry and spectrophotometry (e.g., inductively coupled plasma mass spectroscopy (ICP-MS), inductively coupled plasma atomic emission spectroscopy (ICP-AES), atomic absorption spectroscopy (AAS), X-ray fluorescence spectroscopy (XFS), and capillary electrophoresis (CE)). Such techniques display precise determination of nickel concentration; however, they are relatively restrictive in terms of transportability, and thus, hardly applicable to be used for targeted on-site field applications [17,18]. Besides, despite being highly sensitive, these techniques often suffer from several limitations including expensive instrumentation, high running-cost, tedious sample preparation, and professional maintenance [19–21]. As an illustration, $Ni^{2+}$ detection using flame AAS technique requires extended analysis time, a large volume of analytes, and extensive sample pre-processing steps [22]. On the other hand, ICP-MS method involves substantially skilled technicians while the running cost of the equipment is also remarkably high [22,23]. Furthermore, many electrochemical $Ni^{2+}$-responsive methods such as anodic stripping voltammetry (ASV) suffer from low specificity, low accuracy, as well as a narrow range detection of analytes [24,25]. As an alternative, to meet the requirements of an accurate, highly sensitive/selective, rapid and on-site field applicable detection methods for $Ni^{2+}$ in different media, a diverse panel of optical sensors based on colorimetry, fluorescence, Surface

Plasmon Resonance (SPR) and Surface Enhanced Raman Scattering (SERS) have been designed and exploited [17,26–28]. Indeed, optical methods have emerged as extremely simple and advantageous platforms due to their multiple promising merits such as good accuracy, high sensitivity, low-cost instrumentation, rapid response, and low detection limits [29]. Additionally, a portable device based on colorimetric and fluorogenic sensors can easily and reasonably be employed for on-site analyte detection purposes [30,31].

In this context, a large number of optical sensors such as metal-organic frameworks (MOFs) [32–34], nanomaterials (nanoparticles (NPs), quantum dots (QDs), carbon dots (CDs), etc.) [35–38], small synthetic organic [18,39,40] as well as inorganic sensors [41–43] have been developed to detect trace amount of $Ni^{2+}$ ions. While numerous small molecular sensors based on organic dyes (e.g., fluorescein, rhodamine) have been successfully used for optical detection of $Ni^{2+}$, they often face recurring drawbacks, including poor water solubility, weak absorption coefficients, toxicity, and low stability due to photobleaching effects [17,27,44]. On the other hand, optical nanosensors (colorimetric or fluorogenic) tend to exhibit improved analyte sensing as well as selectivity properties, better stability, and biocompatibility. They have become very popular worldwide because of their advantageous nano-scale features beyond the small size, notably a high surface to volume ratio, compactness, quantum confinement effects, surface functionalization, high adsorption (physio- and chemi-sorption) ability, on-site detection ability [45,46]. Furthermore, the sensitivity and selectivity of nanosensors for a specific metal ion can finely and accessibly be tuned *via* their functionalization with a variety of suitable organic ligands or biomolecules [47,48]. In recent years, many optical organic (organic NPs, CDs, etc.) and inorganic-based nanomaterials (metal nanoparticles (MNPs), semiconductor QDs, core shell quantum dots (CSQDs), etc.) have been exploited for the detection of heavy metal ions [49,50].

This study aims to emphasize the recent advancements in the development of nanomaterial-based optical sensors (such as MNPs, QDs, CSQDs, CDs, etc.) for the selective detection of $Ni^{2+}$ ions. Though some reviews have addressed optical recognition of $Ni^{2+}$, they mainly focus on either small synthetic organic/inorganic probes or the detection of multiple heavy metal ions [27,51,52]. To the authors' knowledge, a detailed review exclusively on nanomaterial-based optical sensors for $Ni^{2+}$ detection has not been previously compiled. In this context, a collection of $Ni^{2+}$-responsive optical nanosensors has been systematically gathered, with careful attention to optical detection methods and the subsequent classification of nanosensors. These sensors are organized in order of increasing nanosensors's design complexity, from nanoparticles to QDs to CSQDs.

## 2. $Ni^{2+}$ ion: Sources and impacts

Among group 10 metals, Ni is the most abundant element while Pd and Pt are relatively scarce. The industrial demand of nickel has dramatically expanded over the last decade, and the annual Ni production worldwide has approached *ca*. 3.0 billion tons [53]. As a result, the probability of human exposure to nickel ions is relatively much higher and still increasing day-by-day. Nickel is often not considered as a toxic element, and it is assumed to be sustainable and green alternative to heavy toxic metals. Indeed, it is typically referred as an essential element for the living beings whereas Pd and Pt are commonly discussed in terms of toxicity. Relatedly, many researchers have shifted their focus on the development of Ni-based catalysts as replacement alternative to more toxic metals like Pd and Pt [54–56].

However, recent studies have revealed a surprising and unexpected picture on this otherwise benign metal. Over the years, there has been a significant rise in the number of research articles focusing on Ni toxicity and correlated adverse effects on human health and plants. A general trend observed from the research works available on Ni, Pd, and Pt toxicity (*Web of Science* database) (Fig. 1) clearly indicates that the nickel exposure to humans is eventually leading to significant toxic effects comparable to conventionally considered toxic heavy metals such as Pd and Pt. Although the consequences of nickel homeostasis to human health and its related diseases have not been fully explored, excessive nickel exposure has been linked to severe health issues, including pneumonitis, asthma, respiratory-system cancer, and neurological dysfunction [57–59].

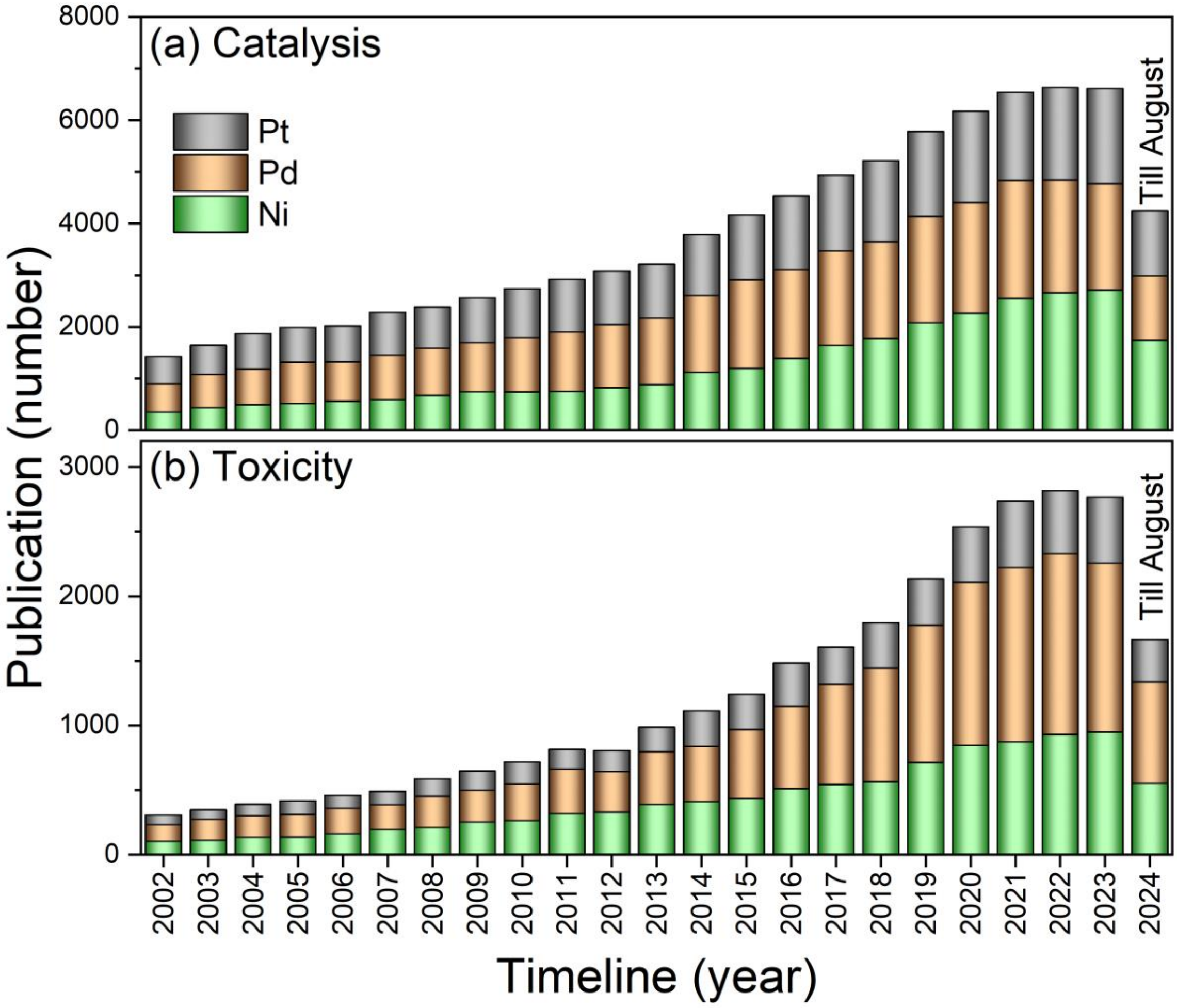

**Fig. 1.** Approximate number of publications dealing with group 10 metal ions (a) catalysis, and (b) toxicity in the last 20 years. Adapted from ISI *Web of Science* (WoS), dated 10-09-2024.

Nickel contamination primarily originates from industrial activities, chemical processes, and medical disposals (Fig. 2), and humans are commonly exposed to nickel through ingestion, inhalation, and dermal absorption. [27,60–62]. The burning of fossil fuels has markedly led to an increase of small Ni particles in ambient air in urban areas [63]. Additionally, Ni is present in tobacco and cigarettes, and nearly 10-20% of Ni inhaled is in gaseous phase. Nickel toxicity has become a huge concern in many countries, especially North America, Denmark, China and Pakistan [57,64,65]. It can cause oxidative damage, modify the enzyme/protein structures, and inhibit root and shoot growth in plants, and also interferes in the photosynthesis by displacing magnesium (Mg) in the chlorophyl [66].

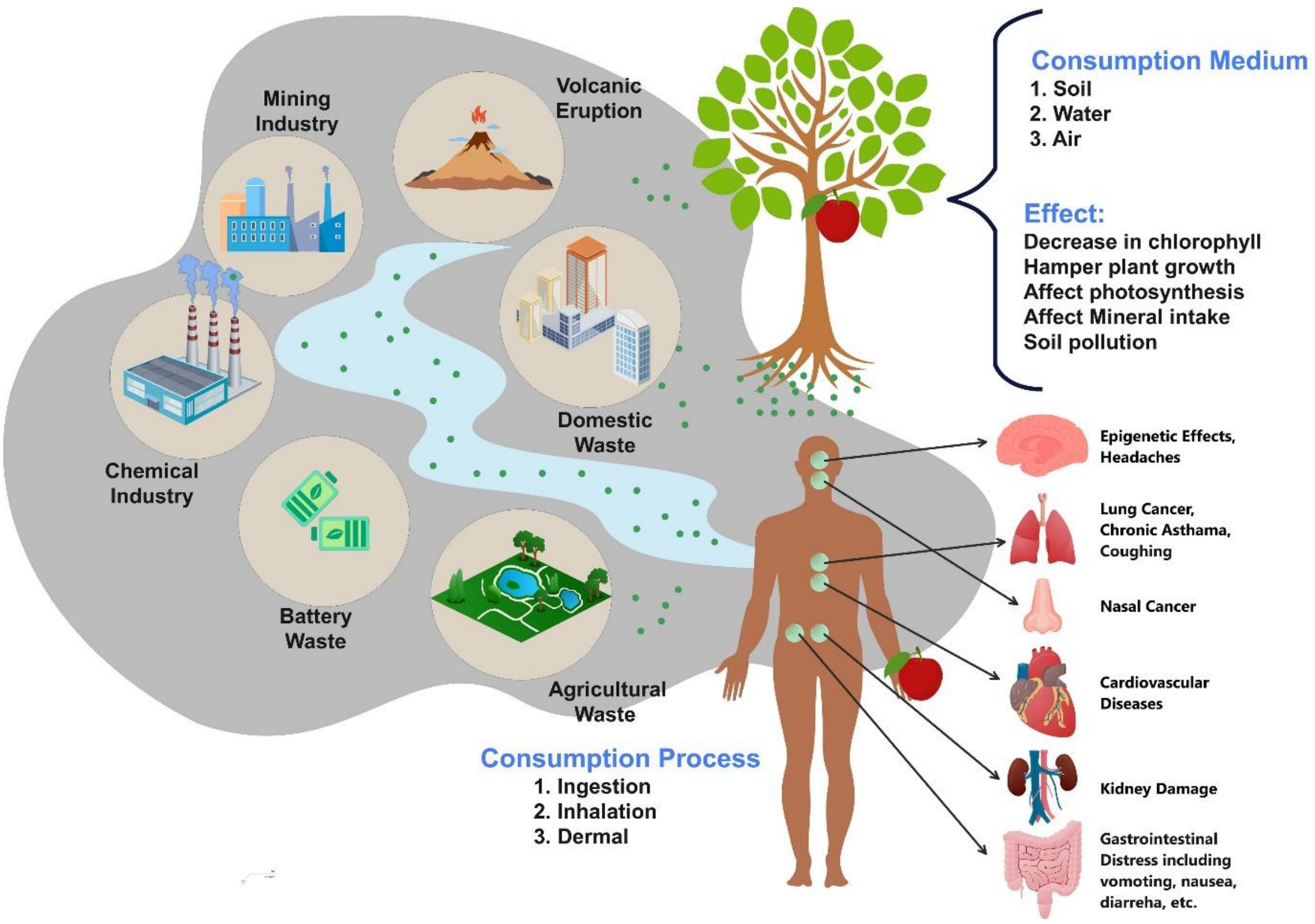

**Fig. 2.** Sources and impact of $Ni^{2+}$ on human health and environment.

As a result, spotting $Ni^{2+}$ in the environment and biological systems is evidently critical. Functionalized nanomaterials including MNPs, QDs, CSQDs and CDs have emerged as viable platforms for optical detection and quantification of $Ni^{2+}$ in a wide range of samples, from cell organelles to all water bodies [67]. Beyond the detection, there is also significance and necessity to understand both the physiological and psychological aspects and related biomedical impacts of $Ni^{2+}$ ions.

## 3. Nanomaterials based sensors

The present era is widely acknowledged as the "Nano Era" where nanoscience and nanotechnology encompass various disciplines of science, engineering and technology to develop effective solutions for benefits of mankind [68]. The concept of nano was introduced by renowned scientist Richard Feynman in the realm of science through a lecture entitled “There is Plenty of Room in the Bottom” [69]. This ultimately opened up a new pathway in advancement of various scientific fields including water treatment, catalysis, sensors, medicines, energy storage, electronics, textiles, cosmetics, food industries, etc., to name but a few [70–72]. The dramatic expansion and implementation of nanoscience and technology across various industries have significantly impacted global economic sectors [73]. The growing fascination with nanomaterials lies behind their unique properties which can be fine-tuned by altering their size, shape, compositions, surface functionalities, and synthetic strategies [74,75].

Over the years, the development of nanomaterials has rapidly advanced, leading to significant progress in nanosensors [76] for selective and sensitive detection of cations, anions, biomolecules, antibiotics, viruses, etc., especially in environmental and biological settings [77–82]. Nanomaterial-based sensor can also effectively carry signal-generating molecules as the size of nanosensors (< 100 nm) is smaller than the opening size and pores of vasculature and tissues which aids in intracellular imaging of analytes [83]. Among different types of nanosensors, optical sensors offer several advantages, including ease of fabrication, simple setup, rapid response time, miniaturization, and suitability for point-of-care applications. These sensors can be broadly categorized into inorganic and organic types, each interacting with target analytes (such as $Ni^{2+}$ in this study) in specific ways that produce distinct colorimetric or fluorogenic responses. These nanosensors can be categorized into two main types based on the optical response triggered by the analyte: (i) colorimetric sensors- primarily utilizing noble MNPs which induce a shift in the SPR band, resulting in a visible color change (easily detected using UV-Vis spectroscopy) and (ii) fluorogenic sensors- typically employing semiconductor QDs or CDs which detect $Ni^{2+}$ through the formation of stable ground-state complexes.

The following sections focus specifically on a range of colorimetric and fluorometric nanosensors developed for detecting $Ni^{2+}$ ions. Detailed information is provided on the fluorescence enhancement and quenching properties, color changes due to shifts in surface plasmon resonance, detection limits, and the overall sensing range of these nanosensors. Additionally, the underlying mechanisms driving colorimetric and fluorometric detection using various nano-based sensors are thoroughly examined.

## 4. Colorimetric $Ni^{2+}$ nanosensors

The colorimetric sensing ability of NPs, typically MNPs, relies on the changes observed in their SPR absorption upon addition of a guest analyte [36,84–87]. The SPR phenomenon occurs when the nanoparticles are excited by an electromagnetic radiation of a certain wavelength, causing the collective oscillation of free electrons in the conductive band of NPs. The collective oscillation of conduction electrons is known as ‘Plasmon’. The SPR feature arises when interaction of electromagnetic waves such as light causes the plasmons to resonate with a specific frequency, provided the wavelength of incident light is much higher than the size of NPs [86]. When this collective oscillation occurs in a particle with a finite volume, it is commonly referred to as localized surface plasmon resonance (LSPR) [88]. The electromagnetic field of incident light polarises the surface free electrons to induce an electric dipole. The direction of electric field generated by the electric dipole inside the NPs ($E_D$) is opposite to the electric field generated by the light ($E_L$) which helps in restoring the equilibrium position of the electrons in NPs (Fig. 3) [89].

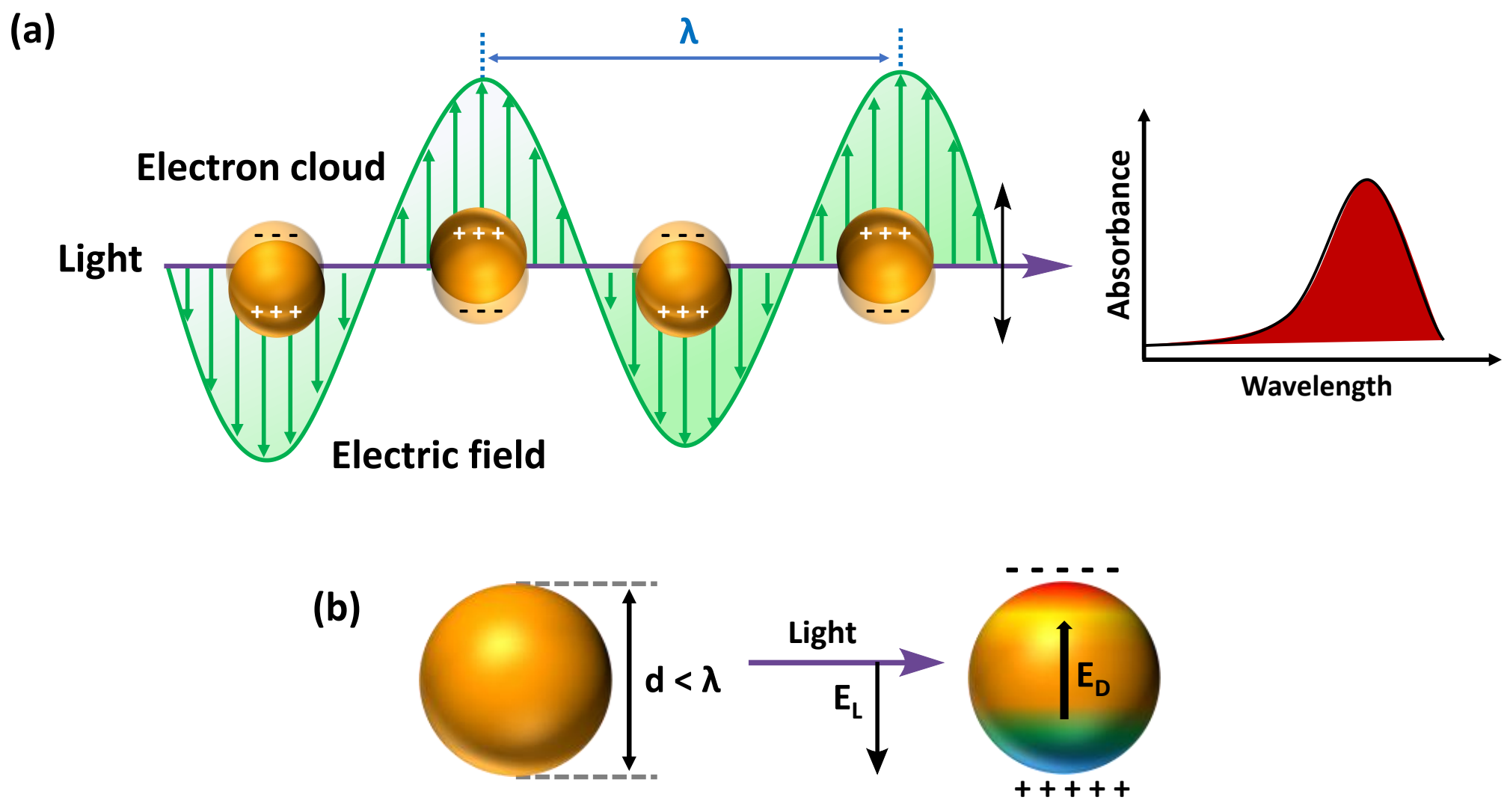

**Fig. 3.** (a) Schematic illustration of the interaction of electromagnetic wave with MNPs and (b) Interaction of light with the particle having size smaller than its wavelength.

It is worth mentioning here that the anisotropic nanostructures may possess multiple SPR absorptions relying on the direction of oscillation of electrons (geometry of nanostructure). For example, a single SPR absorption is observed for gold nanospheres, while both longitudinal and transverse resonances are possible in case of gold nanorods [90]. Interestingly, the plasmonic phenomena mostly occurs when the size of the particle lies in the range of 2-100 nm. If the diameter of particle is too big (>$10^4$ Å {1000 nm}) then the bulk properties dominate. On the other hand, below 2 nm the Kubo gap is very large, and the NPs display insulating behaviour i.e., electrons are bound and therefore prevent the existence of surface plasmons [91]. The SPR for spherical NPs was successfully calculated by Mie's theory (1908) in which the extinction coefficient $\sigma_{ext}$ can be calculated as (Eq. 1) [92]:

$$\sigma_{ext} = 9 * \frac{\omega}{c} \varepsilon_m^{\frac{3}{2}} V \frac{\varepsilon_2}{[\varepsilon_1 + 2\varepsilon_m]^2 + \varepsilon_2{}^2} \qquad \text{Eq. 1}$$

Where V stands for particle's volume, c is the speed of light, ω is the angular frequency of the exciting light, $\varepsilon_m$ is the dielectric constant, $\varepsilon_1$ and $\varepsilon_2$ are the real and imaginary parts of the dielectric function of metal NPs.

As per Eq. 1, the condition for the plasmon occurrence is $\varepsilon_1 = -2\varepsilon_m$ while the plasmon position is dependent on the intrinsic dielectric properties of metal NPs. For instance, the SPR absorption of Al, In, Mg, and Ga lies in the UV region whereas it appears in the visible region for noble metals such as Ag, Cu and Au [89,93,94]. Interestingly, the color of the MNPs, and the characteristic SPR band can easily be tuned by varying different parameters such as the size and shape of the particles, nature of the surface ligands (i.e., surfactants) or adsorbed species, the dielectric properties of the medium, and/or the distance between particles [87,91,95]. This suggests that these noble metals based plasmonic materials can be judiciously chosen for optical sensing applications. Several mechanistic pathways have been demonstrated for the recognition of heavy metal ions using nanosensing platforms. The recognition process may include metal ion induced aggregation, binding, releasing (of some components from the surface of NPs), and etching of NPs (Fig. 4) [96].

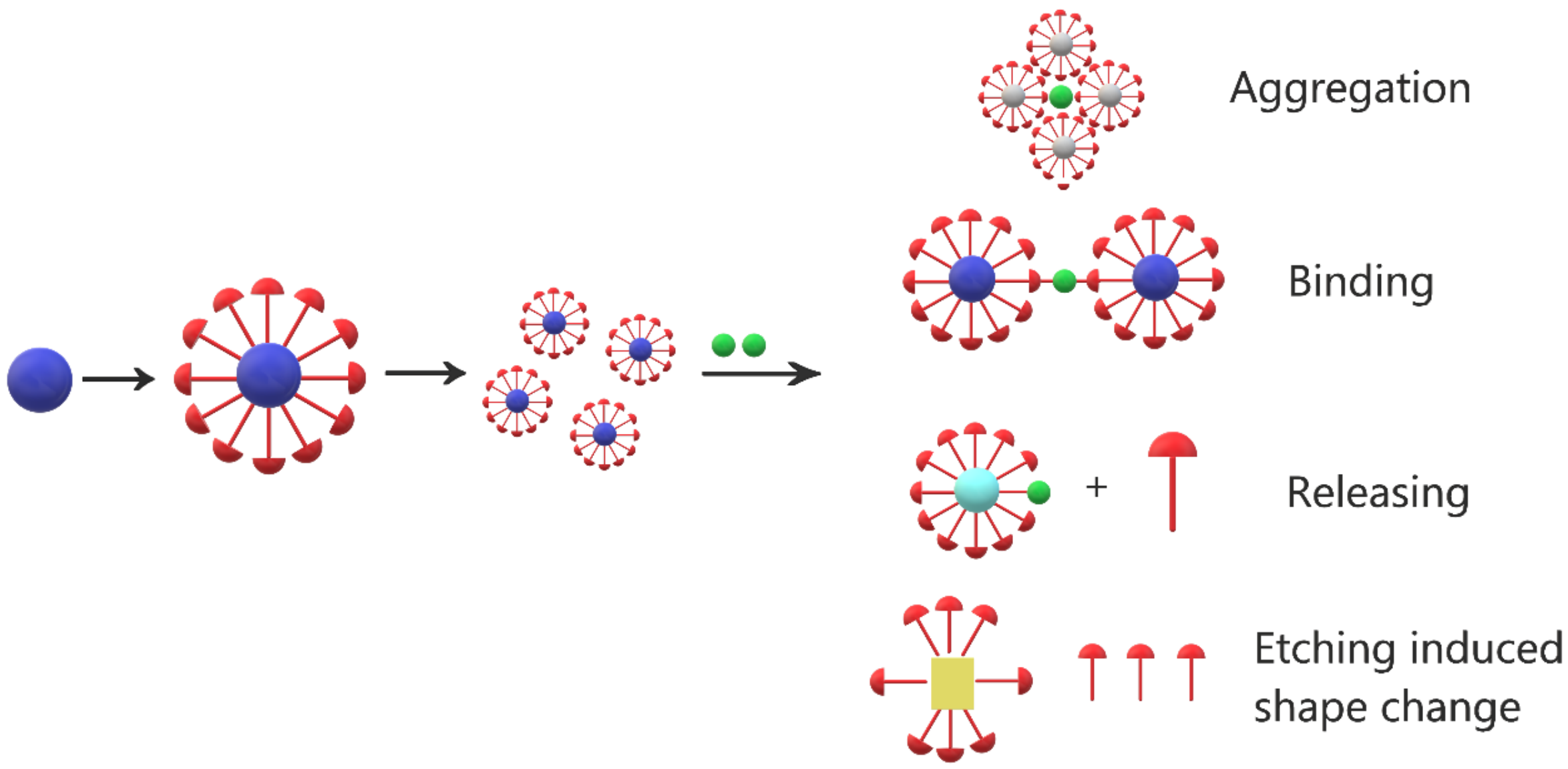

**Fig. 4**. Functioning principle of nanoparticles (NPs) acting as colorimetric sensors for heavy metal ions (or target analyte; shown as green colored sphere).

So far, the mostly utilized colorimetric nanosensors for $Ni^{2+}$ ions are composed of Au and Ag and they rely on aggregation and etching induced shape change for detection. These MNPs exhibit high extinction coefficient, chemical stability, biocompatibility, strong SPR band, resistance against oxidation, and tuneable functionalities making them ideal for optical detection [86,91,95] . As evidence of the importance of gold and silver nanoparticles, we can refer to the wide attention and applications of the latter through the ages. As an illustration, the 4th-century *Lycurgus Cup* showed different colors depending on the incident light intensity [97]. The related color change was then associated to the small amount of gold and silver nanoparticles.

In recent years, many groups including but not limited to Amirjani [86], Yu [98], Garcia [91], and Lu [99] have presented elegant reviews on the use of MNPs in optical sensing of different analytes. For instance, Yu *et* al. highlighted many examples on AuNPs-based colorimetric sensors displaying obvious color changes in presence of different analytes. In year 2022, Chatterjee [95] analysed and discussed surface functionalization dependent sensing behaviors of Ag and Au nanoparticles. However, to the best of author's knowledge, the present work is the first study to cover exhaustively and systematically the use of NPs together with QDs as optical sensors to specifically detect $Ni^{2+}$ ions.

### *4.1. Silver nanoparticles (AgNPs)*

AgNPs based sensors are reported to be highly sensitive having excellent molar extinction coefficients with sharp resonance peaks; however, in comparison to their AuNPs counterparts, they are less popular [100,101]. This is because of the related stability constraints, more toxicity, and less surface explored chemistry and biocompatibility which makes it challenging to employ them in complex biological media [101]. The commonly adopted strategy to develop AgNPs based colorimetric sensory probes is the modification of the surface ligand shell of the nanoparticles [102]. Surface modifications of NPs have usually been performed by apposite functionalization employing different suitable organic ligands, which also implicated the stability of the resultant NPs in various solvents. For example, Glutathione (GSH), a tripeptide having -SH groups, has been used by several groups as a protecting ligand of metal NPs [35,103]. Besides, the functional groups such as $-NH_2$ and -COOH, present in this biomolecule show strong binding affinities towards metal ions. In view of this, many GSH stabilized NPs have thus been developed for in-depth understanding of their applications in sensing of different analytes. The judicious selection of the surface organic ligands is crucial for the selective detection and complexation of target cationic species, and implicitly correlated to the nature of the metal ion target; hence, the HSAB (hard and soft acids and bases) principle, the electronic configuration, geometry and the coordination number of metal ion should be taken into account [104].

Li *et* al. pioneered the development of $Ni^{2+}$-responsive silver nanoparticles (AgNPs) in year 2009 [35]. They demonstrated a colorimetric nanosensor **1** based on glutathione-stabilized Ag nanoparticles (GSH-AgNPs) for the detection of $Ni^{2+}$ ions in aqueous media (Fig. 5). The GSH-AgNPs could easily be prepared by the reduction of $AgNO_3$ with sodium borohydride ($NaBH_4$) in presence of glutathione as stabilizer. The capping of GSH over the surface of **1** was evidenced by FT-IR analysis where the absence of any characteristic peak of -SH group (near 2524 $cm^{-1}$ in free GSH) indicated the functionalization of GSH onto the probe **1**. The appearance of yellow color upon formation of GSH-AgNPs was most likely due to a typical intense SPR absorption band observed near 396 nm. Upon addition of $Ni^{2+}$, the yellow color of NPs turned to deep orange accompanying a red-shift (from 396 nm to 500 nm) and broadening in the plasmon band. On the other hand, these NPs displayed a negligible sensitivity for other competitive metal ions such as $Li^{+}$, $Na^{+}$, $K^{+}$, $Mg^{2+}$, $Ca^{2+}$, $Sr^{2+}$, $Ba^{2+}$, $Mn^{2+}$, $Co^{2+}$, $Cu^{2+}$, $Zn^{2+}$, $Cd^{2+}$ and $Hg^{2+}$ ions. Both the free $-NH_2$ groups from the glutamate unit and the terminal -COOH groups from the glycine moiety participated in binding $Ni^{2+}$ ions, eventually leading to a cross-linking event. $Ni^{2+}$ addition to **1** induced the aggregation process which could be disassembled upon the addition of 1,2-ethylenediamine. Though this probe could be successfully employed for $Ni^{2+}$ detection in aqueous medium, unfortunately the limit of detection (LoD) was depicted to be $7.5 \times 10^{-5}$ mol $L^{-1}$ (or 75.0 μM) which is relatively very high in comparison to the permissible limit of $Ni^{2+}$ in drinking water set by WHO (i.e., 0.34 μM).

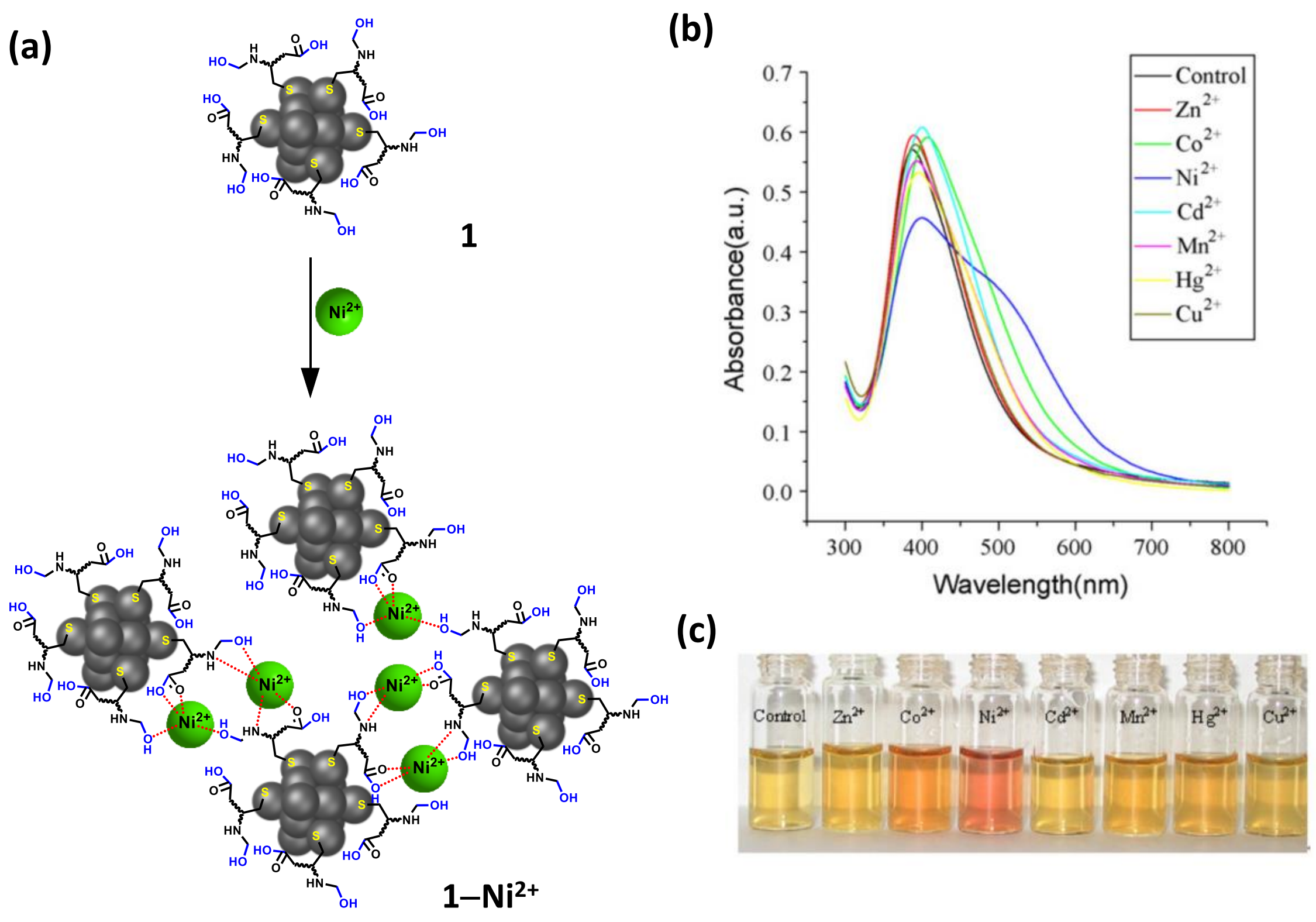

**Fig. 5.** (a) Binding mode of $Ni^{2+}$ in **1**-$Ni^{2+}$ followed by the aggregation process; (b) UV-Vis spectra of **1** with various cations and (c) Visual pictures of **1** with various metal ions. Adapted with permission from Ref [35]. Copyright 2009, Elsevier.

The ligands employed for the capping of MNPs significantly impact their structural, electronic, and optical properties. Furthermore, the sensing characteristics of such NPs based probes can be influenced by an appropriate selection of the ligands. Therefore, it is important to develop sensors which showcase versatile sensing ability, that can be achieved by tuning nanomaterial characteristics like size, shape, composition, functionality, and morphology. Over the preceding years, thiol-based ligands have seen gradual growth to give stable and monodispersed MNPs for various applications. Among such ligands, *N*-acetyl-*L*-cysteine (NAC) has been recently employed as a surface modifier in the development of several QDs and MNPs [105,106]. NAC, a strong antioxidant, is well known as the mucolytic agent as well as the antidote for the poisoning of paracetamol [107,108]. It is a non-toxic and cost-effective thiol-based biomolecule, and its usage as a NPs stabilizer may offer the development of new materials as eco-friendly sensors. In 2012, Shang *et* al. [108] prepared a NAC-templated AgNPs-based colorimetric sensor (**2**) for the detection of $Ni^{2+}$ ions (Fig. 6). The sensor **2** exhibited highly selective colorimetric response for $Ni^{2+}$ over other metal ions (e.g., $Na^+$, $K^+$, $Ba^{2+}$, $Ca^{2+}$, $Mg^{2+}$, $Al^{3+}$, $Cr^{3+}$, $Mn^{2+}$, $Fe^{2+/3+}$, $Co^{2+}$, $Cu^{2+}$, $Zn^{2+}$, $Pb^{2+}$, $Hg^{2+}$, $Ag^+$ and $Cd^{2+}$). The AgNPs have been prepared by the reduction of $AgNO_3$ with $NaBH_4$ in the presence of *N*-acetyl-*L*-cysteine. The formation of NAC-templated AgNPs was evident from the disappearance of the sulfhydryl group (2552 $cm^{-1}$) in FT-IR spectrum, indicating the capping of NAC over the surface of AgNPs. Upon successive addition of $Ni^{2+}$ (2.0 mM - 48.0 mM) to **2**, the absorption band at 390 nm decreased with the concomitant formation of new band near 550 nm. Besides, $Ni^{2+}$ addition induced a rapid aggregation of **2** with a drastic solution color change from yellow to deep orange. The absorbance intensity ratio ($A_{550}/A_{390}$) was found to be linear with the increasing concentration of $Ni^{2+}$. The reported probe could be successfully applied to trace $Ni^{2+}$ in a series of tap water samples collected from the author's institute (i.e., Nanchang University, China) with a recovery rate of 92-106%. Interestingly, in this study, the limit of detection has been depicted as 0.23 µM, which is considerably lower than that of probe **1** (75 µM). The improved detection limit may be attributed to the ligands employed for functionalization, eventually leading to the uniformly distributed MNPs.

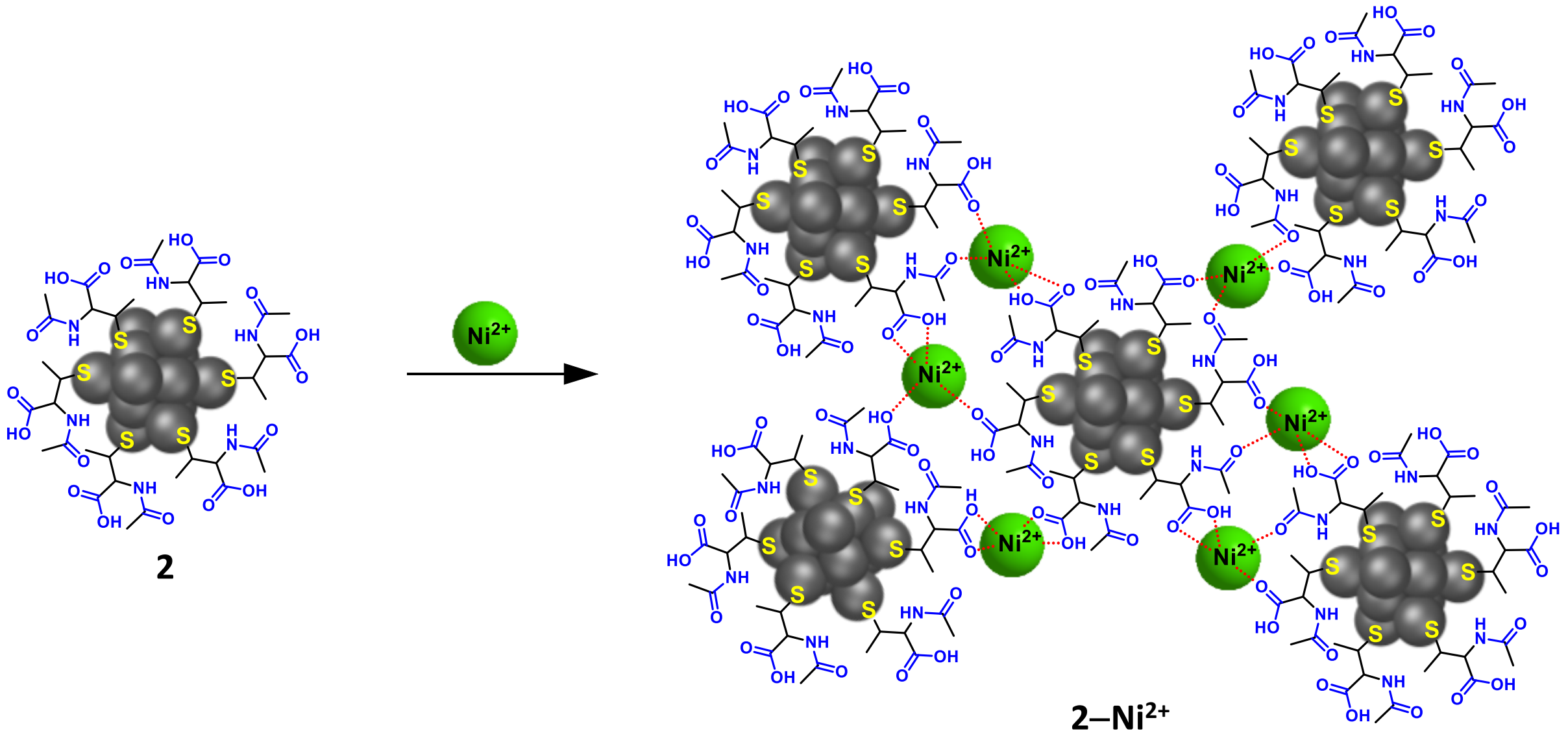

**Fig. 6.** Binding mode of $Ni^{2+}$ in **2**-$Ni^{2+}$ followed by the aggregation process.

AgNPs-based sensors have a unique advantage over other sensors due to their relatively deeper penetration into environmental samples where different organic and inorganic molecular sensors cannot approach; however, many of such sensors are typically subjected to complicated synthetic and post-modification routes. Notably, the employment of these sensors as solid-state paper strips has been rarely explored, which can be a viable, economical and cost-effective detection method. For an applied potential, the metal ion sensing using test paper kits is more attractive than that of only colorimetric sensing employed in the solution phase. The reason lies behind the convenient use of test paper kits to achieve real time monitoring *via* on-site field applications. In this effort, Biswas's group [109] (in year 2014) demonstrated an easily synthesizable AgNPs-based probe **3** using 3,6-di(pyridin-2-yl)-1,2,4,5-*s*-tetrazine (pytz) as a capping agent (Fig. 7), and investigated the metal ion sensing ability of **3** onto test-paper strips. Probe **3** was easily prepared with the reduction of $AgNO_3$ using 3,6-di(pyridin-2-yl)-1,4-dihydro-1,2,4,5-*s*-tetrazine ($H_2$pytz) in ethanolic solution. The ligand $H_2$pytz showed dual functionality by reducing $Ag^+$ ions as well as capping the nanoparticles formed. UV-Vis spectrum of **3** displayed three absorption bands centred at 250 nm, 300 nm and 522 nm that could be ascribed to $\sigma$–$\pi$*, $\pi$–$\pi$* and $n$–$\pi$* transitions, respectively. A broad SPR band for **3** at 410 nm suggested the formation of the spherical nanoparticles. The transmission electron microscopy (TEM) analysis further confirmed the spherical shape of NPs with an average particle size of 10-11 nm. The addition of $10^{-3}$ M of $Na^+$, $Pb^{2+}$, $Mn^{2+}$, $Fe^{3+}$, $Co^{2+}$, $Mg^{2+}$, $Hg^{2+}$, $Zn^{2+}$, $Cr^{3+}$, and $Cd^{2+}$ resulted in negligible changes in the dispersed aqueous solution of **3**. On the other hand, $Cu^{2+}$, $Ni^{2+}$, and $Ag^+$ showed a colorimetric response as the pink solution of **3** turned green (for $Cu^{2+}$), yellow (for $Ni^{2+}$) and brown (for $Ag^+$) within 3 mins. The color change was also accompanied by the appearance of a new absorption peak at 350 nm, 360 nm, and 380 nm in case of $Cu^{2+}$, $Ni^{2+}$, and $Ag^+$, respectively (Fig. 7b). The aggregation of the silver nanoparticles in the presence of $Cu^{2+}$, $Ni^{2+}$, and $Ag^+$ was clearly evidenced by TEM analysis, and this aggregation process was attributed to the interaction between the target metal ion and the nitrogen atoms of the capping agent. The LoD for $Ni^{2+}$ was reported to be 9.4 μM. To assess its applied potential, probe **3** was successfully immobilized over cellulose paper strips to construct a paper-based assay for the recognition of $Cu^{2+}$, $Ni^{2+}$, and $Ag^+$ ions. A color change from original pink to yellow-green for $Cu^{2+}$, yellow for $Ni^{2+}$ and brown for $Ag^+$ could be noticed without the addition of any external reagent (Fig. 7c).

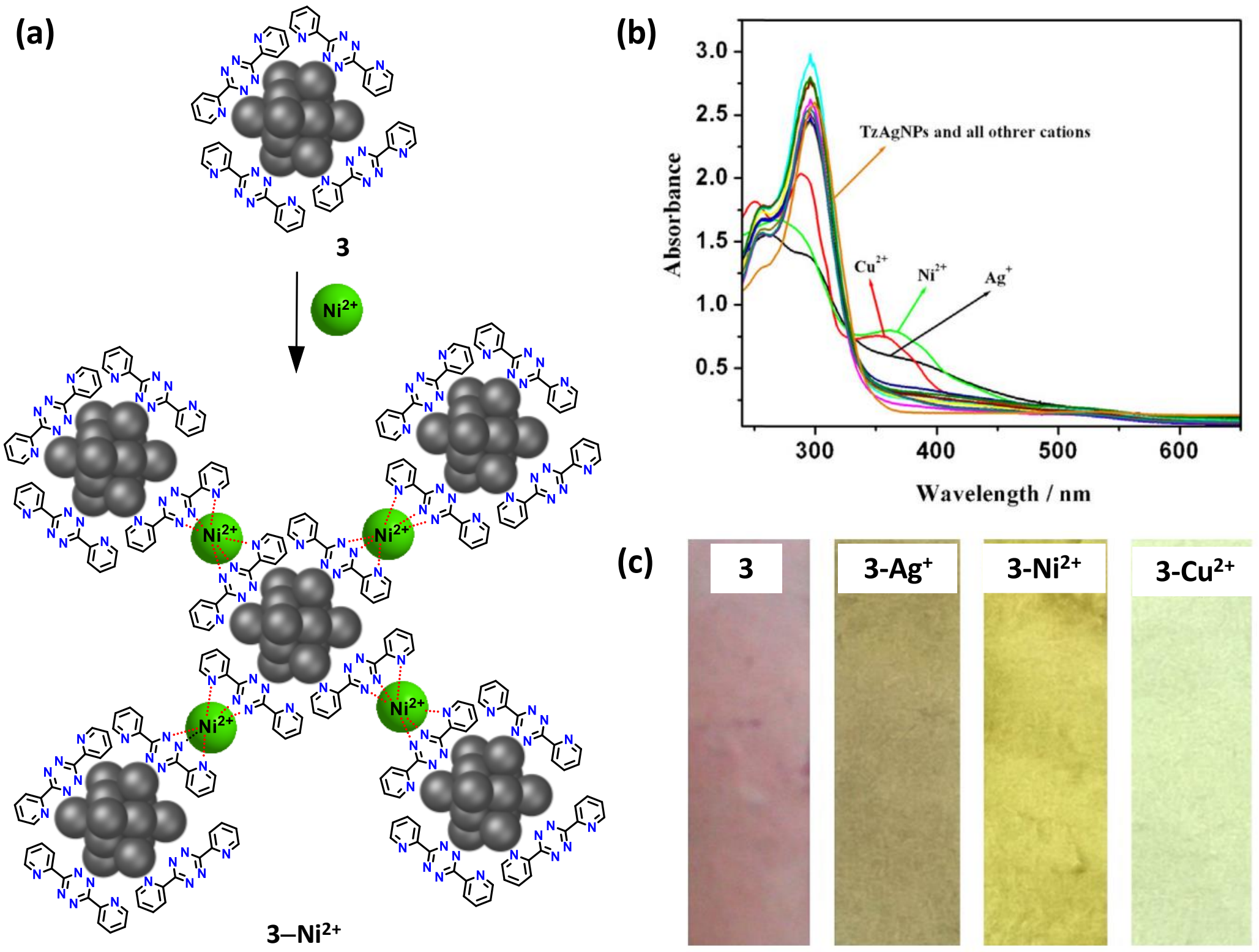

**Fig. 7**. (a) Binding mode of AgNPs-based probe **3** with $Ni^{2+}$ followed by the aggregation of NPs; (b) UV-Vis spectra of probe **3** in presence of assorted metal ions and (c) Photographs of test paper strips with $Ag^{+}$, $Ni^{2+}$ and $Cu^{2+}$ ions. Adapted with permission from Ref [109]. Copyright 2014, Elsevier.

Taking advantage of silver ions reducing characteristic and strong metal ion binding ability of phenols, Anthony's group [110] (in 2015) used a phenol-based *N*-(2-hyrdoxybenzyl)-isopropylamine ligand to synthesize AgNPs probe **4** (Fig. 8a). Interestingly, this ligand served both as a reducing and surface functionalizing agent to provide AgNPs-based colorimetric assays for metal ions and anions. UV-Vis spectrum of **4** in aqueous phase displayed the typical SPR band at 408 nm. High resolution TEM (HR-TEM) analyses clearly revealed the presence of spherical polydispersed NPs with an average particle size ranging from 10 nm to 30 nm. The colorimetric sensing ability of probe **4** was tested against $10^{-3}$ M of different metal ions (e.g., $Cr^{3+}$, $Al^{3+}$, $Fe^{3+}$, $Co^{2+}$, $Ni^{2+}$, $Cu^{2+}$, $Zn^{2+}$, $Ca^{2+}$, $Mg^{2+}$, $Cd^{2+}$, $Hg^{2+}$ and $Pb^{2+}$) as well as anions (e.g., $F^-$, $Cl^-$, $Br^-$, $I^-$, $NO_2^-$, $NO_3^-$, $SO_3^{2-}$, $SO_4^{2-}$, $CO_3^{2-}$, $(COO)_2^{2-}$, $H_2PO_4^-$, and $SCN^-$). It is worth mentioning here that some anions may also act as interfering species, especially when the metal ion sensing experiments are carried out in real samples. However, so far, the reported $Ni^{2+}$-responsive nanoprobes have been rarely investigated for anion sensing. The probe **4** exhibited different color changes towards $Ni^{2+}$ (red wine), $Cd^{2+}$ (orange), $Co^{2+}$ (pink), $Ca^{2+}$ (orange), $(COO)_2^{2-}$ (dark grey), $H_2PO_4^-$ (light grey), and $HPO_4^{2-}$ (orange) ions. Interestingly, upon the gradual addition of $Ni^{2+}$ to **4**, the intensity of SPR absorption at 408 nm declined with the appearance of a new intense absorption band near 580 nm. HR-TEM analysis further supported the aggregation of NPs induced by $Ni^{2+}$ ions (Fig. 8b). A similar band at a slightly lower wavelength (i.e., 570 nm) was observed for $Co^{2+}$ but with relatively less intensity as compared to the band observed for $Ni^{2+}$ ions. The absorption band intensity was even weaker for $Ca^{2+}$ and $Cd^{2+}$ ions. Furthermore, the addition of $Co^{2+}$ in the solution of adduct **4**-$Ni^{2+}$ produced no spectral change. In contrast, the addition of $Ni^{2+}$ in **4**-$Co^{2+}$ solution displayed a strong absorption band at longer wavelengths with an obvious solution color change. These data clearly suggested the preferential binding of probe **4** with $Ni^{2+}$ over other assorted metal ions. Therefore, the selective colorimetric response of **4** for $Ni^{2+}$ has been fully attributed to the aggregation event induced by $Ni^{2+}$ coordination.

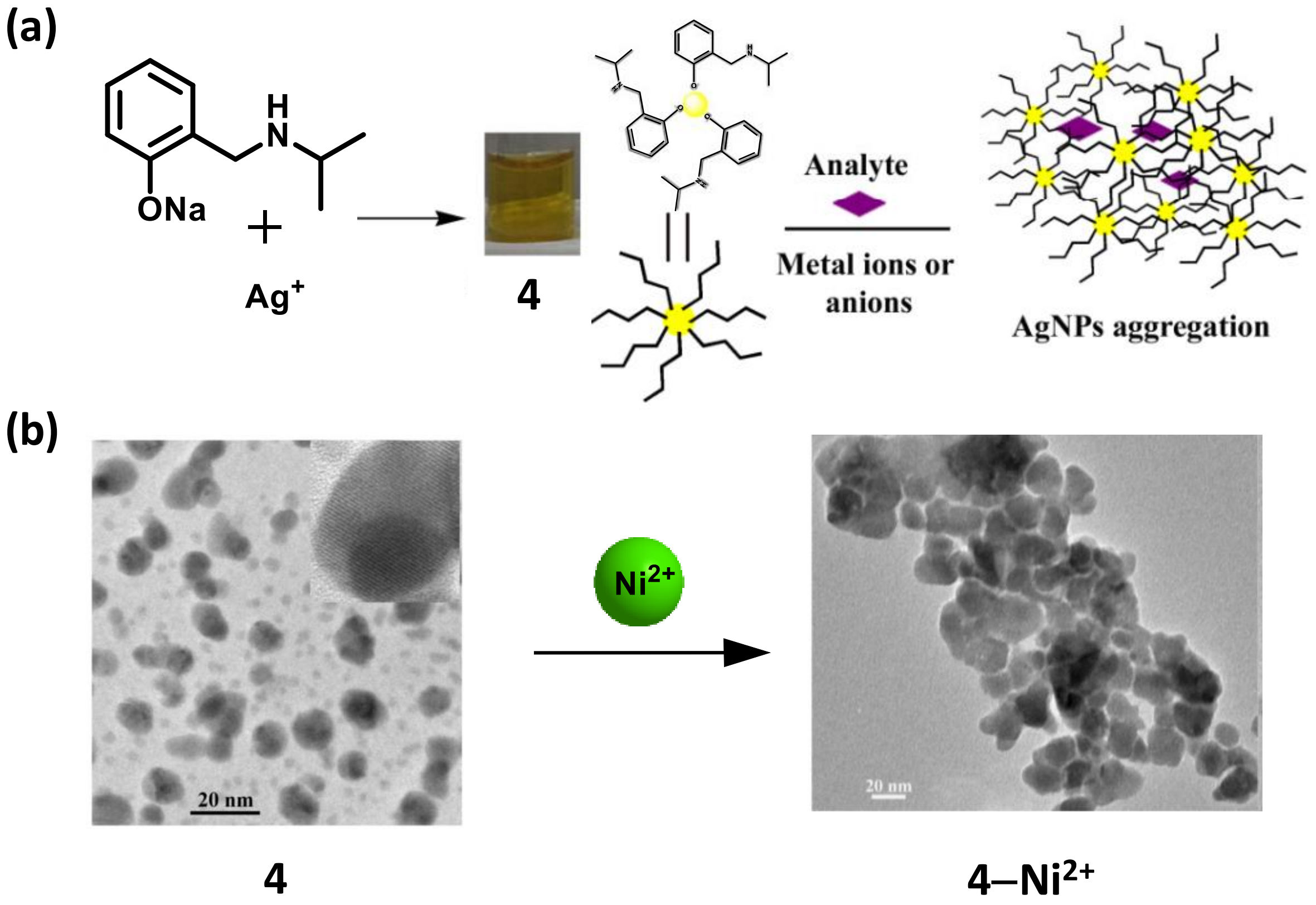

**Fig. 8**. (a) Binding mechainsm of AgNPs-based probe **4** with $Ni^{2+}$ ions followed by aggregation and (b) TEM images of probe **4** with and without $Ni^{2+}$ ions. Adapted with permission from Ref [110]. Copyright 2014, Elsevier.

Over the preceding years, triangular silver nanoprisms (AgNPrs) have also received increasing attention from the researchers for their potential applications in simultaneous sensing of anions and metal ions. Such NPrs contain three sharp vertices that cause these metallic nanoparticles to be unstable due to their high surface energy resulting from Gibbs-Thomson's effect [111]. Some anions, including halides, can etch the sharp vertices of triangular AgNPrs and change their morphology to a circular shape. When $Cl^-$, $Br^-$, $I^-$, $H_2PO_4^-$, or $SCN^-$ ions are added to a solution of triangular AgNPrs, the ions first combine with the facet surface, which has the highest crystal surface energy and is not entirely protected to form a precipitant with $Ag^+$. In this way, triangular AgNPrs are etched to give circular nanodisks. The morphological change caused by this selective etching results in a larger LSPR blue shift. This facet etching effect of triangular AgNPrs has been applied for the determination of several anions ($Cl^-$, $Br^-$, $I^-$, $H_2PO_4^-$ or $SCN^-$) and cysteine [85,112–115]. On the contrary, different metal ions act as anti-etching agents and inhibit the etching effect exhibited by anions to freeze the color and the shape of the nanoprisms. As a consequence, the anti-etching metal ions may easily be visualized by the naked eye or can be analysed by UV-Vis studies [36,85].

In 2016, Chen *et* al. [36] demonstrated triangular-shaped AgNPrs (**5**) stabilized by GSH as a highly selective colorimetric probe for $Ni^{2+}$ ions. Probe **5** displayed the etching effect in presence of $I^-$ ions. This is due to the fact that GSH causes less steric hindrance to $I^-$ which could easily associate with the vertices and edges of **5**, thus leading to AgNPrs etching. The shape of **5** was altered from nanoprism to nanodisk after the etching effect. Addition of $Ni^{2+}$ to **5** promoted the catalytic oxidation of GSH by $O_2$ to afford the dimer GSSG (containing -S-S-bond) that caused significant steric hindrance to $I^-$ ions to ultimately inhibit the etching effect (Fig. 9(a)). UV-Vis spectrum of AgNPrs **5** exhibited three absorption peaks near 330 nm, 450 nm, and 660 nm. Individual addition of either $Ni^{2+}$ or glutathione caused a negligible change in color or absorption bands of **5**, indicating no etching in the presence of these species (Fig. 9(b)). On the other hand, UV-Vis spectrum of iodide-incubated **5** showed the blue-shift of original 660 nm band to 493 nm along with an evident color change from blue to red due to the etching of **5**. The detection limit has been depicted to be 50.0 nM by naked eye visualization and 5.0 nM by UV-Vis spectroscopy, far less than the LoD values of other reported $Ni^{2+}$ probes based on noble MNPs. Although this probe exhibited an excellent LoD and high selectivity for $Ni^{2+}$, the presence of a strong reductive environment (such as $NaBH_4$) led to a remarkably decreased recognition efficiency of probe **5**. The reason lies behind the fact that GSSG reduced back under reductive environment to ultimately induce AgNPrs etching by iodide ions. Hence, the employment of this probe must be avoided under a strongly reductive environment.

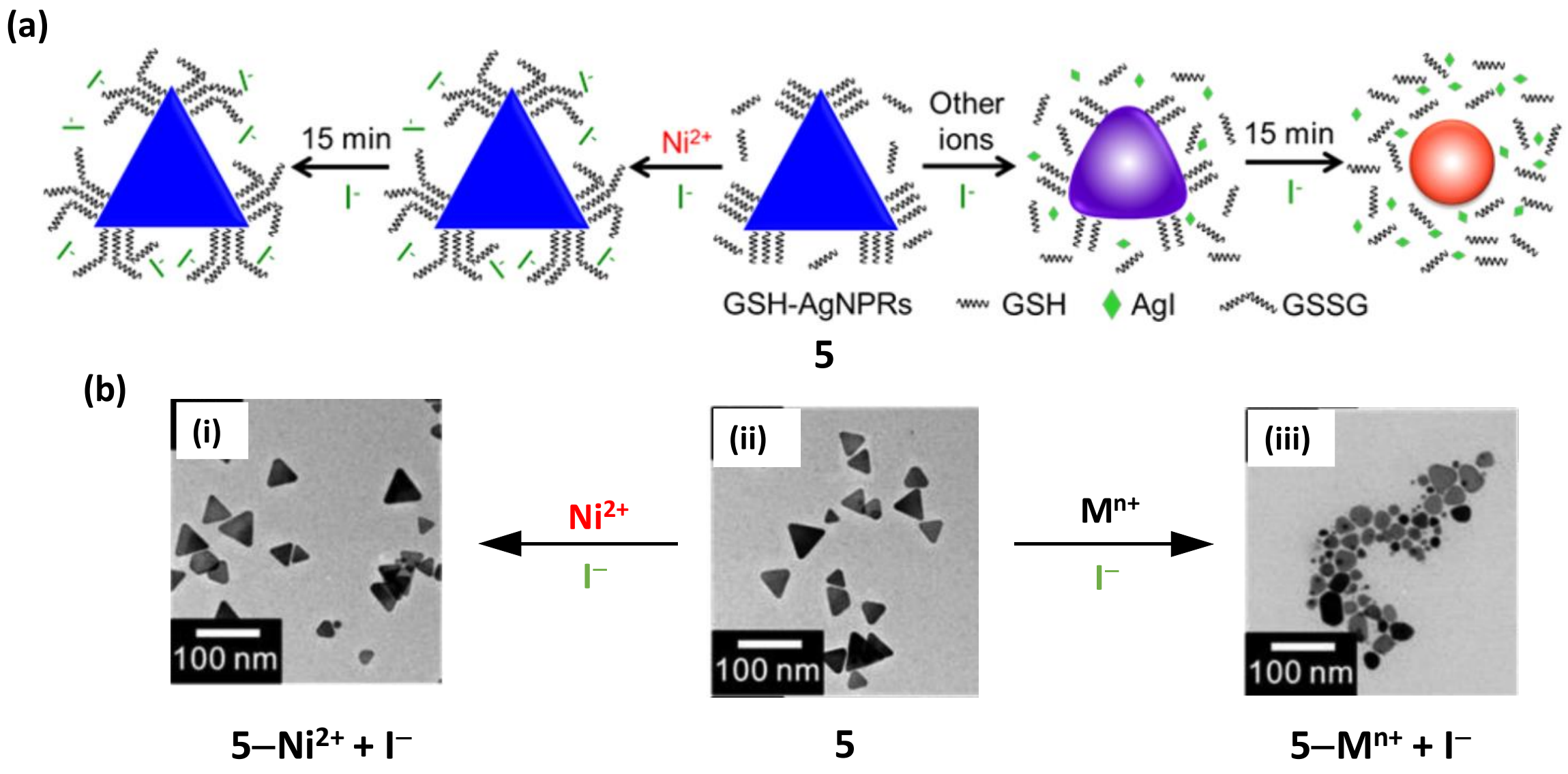

**Fig. 9.** (a) Mechanism of AgNPrs-based $Ni^{2+}$ probe **5** and (b) TEM images of (i) **5** + $Ni^{2+}$ + $I^-$, (ii) **5**, and (iii) **5** + $M^{n+}$ + $I^-$ (where $M^{n+}$ stands for other metal ions). Adapted with permission from Ref [36]. Copyright 2016, American Chemical Society.

So far, only one specific ligand could be employed as a stabilizer and recognition unit to develop $Ni^{2+}$-responsive AgNPs; however, the co-functionalization with more than one ligand may sometimes result in an improved $Ni^{2+}$ sensing ability of AgNPs. For the first time, in 2017, Feng *et* al. [116] developed a AgNPs-based colorimetric nanoprobe **6** for selective recognition of $Ni^{2+}$ using two specific ligands, i.e. AMP (i.e., adenosine monophosphate) and SDS (i.e., sodium dodecyl sulfonate) as stabilizers. Both the ligands bind to NPs *via* sulfonate and amine functionalities, and they are known to display cooperative binding for $Ni^{2+}$ ions. Upon addition of $Ni^{2+}$, the absorption band of **6** at 396 nm (i.e., SPR band) declined and a new peak emerged at 515 nm. These spectral changes were ascribed to $Ni^{2+}$-induced aggregation of **6**, arising most likely from the synergistic electrostatic interaction between SDS and $Ni^{2+}$, as well as the coordinative binding of $Ni^{2+}$ with AMP fragment. This aggregation was also accompanied with a drastic color change in probe's solution (from dark yellow to orange). The average hydrodynamic size of **6**, measured by DLS (Dynamic Light Scattering) analysis, was found to be increased from 9.8 nm to 58.3 nm upon $Ni^{2+}$ addition, which was further corroborated through TEM analysis. Optimization of the sensing experiments revealed the best possible functioning of probe **6** in the concentration range of $Ni^{2+}$ as 4.0-60 μM, and the LoD value was calculated as 0.60 μM. The authors reported that the co-functionalized **6** displayed apparent advantages over the probe grafted solely with SDS or AMP, in terms of metal ion selectivity.

In 2018, Mochi *et* al. [117] synthesized AgNPs-based optical probe **7**, equipped with 3-mercapto-propane sulfonic acid as the capping agent, for selective detection of $Ni^{2+}$ and $Co^{2+}$ ions in aqueous solution (Fig. 10). The sensing ability of **7** was investigated using UV-Vis study by monitoring the changes observed in its SPR band in presence of the above-mentioned metal ions. Upon increasing the concentration of both ($Ni^{2+}$ and $Co^{2+}$) the metal ions in probe's solution, the SPR band red-shifted and broadened. In comparison to $Co^{2+}$, the addition of $Ni^{2+}$ caused relatively more significant changes in the wavelength as well as the shape of SPR band of **7**. It was clearly evident from the remarkable extent of red-shift in **7** upon introduction of metal ions (36.0 nm for 2.0 ppm $Ni^{2+}$; 17.0 nm for 3.0 ppm $Co^{2+}$). The specific detection of **7** towards the target metal ions has been obtained upto 500 ppb. The optical changes were intense enough to produce a colorimetric response (visible to the naked eye), even at 1.0 ppm concentration of target metal ion. Based on the results obtained, the high selectivity of the probe has been ascribed to the strong coordination ability of $Ni^{2+}$ and $Co^{2+}$ (in the range of 0.5-2.0 ppm) with surface functionalities of the probe.

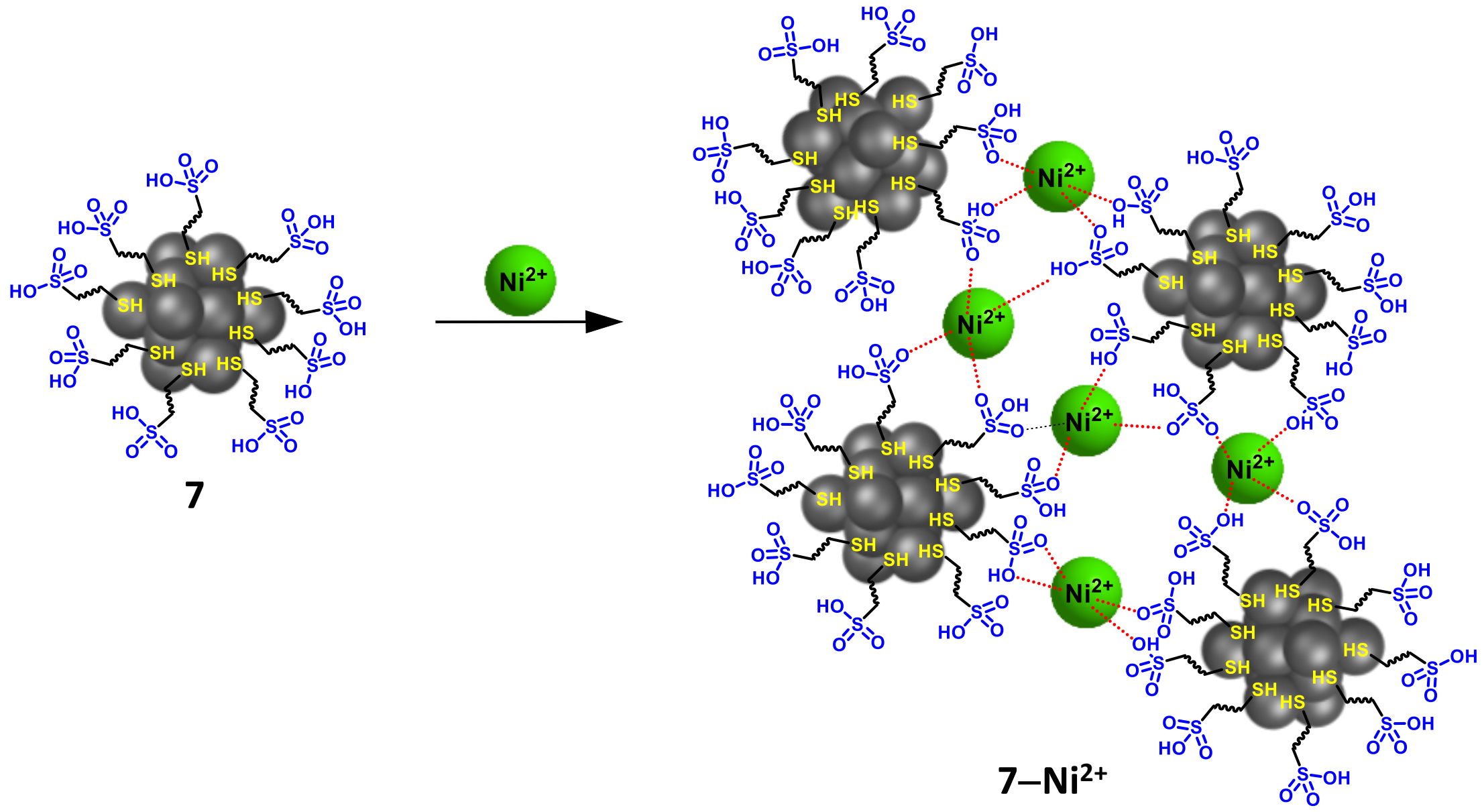

**Fig. 10.** Binding mode of AgNPs-based probe **7** with $Ni^{2+}$ followed by the aggregation process.

In year 2019, Yoon *et* al. [85] developed a highly selective colorimetric method for the detection of $Ni^{2+}$ by employing hydrogen peroxide etching of triangular silver nanoprisms-based probe **8**. The coordination complex of $Ni^{2+}$ with the citrate ions catalyzed the formation of $H_2O_2$ in the colloidal solution of **8**. The obtained $H_2O_2$ then attacks the sharp edges of **8** to eventually afford the circular Ag nanodisks (Fig. 11). UV-Vis spectrum of **8** revealed three bands at 335 nm, 475 nm, and 750 nm. The bands at 335 nm and 475 nm were attributed to the

*out*- and *in*-plane quadrupole plasmon resonances while the band at 750 nm was due to *in*-plane dipole plasmon resonance band. The addition of $Li^{+}$, $Na^{+}$, $K^{+}$, $Ag^{+}$, $Ba^{2+}$, $Ca^{2+}$, $Cd^{2+}$, $Co^{2+}$, $Cu^{2+}$, $Hg^{2+}$, $Ga^{2+}$, $Mn^{2+}$, $Mg^{2+}$, $Pb^{2+}$, $Sn^{2+}$, $Zn^{2+}$, $Al^{3+}$, $As^{3+}$, $Fe^{3+}$, $Ti^{3+}$, $Ge^{4+}$, $Cr^{6+}$, $NO_2^{-}$, $NO_3^{-}$, $PO_4^{-}$, $SO_4^{2-}$ and $F^{-}$ produced no spectral changes in **8**. However, $Ni^{2+}$ addition to **8** induced a substantial change in the absorption bands coupled with an obvious color change in probe's suspension. The absorption band at 750 nm was found to be highly sensitive to the edge length and the thickness of AgNPrs. Consequently, the broad resonance at 750 nm completely disappeared with the successive addition of $Ni^{2+}$ with a subsequent increase in the absorption intensity of the band near 480 nm. These data exhibited that the sensing of $Ni^{2+}$ using probe **8** was found to be particularly sensitive to nano-prism shape. The LoD value for $Ni^{2+}$ was as low as 21.6 nM and other competing species failed to interfere in the selective detection of $Ni^{2+}$ by **8**. The probe was also able to detect $Ni^{2+}$ in spiked tap and pond water samples, and the results were in full agreement with the data obtained from ICP-OES analysis (inductively-coupled-plasma-optical-emission-spectrometry).

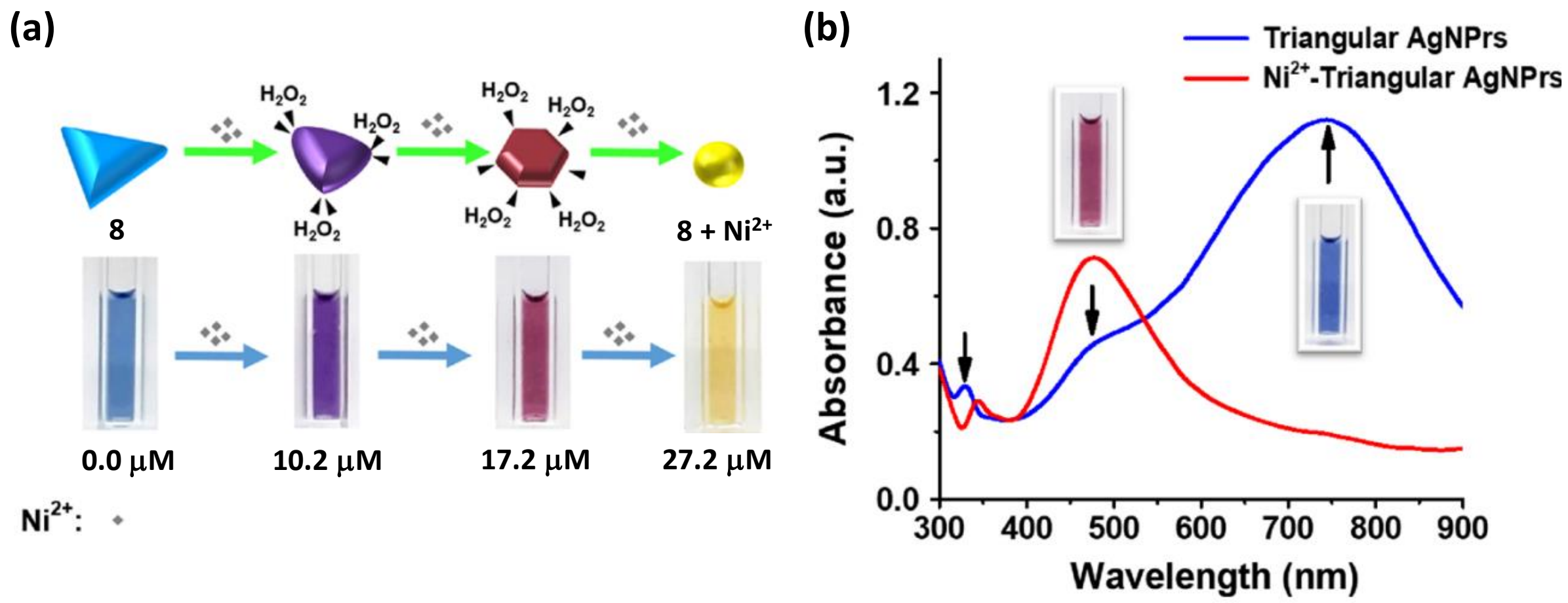

**Fig. 11.** (a) Morphological changes and the colour change in AgNPrs-based probe **8** upon addition of different concentration of $Ni^{2+}$ ions and (b) Variation in the LSPR absorption band of probe **8** in the presence and absense of $Ni^{2+}$ ions. Adapted with permission from Ref [85]. Copyright 2019, Elsevier.

In the same year, Almaquer *et* al. [118] reported the development of a simple colorimetric assay for $Ni^{2+}$ recognition in aqueous solution using citrate-stabilized AgNPs (**9**) aided with carboxylate group (Fig. 12). The process of the nanoparticle synthesis was much simplified with no post-synthesis modifications. The monodispersed nanoparticles **9** displayed a SPR band at 392 nm. The average size of the nanoparticles was calculated as 10-14 nm using full-width-at-half-maximum (FWHM) method, and this size of NPs was further verified by TEM analysis. The metal ions such as $Li^{+}$, $Mg^{2+}$, $Ca^{2+}$, and $Ag^{+}$ exhibited nominal changes in the optical behavior of **9**. However, upon $Ni^{2+}$ addition, a color change could be observed from yellow to orange with a broadening in the SPR absorption of **9**. The changes were attributed to the aggregation of NPs induced by the compelling interaction of $Ni^{2+}$ with the carboxylate and hydroxyl sites of the citrate moiety. At higher concentrations of $Ni^{2+}$, the color change was much more intense, and the SPR band was relatively more broadened due to the highly reduced interparticle distance. The LoD was calculated to be 0.75 mM for $Ni^{2+}$ ions. Probe **9** was also investigated for the detection of $Ni^{2+}$ in real water samples such as spiked tap water and distilled water samples.

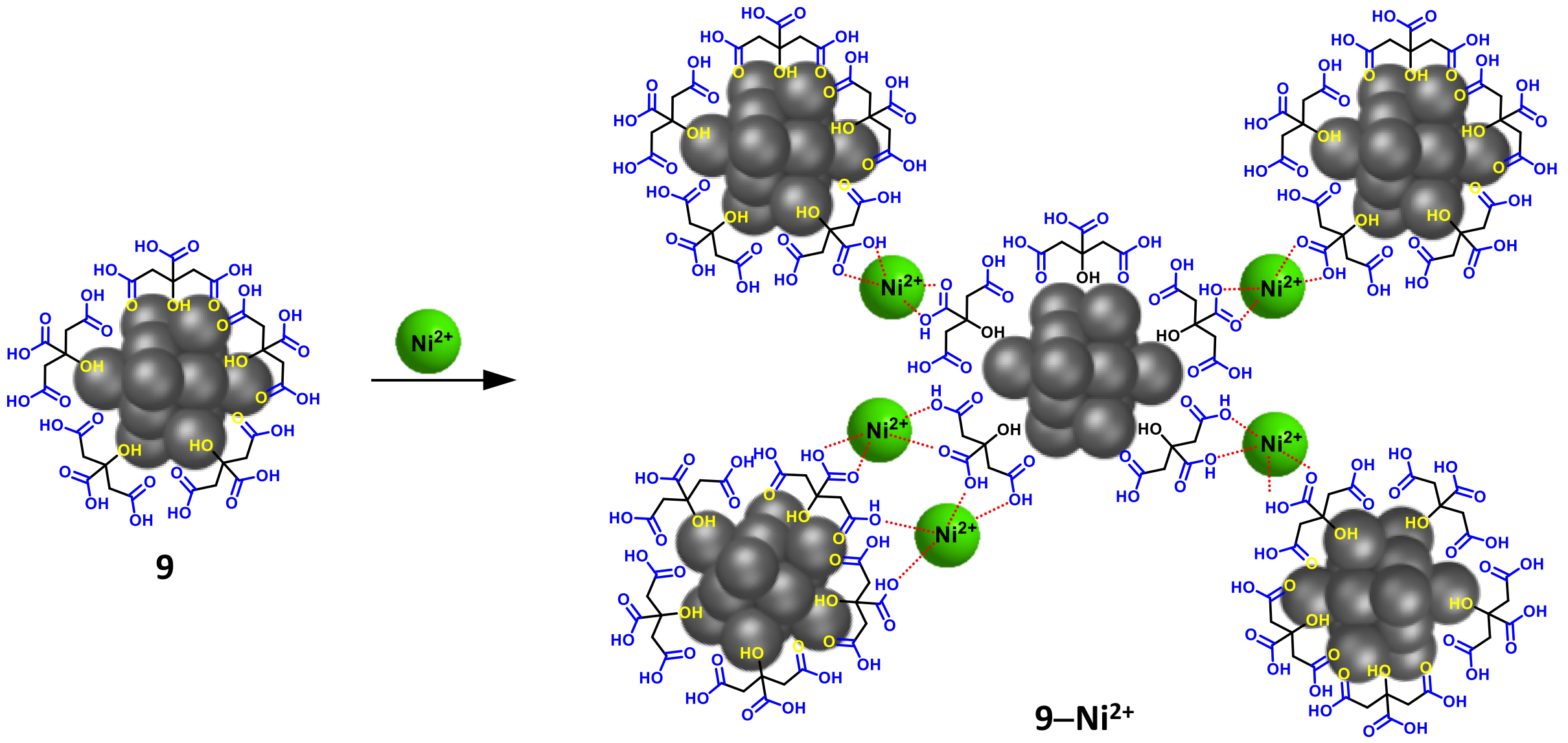

**Fig. 12**. (a) Binding mode of AgNPs-based probe **9** with $Ni^{2+}$ followed by its aggregation.

In recent years, the utilization of microorganisms and plant-based extracts has emerged as a viable approach to design and develop colorimetric nanosensors. The reason lies behind the fact that these natural biomaterials include several important bioactive components (such as bioactive polyphenol, phenolic acid, and alkaloids), which can act both as reducing as well as stabilizing agents for resulting nanomaterials. Such a green synthetic method leads to a reduction of toxic chemicals in synthetic procedures whereby NPs having specific characteristics can be produced. Such an approach displays synthetic-ease, cost-effective, and eco-friendly features. Different factors such as the plant extract component, reductant concentration, reaction time and temperature, and the pH of the medium markedly influence the size, quality, and morphology of the resultant NPs.

*Citrullus lanatus* (watermelon) peel and rind extract, an agricultural waste, is well known to serve as a reductant as well as a stabilizer to afford MNPs. The watermelon's rind and pulp do contain thiamine, niacin, riboflavin, and ascorbic acid. In 2019, Khwannimit *et* al. [119] presented the green synthesis of AgNPs probe **10** using *Citrullus lanatus* extract, which acted as both reductant and the stabilizer. The surface of probe **10** was modified with *L*-cysteine (Cys-AgNPs) to attain the selective detection of metal ions (Fig. 13). The -SH group in *L*-cysteine showed high affinity for AgNPs, whereas -COOH and -$NH_2$ groups were made available to potentially bind the targeted metal ions. The SPR absorption band for **10** appeared at 415 nm. The NPs were spherical in shape, and the average particle size was depicted to be 20-40 nm as evidenced by the SEM analyses. To attain the deprotonation of -COOH and -$NH_2$ groups, the sensing experiments have been carried out at pH 8.0 (slightly alkaline medium) with 10 mins incubation time. Upon addition of $Ni^{2+}$ to the suspension of **10**, a sharp color change from yellow to orange could be realized, which was accompanied by the emergence of a new SPR band at 455 nm. These changes have been ascribed to the typical aggregation event of the NPs in the presence of $Ni^{2+}$. The theoretical calculations have also been performed to gain deep insights into the cation-responsive behavior of probe **10**.

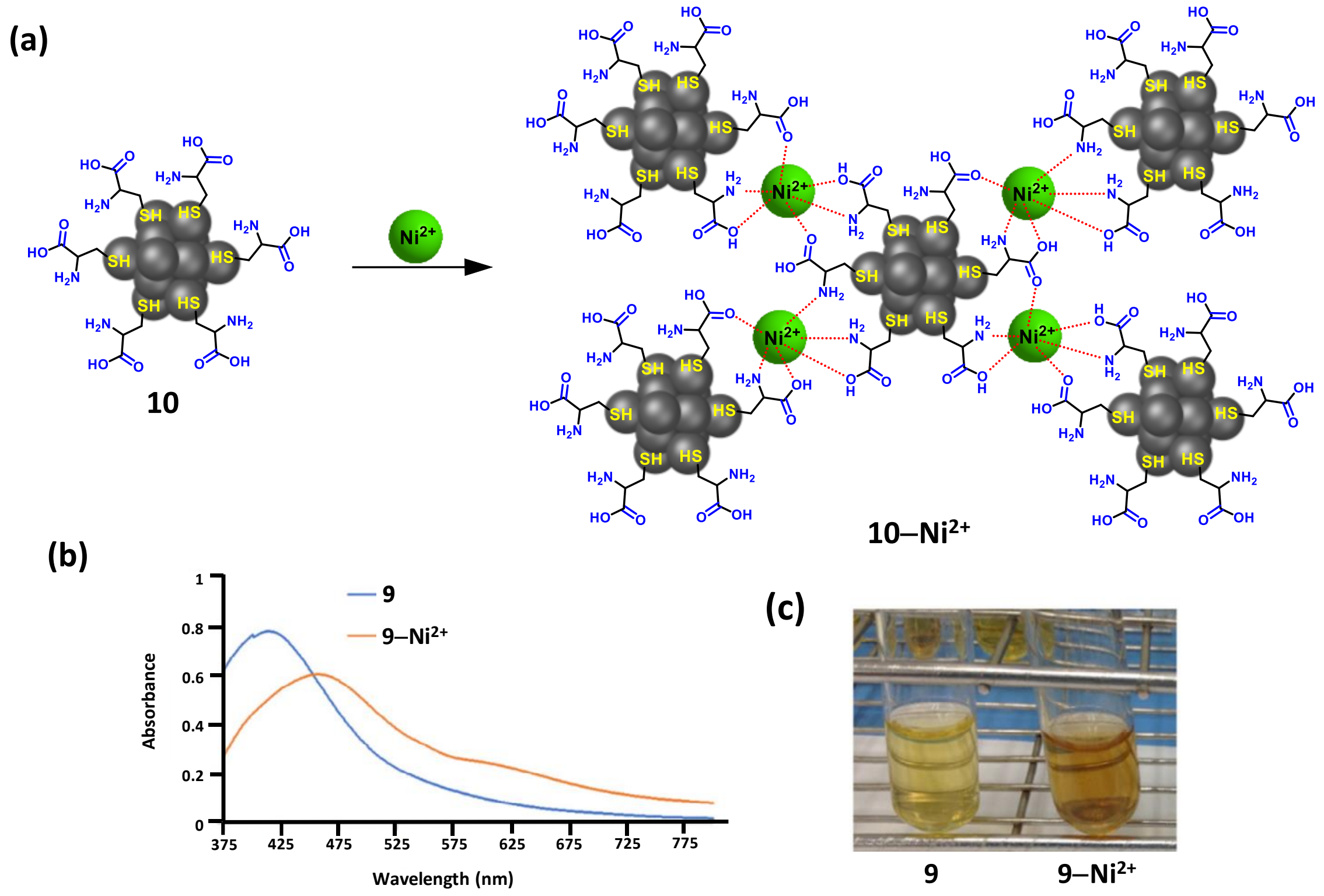

**Fig. 13.**(a) Binding mode of AgNPs-based probe **10** with $Ni^{2+}$ followed by its aggregation; (b) Variation in the SPR band upon addition of $Ni^{2+}$ ion in probe **10** and (c) Colorimetric changes in probe **10** after the addition of $Ni^{2+}$ ions. Adapted with permission from Ref [119]. Copyright 2019, Elsevier.

Acknowledging the fact that the green synthesis of nanoparticles is extremely important and relevant for green and sustainable chemistry, another AgNPs probe **11** was synthesized by Aravind *et* al. [120] utilising cysteine-based *Brassica oleracea var. Italica (BI)* extract. The synthetic process was microwave-assisted, which resulted in an excellent yield of NPs without any external capping agent. *BI* is a vegetable belonging to the Brassica family, and its extract contains different biomolecules such as carotenoids, Vit. C, calcium, fiber, folate, and amino acids. Due to the presence of flavonoids and polyphenols in its extract, *Brassica oleracea var. Italica* has been utilized as an $Ag^{+}$ reductant as well as a stabilizer to give AgNPs probe **11**. The brown aqueous suspension of **11** exhibited a narrow and smooth SPR absorption near 402 nm. The cation sensing experiments for **11** have been carried out at the optimum pH (pH 7.0) and the introduction of various cations including $Cu^{2+}$, $Cr^{3+}$, $Hg^{2+}$, $Zn^{2+}$, $Cd^{2+}$, $Fe^{2+}$, $Pb^{2+}$, and $Co^{2+}$ produced no remarkable changes in UV-Vis spectra of **11**. However, the addition of $Ni^{2+}$ completely diminished the SPR band with a new band emerging at 290 nm. Typically, the spectral alterations have been ascribed to the aggregation of NPs in the presence of $Ni^{2+}$ ions. The aggregation process was further corroborated by TEM studies, where the particle size increased from 12.8 nm to 21.2 nm upon $Ni^{2+}$ addition. A strong colorimetric response was also observed and a striking solution color change of **11** from brown to orange could be realized upon $Ni^{2+}$ addition. Furthermore, the luminescence intensity of **11** exhibited a strong enhancement at 580 nm in presence of $Ni^{2+}$ ions. This probe was also capable of detecting $Ni^{2+}$ in lake water, steel, and electroplating industry with an excellent recovery of 99%, and the limit of detection was depicted to be 1.816 µM.

In subsequent year (2021), Rossi *et* al. [121] demonstrated the synthesis of probe **12** by functionalization of AgNPs with mercaptoundecanoic acid (Fig. 14). The SEM studies revealed that the spherical shaped nanoparticles had an average diameter of 20 nm. The SPR band for **12** shifted from 417 nm to 477 nm upon the addition of $Ni^{2+}$ ions. These spectral alterations were associated with a color change of the probe's solution from yellow to purple (visible to the naked eye). This report suggested the clustering of NPs while $Ni^{2+}$ ions acted as a bridge between the two carboxylate groups of the mercaptoundecanoic acid. Upon further increasing $Ni^{2+}$ concentration, the absorption band at 477 nm intensified, and simultaneously, the original SPR band at 417 nm diminished completely. The bridging of the nanoparticles by $Ni^{2+}$ ions led to the formation of a superlattice that collapsed after reaching a critical mass at a higher concentration of analytes. The clustering or collapse effects of NPs with or without $Ni^{2+}$ could easily be analysed using SEM studies. Other metal ions like $Mn^{2+}$, $Co^{2+}$, $Cd^{2+}$, $Zn^{2+}$, $Fe^{2+}$, $Pb^{2+}$, $Hg^{2+}$, and $Cr^{3+}$ failed to produce any change in UV-Vis spectra of **12**, and the related LoD value for $Ni^{2+}$ was calculated as 2.15 μM.

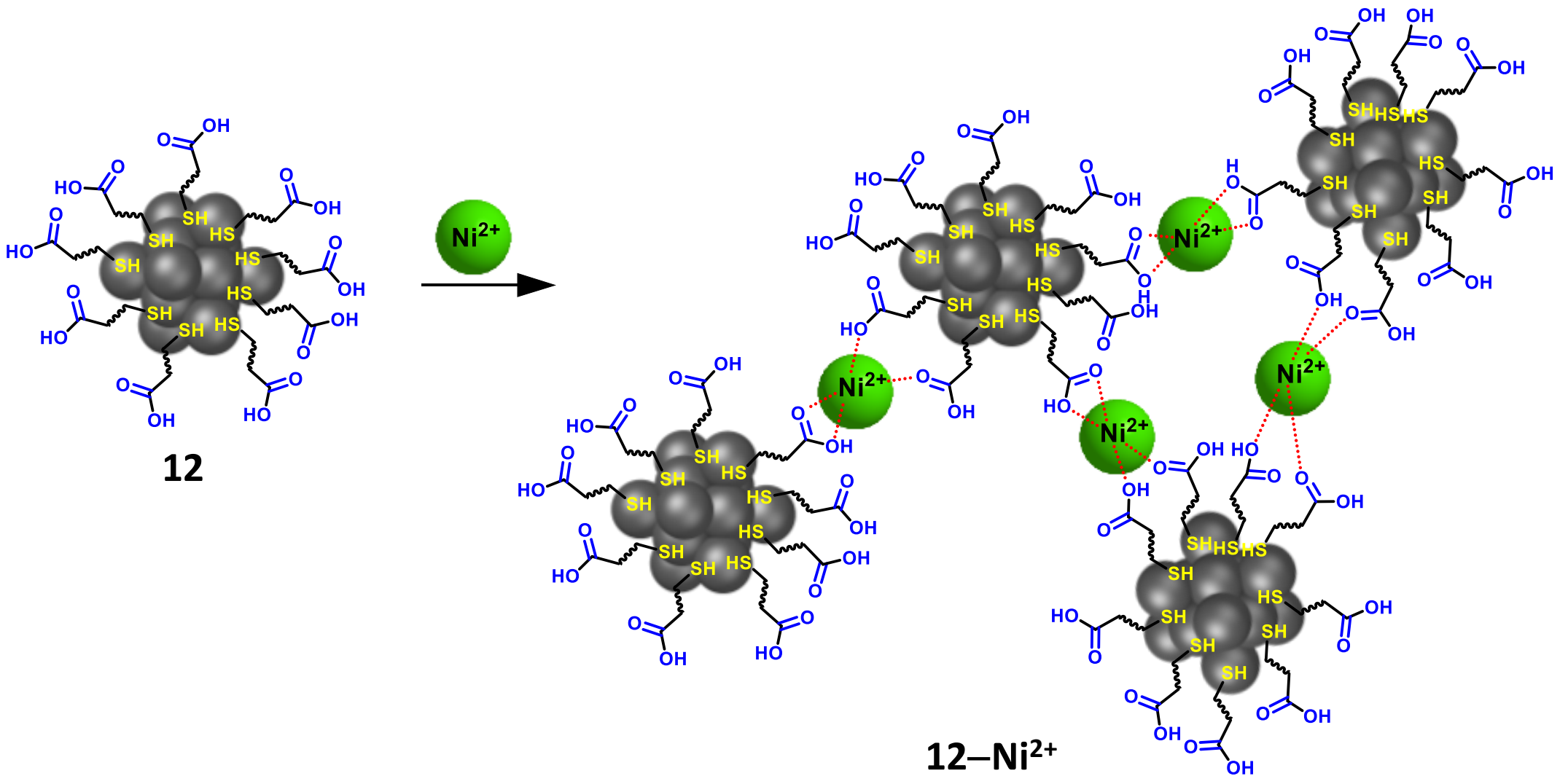

**Fig. 14.** Binding mode of AgNPs-based probe **12** with $Ni^{2+}$ ions followed by its aggregation**.**

In 2022, Ghorbani's group [122] reported a AgNPs-based probe **13** utilizing the extract of green walnut husk (*Juglans Regia L.*) as a reducing and stabilizing agent containing natural phenolic content onto the surface of NPs (Fig. 15). The phenolic groups served as the receptor unit for the detection of target metal ions. UV-Vis analysis of the brownish-yellow aqueous suspension of **13** displayed the SPR absorption centered at 445 nm. The average size of the spherical nanoparticles was determined to be 26.40 nm using TEM and SEM analyses. The addition of $Cd^{2+}$ and $Ni^{2+}$ to **13** led to a decrease in the SPR band along with a bathochromic shift, which could be attributed to the aggregation of NPs. The sedimentation was observed especially in case of $Ni^{2+}$ ions. The polyphenolic moieties present over the surface of NPs could chelate the metal ions through their hydroxyl and carboxyl groups. Other ionic species such as $Na^{+}$, $K^{+}$, $Ca^{2+}$, $Cu^{2+}$, $Mn^{2+}$, $Hg^{2+}$, $Fe^{2+}$, $PO_4^{-}$, $AsO_4^{-}$, $SO_4^{2-}$, $CO_3^{2-}$, $HCO_3^{-}$, $NO_3^{-}$, and $Cl^{-}$ produced only nominal changes in the SPR absorption. The optimum pH for the reaction was reported to be 6.0, and the LoD value was calculated as low as 0.2 nM. As a proof of concept, probe **13** was also used to detect $Ni^{2+}$ ions in the spiked river water with excellent recoveries from 90.57 to 113.61%.

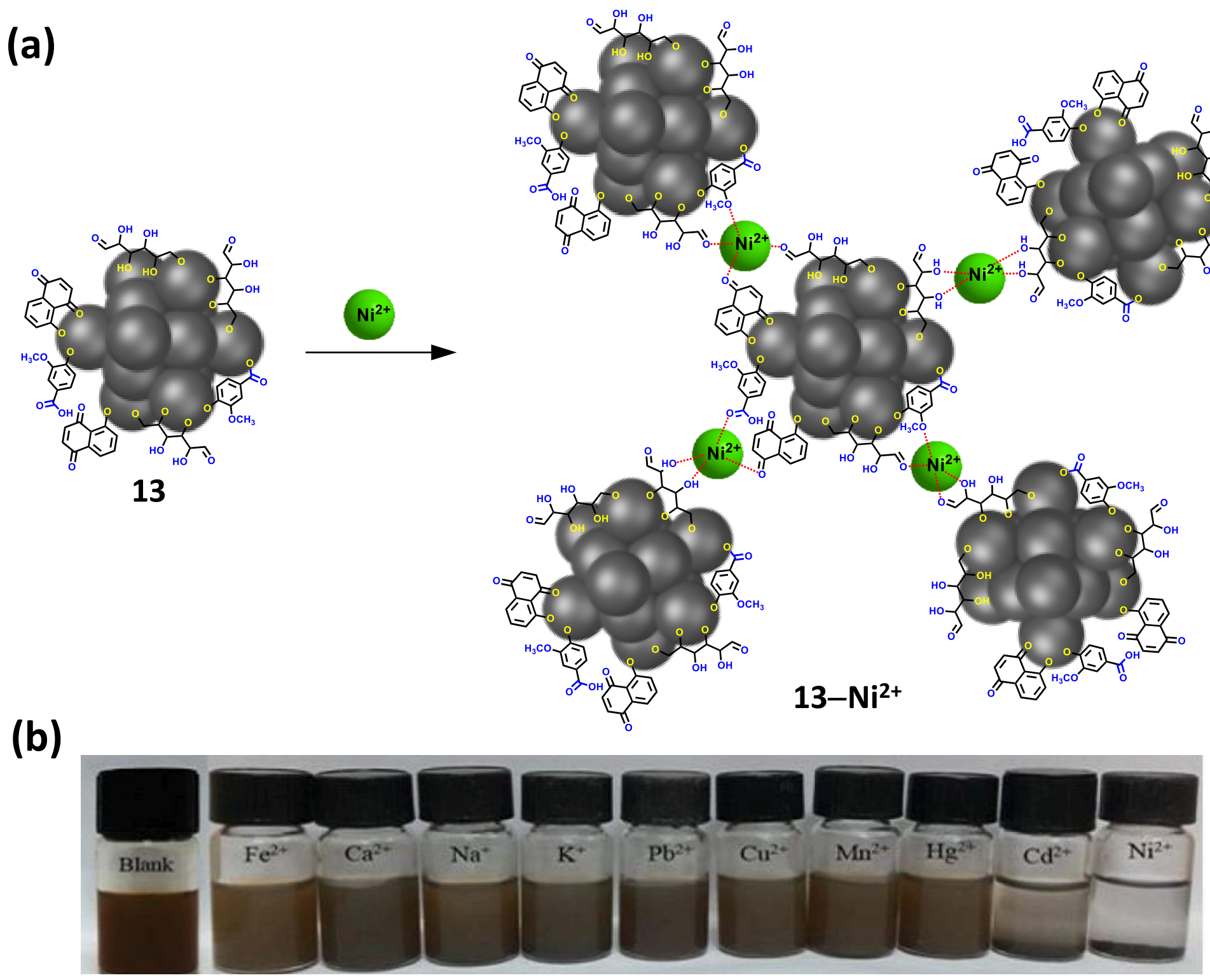

**Fig. 15. (a)** Binding mode of AgNPs-based probe **13** with $Ni^{2+}$ ions followed by aggregation and (b) Coloriemtric changes in AgNPs probe **13** with various metal ions. Adapted with permission from Ref [122]. Copyright 2019, Elsevier.

**Table 1**. Silver nanoparticles (AgNPs) based optical probes for detection of $Ni^{2+}$ ions.

| Probe | $\lambda_{abs}$ (nm) (SPR) | Particle size (nm) | Stabilizing agent | Target ions | LoD (µM) | Colorimetric effect | Solvent | Practical application | [Ref.] |
|---|---|---|---|---|---|---|---|---|---|
| 1 | 396 | 8 | Glutathione | $Co^{2+}$, $\mathbf{Ni^{2+}}$ | 75 | Yellow to Deep-orange | Water | - | [35] |
| 2 | 390 | 12 | *N*-acetyl-*L*-cysteine | $\mathbf{Ni^{2+}}$ | 0.23 | Yellow to Deep-orange | Water | Tap water | [108] |
| 3 | 410 | 10-11 | 3,6-di(pyridine-2-yl)-1,2,4,5-s-tetrazine | $Ag^{+}$, $\mathbf{Ni^{2+}}$, $Cu^{2+}$ | $Cu^{2+}$ = 17.5, $\mathbf{Ni^{2+}}$ **= 9.4,** $Ag^{+}$ = 13.2 | Pink to Yellow | Water | Test strips | [109] |
| 4 | 408 | 25 | *N*-(2-hydroxybenzyl)-isopropylamine | $Co^{2+}$, $\mathbf{Ni^{2+}}$, $Cd^{2+}$, $Ca^{2+}$ | ≈1 | $Ni^{2+}$=Yellow to Red-wine $Co^{2+}$=Yellow to Pink $Cd^{2+}$=Yellow to Orange $Ca^{2+}$=Yellow to Orange | Water | - | [110] |
| **5** | 660 | - | Glutathione | $\mathbf{Ni^{2+}}$ | $5 \times 10^{-3}$ | Blue Color Retained | Water | Tap and lake water | [36] |
| **6** | 396 | 9.8 | Adenosine monophosphate, Sodium dodecyl sulphate | $\mathbf{Ni^{2+}}$ | 0.60 | Yellow to Orange | Water | Tap and lake water | [116] |
| **7** | 402 | 8 | 3-mercapto-1-propanesulfonic acid | $Co^{2+}$, $\mathbf{Ni^{2+}}$ | 500 | Yellow to Light grey | Water | - | [117] |
| **8** | 475 | 40.3 | citrate | $\mathbf{Ni^{2+}}$ | $21.6 \times 10^{-3}$ | Blue to Yellow | Water | Tap and pond water | [85] |
| **9** | 392 | 10.4 | citrate | $\mathbf{Ni^{2+}}$ | $0.75 \times 10^{-3}$ | Yellow to Orange | Water | Tap water | [118] |
| **10** | 415 | 30 | Citrullus lanatus, L-cysteine | $\mathbf{Ni^{2+}}$ | – | Yellow to Orange | Water | - | [119] |
| **11** | 402 | 12.8 | Brassica oleracea var. Italica | $\mathbf{Ni^{2+}}$ | 3.302 | Brown to Orange | Water | Lake water, steel and electroplating industry | [120] |
| **12** | 417 | 173.13 | Mercaptoundecanoic acid | $\mathbf{Ni^{2+}}$ | – | Yellow to Purple | Water | - | [121] |
| **13** | 445 | 26.40 | Green walnut extract of Juglans regia | $Cd^{2+}$, $\mathbf{Ni^{2+}}$ | $0.2 \times 10^{-3}$ | Brown-yellow to Pale-grey | Water | River water | [122] |

### *4.2. Gold nanoparticles (AuNPs)*

The choice of AuNPs to fabricate optical probes is always justified by their high stability toward oxidation, shape-dependent optical features, stability of dispersion, remarkable tunable size, relative ease of synthesis, and facile modification of surface with desirable functionality [98]. All such properties enable AuNPs to be a good choice among other metal-based NPs for a diverse range of applications, including optical device fabrication, SERS effect, bio-imaging and labeling, drug delivery, catalysis, etc. [91]. This section mainly covers most of the works reported on the usage of AuNPs for colorimetric detection of $Ni^{2+}$ ions. Colorimetric detection of metal ions using AuNPs has become a field of intense research activity since AuNPs offer detection of very low concentrations of target analytes. Color change of Au colloidal dispersion (visible to the naked eye) is directly associated with the change in the SPR band of AuNPs that typically occurs due to the variation in particle size, interparticle distance, or any surface modification during the aggregation process [95].

To the best of our knowledge, Graham's group reported [123], for the first time in 2012, the use of AuNPs for sensing $Ni^{2+}$ ions in the aqueous phase. They investigated the influence of the nanoparticles' size on the colorimetric and SERS based sensing of $Ni^{2+}$ ions. Probe **14** (average particle size: 14.0 nm) was stabilized by carboxylate containing citrate and thioctic acid groups, which further allowed a covalent linking of nitriloacetic and *L*-carnosine over the surface of AuNPs (Fig. 16). For comparison, a set of these nanoparticles was also grown up to 45.0 nm in diameter by seeded growth. In absence of $Ni^{2+}$, the nitrilotriacetic acid and **14** exhibited no affinity towards each other. Nevertheless, $Ni^{2+}$ addition induced a rapid aggregation of NPs changing the SERS intensity, responsible for the colorimetric response. The *L*-carnosine with nitrilotriacetic acid could coordinate with $Ni^{2+}$ using six histidine residues provided by *L*-carnosine [123,124]. The addition of 5.0 mM $Ni^{2+}$ to 40.0 pM dispersed aqueous solution of **14** resulted in a sharp color change from raspberry-red to purple. UV-Vis analysis was employed to investigate the response of $Ni^{2+}$ towards the probes of different sizes (14.0 nm and 45.0 nm) *via* monitoring the change in the absorption band at 700 nm. The maximum change was exhibited by 45.0 nm nanoparticles which was further confirmed by plotting the $Abs_{700}/Abs_{SPR}$. The change in concentration of $Ni^{2+}$ from the dimer formation at 5.0 ppm to large tetramer cluster at 50-60 ppm was clearly revealed by SEM analysis. The selective binding of **14** for $Ni^{2+}$ was further investigated in presence of competing cationic species such as $Co^{2+}$, $Fe^{3+}$, $Zn^{2+}$, and $Cu^{2+}$ ions. Except for $Cu^{2+}$, no ions displayed remarkable interference with the sensing process. The LoD value for $Ni^{2+}$ was reported as 0.5 ppm in this study.

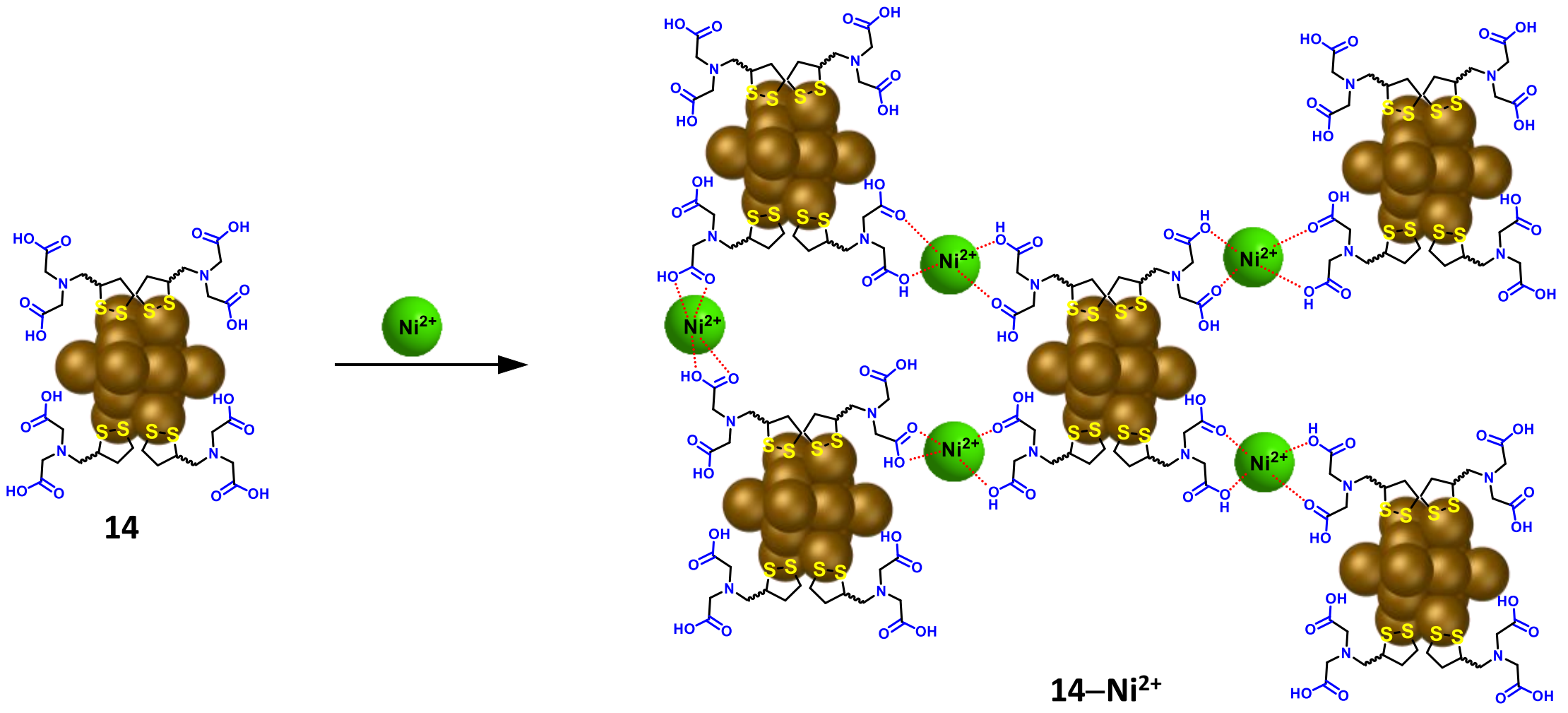

**Fig. 16.** Binding mode of AuNPs-based probe **14** with $Ni^{2+}$ ions followed by aggregation.

In the same year, a peptide-functionalized citrate-AuNPs probe **15** was reported by Ye and co-workers [125] which could simultaneously detect $Cd^{2+}$, $Ni^{2+}$, and $Co^{2+}$ ions. The binding between the AuNPs and the peptide was established *via* a thiol group present in the side-chain of cysteine (Fig. 17a). The aqueous suspension of AuNPs was red in color with a strong SPR absorption at 520 nm, as evidenced by UV-Vis measurements. The addition of $Cd^{2+}$, $Ni^{2+}$, and $Co^{2+}$ induced a color change from wine red to purple-blue along with a red-shift in the absorption in the range of 560–600 nm (Fig. 17b). Rest of the cations like $Ag^{+}$, $Na^{+}$, $K^{+}$, $Mg^{2+}$, $Zn^{2+}$, $Co^{2+}$, $Mn^{3+}$, $Ni^{2+}$, $Cu^{2+}$, $Hg^{2+}$, $Pb^{2+}$ and $Ba^{2+}$ did not show much change in the spectra and exhibited no colorimetric response. In comparison to $Ni^{2+}$ and $Co^{2+}$, the colorimetric response for $Cd^{2+}$ was relatively stronger, which could be measured by the $Abs_{600}/Abs_{520}$ ratio (Fig. 17c). The probe **15** was also able to detect $Cd^{2+}$, $Ni^{2+}$ and $Co^{2+}$ in the river water with a recovery of above 95%. For individual detection of cations, masking agents like GSH, imidazole, and EDTA were utilized to mask $Cd^{2+}$, $Ni^{2+}$ and $Co^{2+}$, respectively. With the aid of masking agents, simultaneous detection of cations could also be successfully achieved in the buffer solution as well as in the spiked river water. The LoD value for $Ni^{2+}$ ion was computed as 0.3 µM in this report.

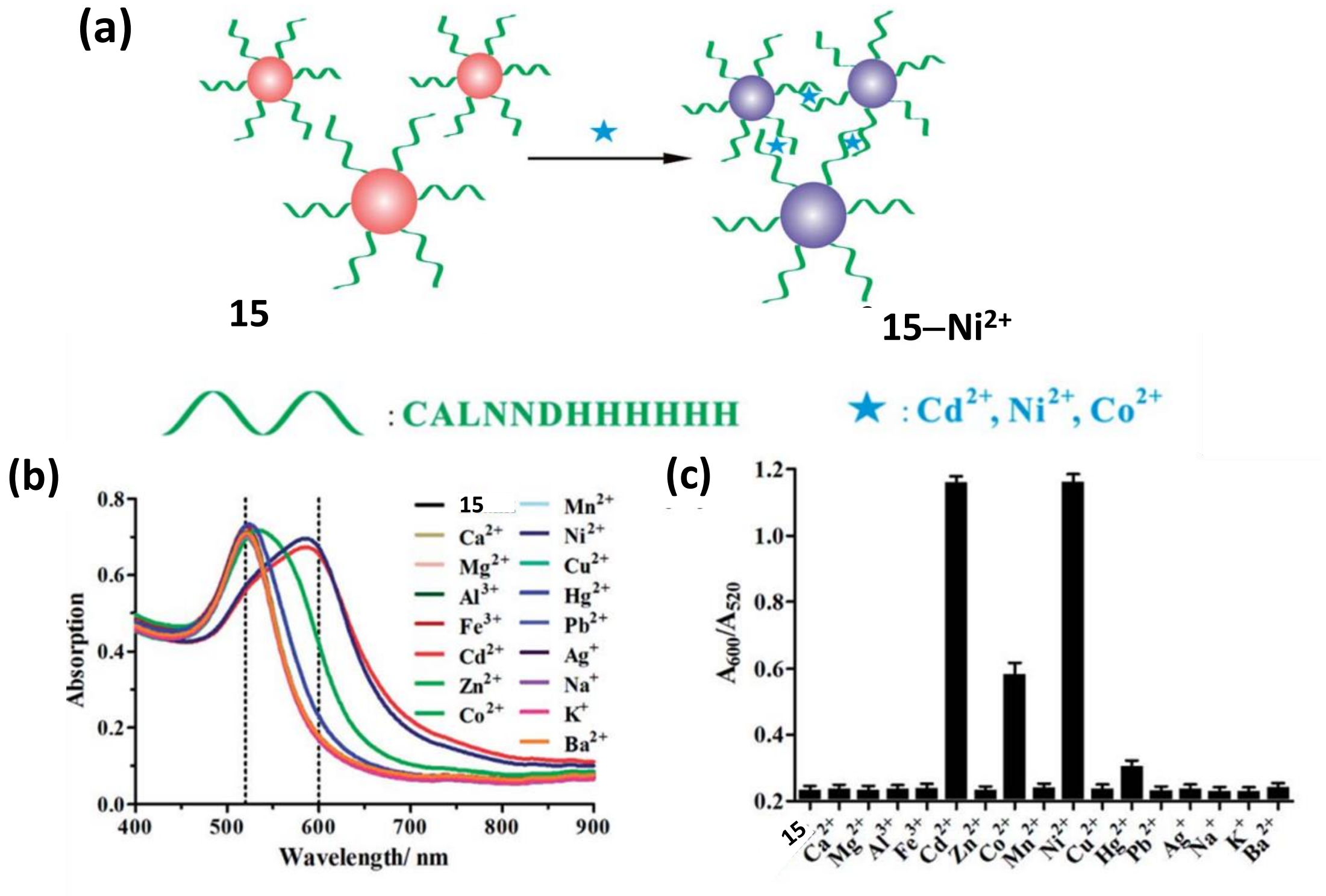

**Fig. 17**. (a) Sensing mechanism of peptide-functionalized AuNPs **15**; (b) UV-Vis spectral analysis for probe **15** with different metal ions and (c) Absorption ratio ($Abs_{600}/Abs_{520}$) of the probe **15** in presence of different metal ions. Adapted with permission from Ref [125]. Copyright 2012, The Royal Society of Chemistry.

Later in 2014, Azzam *et* al. [126] developed a AuNPs-based probe **16** using dithiol surfactants for the detection of $Ni^{2+}$ and $Zn^{2+}$ ions (Fig. 18). TEM studies revealed spherical-shaped nanoparticles with an average diameter of 20.0 nm. The alkyl chains in the surfactants adsorb over the surface of NPs to improve their dispersion behavior, eventually leading to prevention of the aggregation process. The effect of the change in the length of alkyl fragments on probe **16** has also been scrutinized in this study. Variations in the alkyl chain length resulted in a decrease in surface energy which reduced the reactivity of the particles towards the metal ions. The SPR absorption of **16** in water was observed at 520 nm that declined upon the addition of $Ni^{2+}$ and $Zn^{2+}$ ions. The increase in the alkyl chain length from five to eight carbon atoms also assisted in the binding of the metal ions, which led to the drastic changes in the UV-Vis spectra of **16**. The negative charge in the periphery of **16** attracted the cations and assisted in binding to the thiol group. This response of **16** was found to be improved as compared to the individual surfactants.

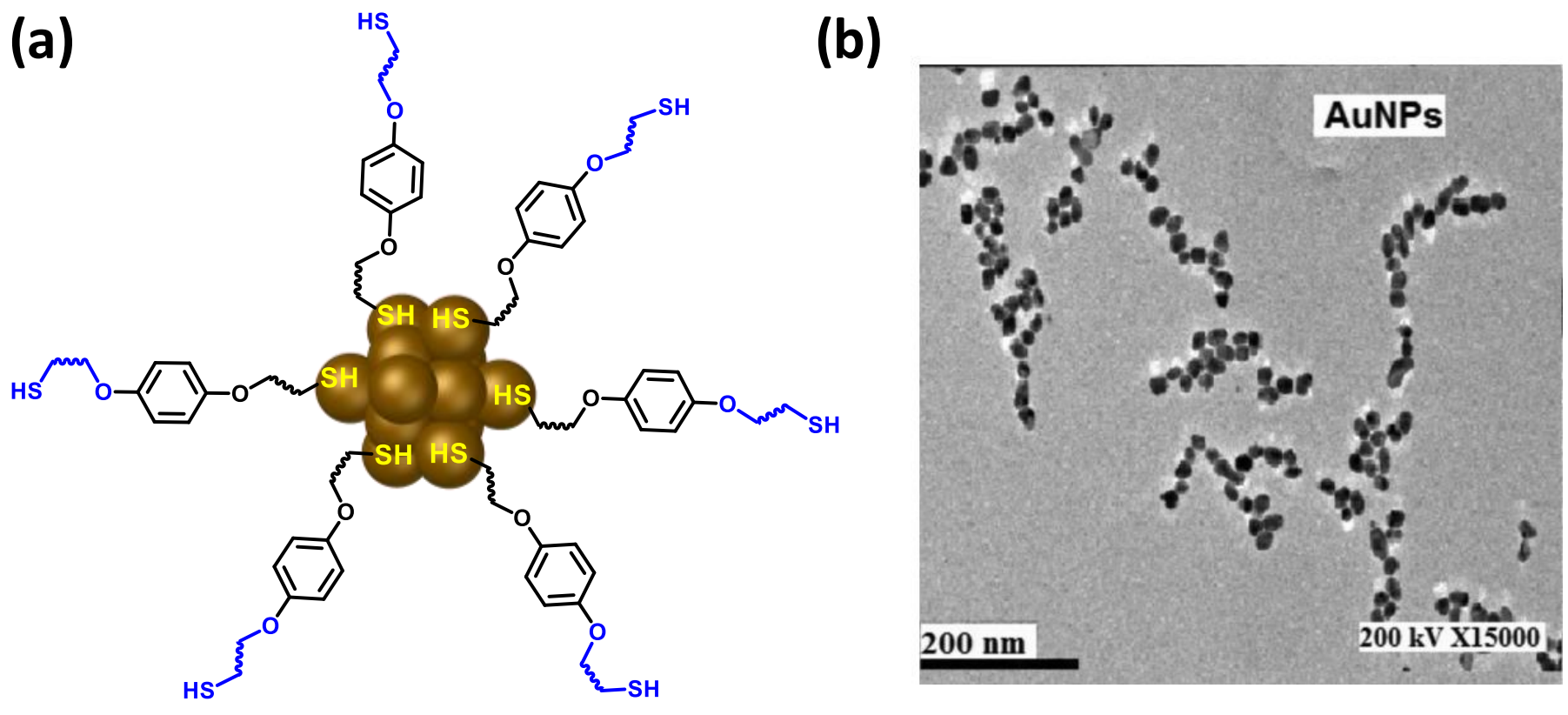

**16**

**Fig. 18.** (a) The schematic for probe **16** and (b) TEM image of probe **16**. Adapted with permission from Ref [126]. Copyright 2014, Elsevier.

Though the reducing agents are widely applied for the synthesis of the metal-based nanoparticles, their excessive usage is linked to adverse effects on human health and environment [127,128]. And thus, the employment of green synthetic routes to develop nanoparticles is of huge interest to the researchers. In 2015, Rajendiran's group [128] synthesized a AuNPs-based probe **17** with the help of sunlight and eco-friendly *N*-cholyl-*L*-valine as a capping agent (Fig. 19). UV-Vis studies of **17** revealed the highest intensity SPR band at 520 nm in aqueous medium, while the concentration of the valine moiety was fixed at $3.3 \times 10^{-4}$ M. Probe **17** was found to be stable in the pH range of 8.0-11.0, and therefore, pH 9.0 was employed for the detection study of cations. Once formed, **17** was stable even after 6 months of storage at room temperature due to the valine moiety, which could reduce/stabilize **17** by providing various hydrophobic and hydrophilic interaction sites. The average size of these AuNPs could be easily tuned from 8.0 nm to 40.0 nm by varying the concentration of the capping agent. At low valine concentrations ($1.0 \times 10^{-4}$ M), the particles were spherical with 40.0 nm particle size whereas at higher valine concentration ($2.3 \times 10^{-4}$ M), the NPs adopted spherical shape with an average particle size of 8.0 nm. The addition of 200 nM of assorted metal ions (e.g., $Ba^{2+}$, $Ca^{2+}$, $Cd^{2+}$, $Cr^{3+}$, $Cu^{2+}$, $Fe^{3+}$, $Hg^{2+}$, $Mg^{2+}$, $Mn^{3+}$, $Na^{+}$, $Pb^{2+}$, and $Zn^{2+}$) showed nominal changes in absorption spectra of **17**; however, $Ni^{2+}$ addition induced a remarkable decrement in the SPR absorption followed by a bathochromic shift (from 524 nm to 543 nm). The spectral changes have been attributed to the aggregation of AuNPs, and an obvious solution color change of **17** from pink to violet has been noticed. The LoD value for $Ni^{2+}$ was depicted as 10 nM in the aqueous solution. The detection of $Ni^{2+}$ was also demonstrated in the spiked river water at even 15 nM concentration. The addition of $Co^{2+}$ to **17** also showed a color change from pink to light pink with a decrease in SPR intensity, but no shift in this band has been noticed. Other cations displayed negligible colorimetric response and failed to interfere in $Ni^{2+}$ sensing by **17**.

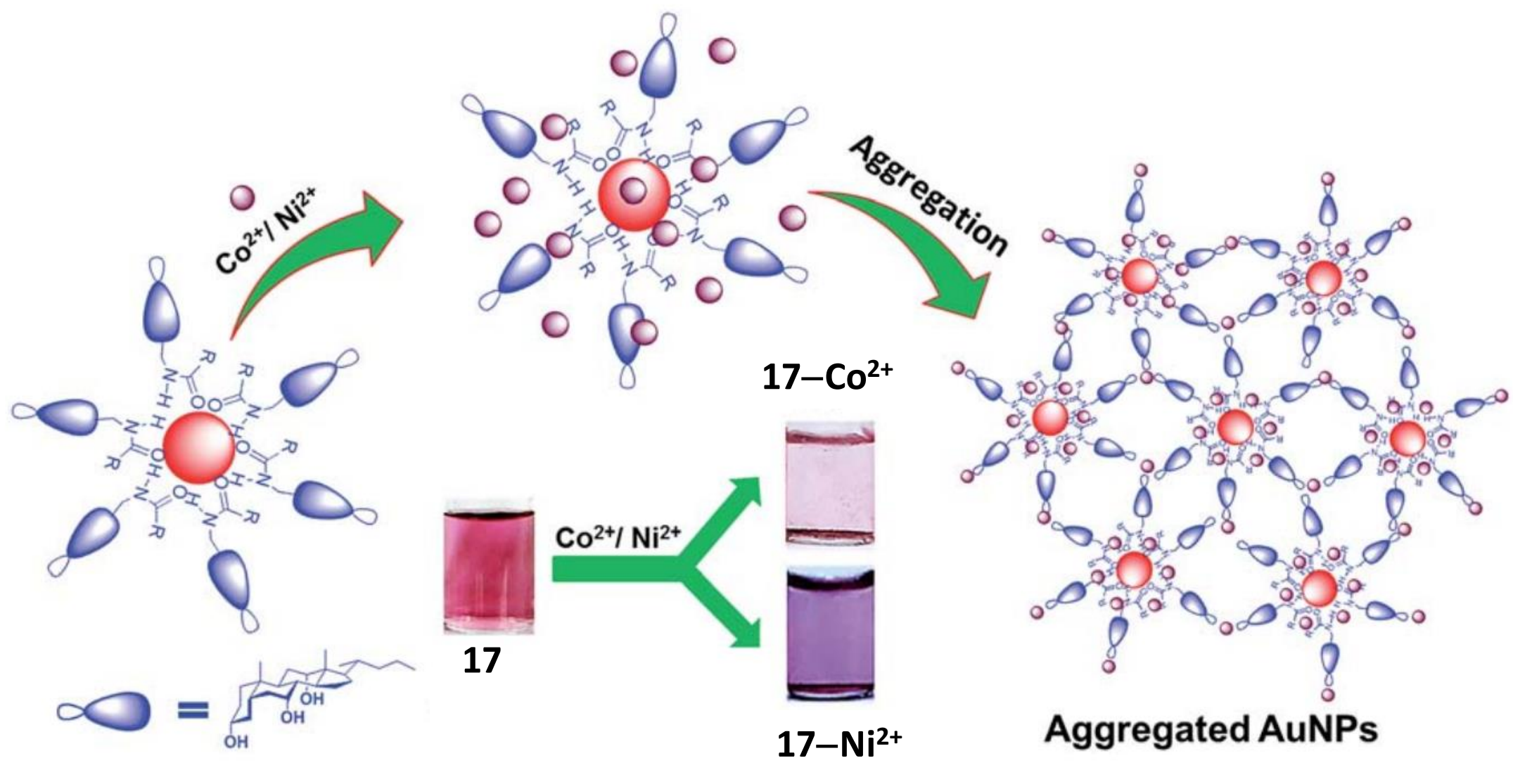

**Fig. 19.** Schematics for the detection of $Ni^{2+}$ and $Co^{2+}$ using AuNPs based probe **17**. Adapted with permission from Ref [128]. Copyright 2015, The Royal Society of Chemistry.

Two years later, Shrivas *et* al. [129] synthesized another AuNPs-based probe **18** capped by the malonate groups (Fig. 20). The TEM study in agreement with DLS analysis revealed an average size of NPs as 12.5 nm. FT-IR analyses corroborated the capping of the malonate groups over the surface of the gold nanoparticles. The self-aggregation of particles at low pH (pH < 5.0) and the decrease in absorbance ratio ($Abs_{525}/Abs_{610}$) at higher pH (pH > 8.0) prompted the authors to conduct pH optimization studies. The optimum ratio of absorbance could only be obtained when the pH of the probe's solution was maintained between 7.0 and 8.0. No change in solution's color was noticed with $Na^+$, $Mg^{2+}$, $Al^{3+}$, $Cr^{3+}$, $Fe^{3+}$, $Co^{2+}$, $Cu^{2+}$, $Zn^{2+}$, $As^{3+}$, $Cd^{2+}$, and $Hg^{2+}$ ions; however, $Ba^{2+}$ and $Ni^{2+}$ turned the solution's color from pink to blue. The SPR absorption of **18** at 525 nm was red-shifted to 610 nm after the addition of $Ba^{2+}$ and $Ni^{2+}$ ions. Many competing cations as well as anions did not interfere in the selective sensing of probe for $Ni^{2+}$ and $Ba^{2+}$ ions. The reason behind the band shift and the color change has been ascribed to the aggregation of **18** that occurred *via* coordinative interaction between the carboxylate groups of the malonate moiety and the targeted metal ion. The drastic increase in the localized electric field between two particles caused by the aggregation of NPs resulted in the shift of SPR band. The LoD value for $Ni^{2+}$ and $Ba^{2+}$ was reported to be 3.0 ng/mL and 5.0 ng/mL, respectively. The cation sensing ability of probe **18** was also studied in the real water sources, including tap water, pond water, and river water.

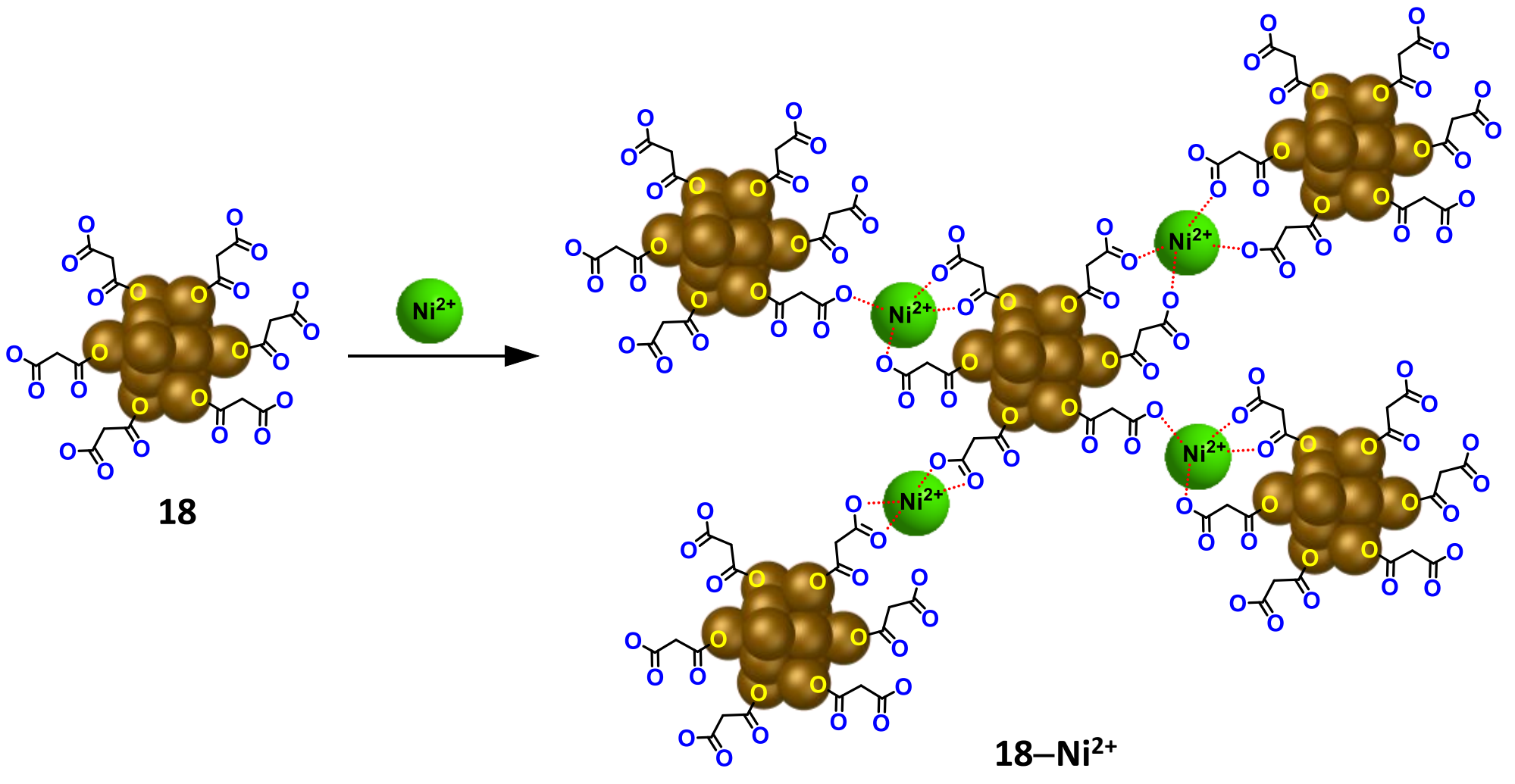

**Fig. 20.** Binding mode of AuNPs-based probe **18** with $Ni^{2+}$ followed by the aggregation process.

In 2018, Salimi *et* al. [130] reported probe **19**, the ribavirin-based gold nanoparticles, for selective detection of $Cr^{3+}$ and $Ni^{2+}$ ions in aqueous medium (Fig. 21). The average particle size of **19** was found in the range 15–20 nm, as evidenced by TEM analysis. UV-Vis spectrum of **19** in aqueous phase displayed its characteristic SPR band at 530 nm. The addition of $Cr^{3+}$ and $Ni^{2+}$ to **19** red-shifted the SPR absorption to 670 nm, most likely due to the induced aggregation of NPs, and this event was accompanied by an intense change in color of the probe's solution from red to blue. Other metal ions such as $Mg^{2+}$, $Mn^{3+}$, $Na^{+}$, $Ag^{+}$, $Al^{3+}$, $Ca^{2+}$, $Cd^{2+}$, $K^{+}$, $Cu^{2+}$, $Fe^{2+}$, $Fe^{3+}$, $Hg^{2+}$, $Co^{2+}$, $Pb^{2+}$, and $Zn^{2+}$ could not produce any changes in the spectra. Probe **19** displayed excellent stability and could function at pH 3.0–8.0. The LoD value for $Ni^{2+}$ and $Cr^{3+}$ were calculated to be 25.2 nM and 30.5 nM, respectively. The probe could also detect the cations in tap water and drinking water samples by monitoring the changes observed in absorption ratio (i.e., $A_{670}/A_{530}$).

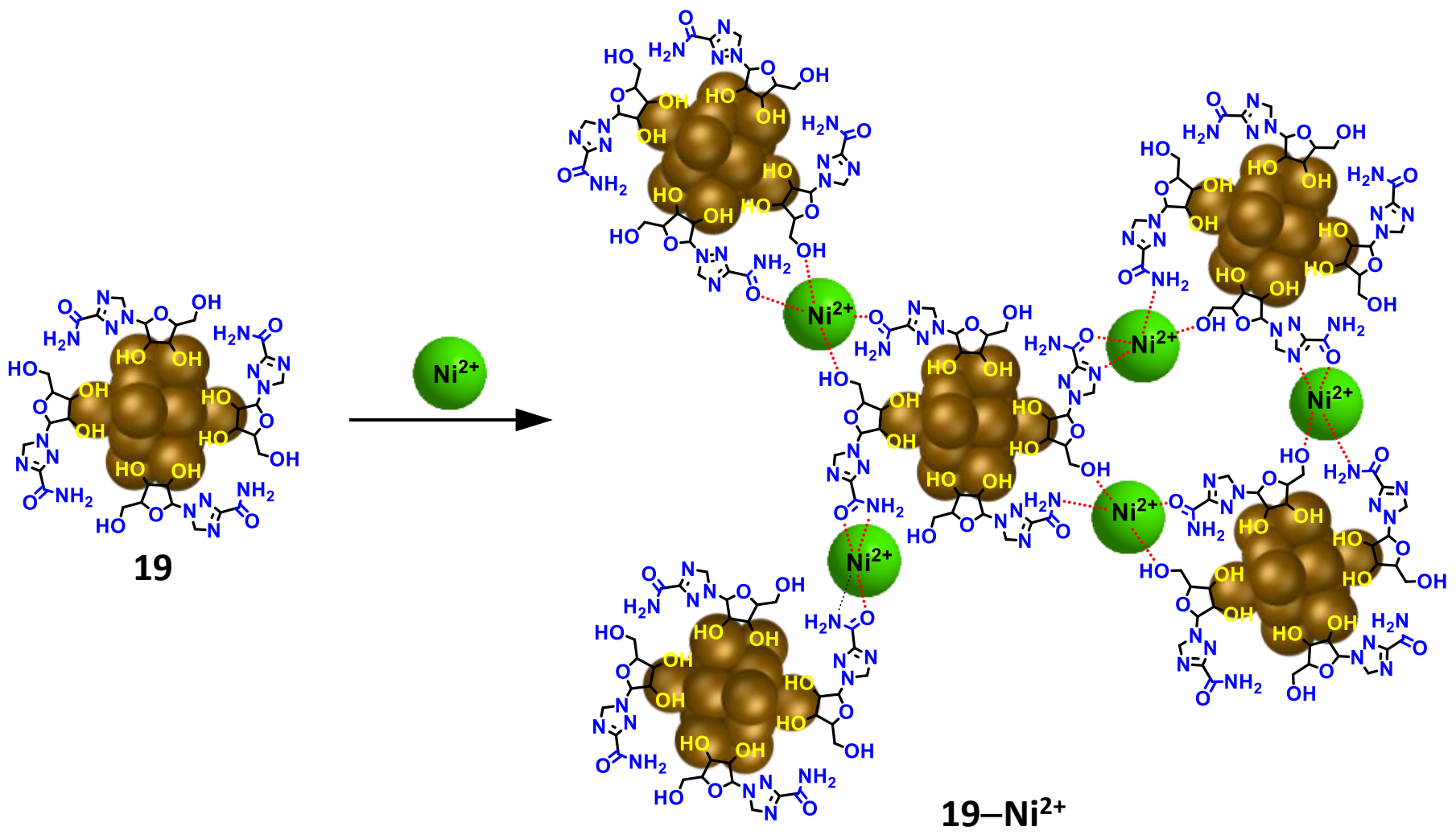

**Fig. 21.** Binding mode of AuNPs-based probe **19** with $Ni^{2+}$ followed by the aggregation process.

Despite the successful employment of various AuNPs in the metal ion sensing methods, the detection of $Ni^{2+}$ in real specimens, e.g., urine, seawater, soils, etc. is still facing issues and challenges. Among the salient reasons is the complexity of such samples having a high-level of salts and other interfering species in the sample medium. In many cases, if a small-analyte responsive ligand, also serving as capping agent for nanoparticles, is incapable of conferring the colloidal stability that could hold-up the matrices, non-desirable aggregation of NPs may occur before the addition of the target analyte [131,132]. To address these issues, surface modification of AuNPs with specific ligands that are stable enough during the sensing process is highly required. Surface coating of gold-NPs with zwitterionic compounds, e.g., cysteine betaine, polybetaines and other mixed-charge molecules has emerged as an efficient method to fabricate stable AuNPs [38]. Surface coating of AuNPs with zwitterionic self-assembled monolayer allows the water layer formation around AuNPs due to the string affinity of zwitterions towards water. The resultant water layer will promote hydration repulsion to eventually resist the aggregation of NPs.

In an elegant study, Surareungchai's group [38] (in 2018) demonstrated the zwitterionic-polypeptide capped AuNPs (**20**) as a highly selective and sensitive probe for $Ni^{2+}$ recognition (Fig. 22a). The characteristic absorption maximum for **20** was observed at 522 nm. These nanoparticles displayed self-aggregation at pH =7.5 or below, and therefore, pH above 8.0 was chosen for metal ion sensing studies. The ionic species such as $K^{+}$, $Mg^{2+}$, $Ag^{+}$, $Ca^{2+}$, $Mn^{3+}$, $Fe^{3+}$, $Co^{2+}$, $Ni^{2+}$, $Cu^{2+}$, $Zn^{2+}$, $Cd^{2+}$, $Sn^{2+}$, $Hg^{2+}$, $Pb^{2+}$, and $Fe^{2+}$ were added to the red solution of **20**, and the colorimetric and spectral changes (UV-Vis) were monitored at different pH values. The probe could detect $Ni^{2+}$, $Hg^{2+}$, $Zn^{2+}$, and $Cd^{2+}$; however, changing the pH to 9.0 suppressed the detection of $Cd^{2+}$, $Hg^{2+}$ and $Zn^{2+}$, and the optimum pH was thus fixed at 9.0. $Hg^{2+}$ ions promoted aggregation at pH 9.0 if the solution mixture was kept beyond 15 min therefore, the detection process was further optimized by time. The aggregation was monitored using the $Abs_{650/520}$ ratio, and the LoD value was 4.92 μM (Fig. 22b). In order to further increase the sensitivity toward the value prescribed by WHO (0.34 μM), the ionic strength was increased by adding 100 mM of NaCl, which resulted in a LoD value of 2.56 μM. Finally, the detection limit value of 34.0 nM could be achieved when the particle size of **20** was optimized and increased from 13.0 to 40.0 nm. The sensing behavior of **20** for $Ni^{2+}$ has been attributed to the coordinative binding of the terminal -$NH_2$ group with $Ni^{2+}$ ions. The probe was also found to be reversible, as the aggregated particles were completely dispersed after addition of 100 mM NaOH (pH > 10.0) (Fig. 22c). This dispersion was due to 'like-charge repulsion' as the basic environment supports the facile deprotonation. $Ni^{2+}$ detection was also monitored in the environmental and biological samples such as urine, soil, sea, drinking, and tap water.

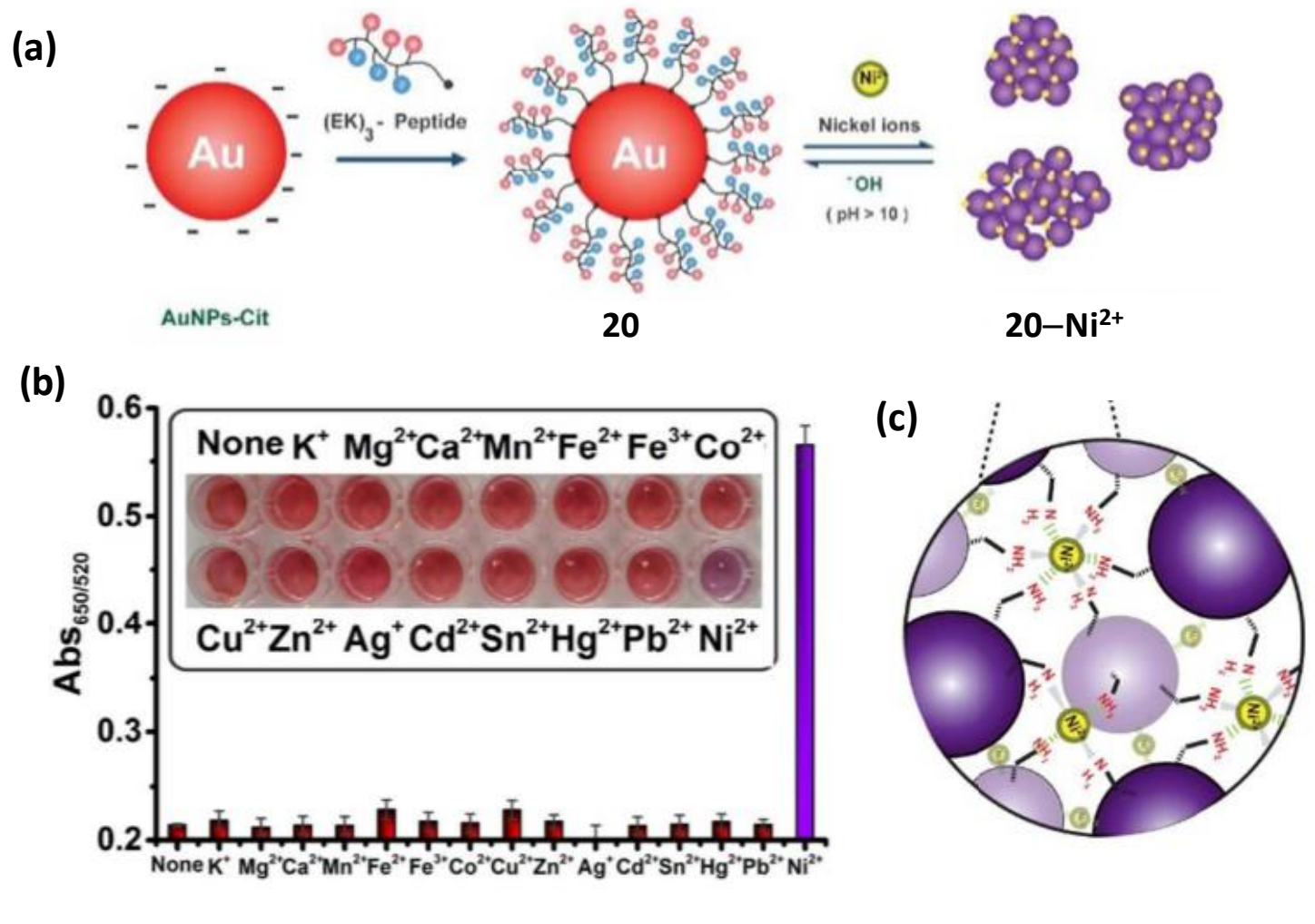

**Fig. 22.** (a) Binding mode of peptides-capped AuNPs **20** with $Ni^{2+}$ ions; (b) Absorption ratio ($Abs_{650/520}$) of probe **20** with different metal ions together with the colorimetric response and (c) Aggregation of nanoparticles induced upon $Ni^{2+}$ addition. Adapted with permission from Ref [38]. Copyright 2018, The Royal Society of Chemistry.

In 2019, Zinjarde's group [133] synthesized a biogenic AuNPs probe **21** using actinomycetes genus called *Nocardiopsis dassonvillei NCIM 5124*. The SPR absorption for **21** was measured at 530 nm in aqueous medium. The addition of $Ni^{2+}$ to **21** destabilized the AuNPs, causing an aggregation, which resulted in a red-shift in the SPR band from 530 nm to 660 nm. A sharp color change from ruby red to purple could also be observed to the naked eye. The spectral changes were nominal when $Zn^{2+}$, $Cd^{2+}$, $Cr^{3+}$, $Mo^{2+}$, $Cu^{2+}$, $Pd^{2+}$, and $Hg^{2+}$ were targeted. An optical sensing tool was also developed by generating light from a tungsten halogen lamp, which was introduced and transferred to the reaction mixture connected to a spectrometer. The spectral response significantly decreased at higher concentration of $Ni^{2+}$ due to the aggregation event. The LoD value has been reported to be 2.0 ppm for $Ni^{2+}$ in this study. Two years later (in 2021), Zhang *et* al. [134] reported AuNPs-based probe **22** with non-toxic and biologically important phytic acid groups acting as a stabilizer and reductant to the NPs. This probe exhibited a highly selective colorimetric response toward $Ni^{2+}$ in aqueous phase (Fig. 23a). HR-TEM studies revealed an average particle size of **22** as 25.6 ± 3.1 nm. UV-Vis spectra of **22** in dispersed aqueous medium displayed a typical SPR band near 523 nm that was red-shifted to 650 nm upon $Ni^{2+}$ addition (Fig. 23b). Authors speculated that the addition of $Ni^{2+}$ reduced the extent of the surface negative charges, and thus, typically lead to the aggregation of **22**. The optimum pH was chosen as 3.6, where the nanoparticles were found stable with an optimized reaction time of 15 min (for a maximum $A_{650/523}$ ratio). The addition of other ions e.g., $Fe^{3+}$, $Fe^{2+}$, $K^{+}$, $Cu^{2+}$, $Mn^{3+}$, $Cd^{2+}$, $Co^{2+}$, $Pb^{2+}$, $Zn^{2+}$, $Mg^{2+}$, $Ca^{2+}$, $Al^{3+}$, and $Cr_2O_7^{2-}$ showed negligible changes in the absorption spectra whereas, $Ni^{2+}$ resulted in a drastic spectral change. The LoD value for the $Ni^{2+}$ detection was calculated as 0.216 µM. This probe was also tested against real samples, such as drinking water spiked with $Ni^{2+}$ ions. The probe detected $Ni^{2+}$ when incubated at pH = 3.6 for 15 min with an excellent recovery value from 98.4% to 112%.

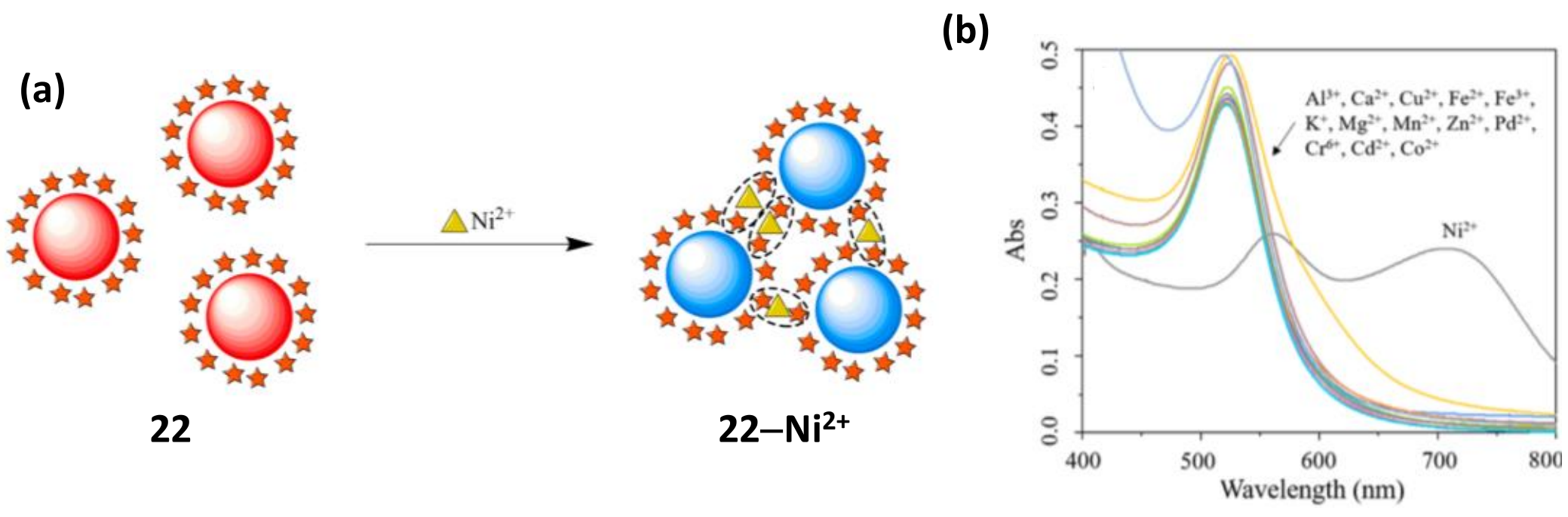

**Fig. 23**. (a) Sensing of $Ni^{2+}$ exhibited by AuNPs-based probe **22** and (b) UV-Vis spectra of probe **22** in presence of assorted metal ions. Adapted with permission from Ref [134]. Copyright 2021, Elsevier.

Recently, Conkova *et* al. [26] synthesized AuNPs-based probe **23** capped with a Schiff base ligand for the detection of $Fe^{2+}$, $Cu^{2+}$, and $Ni^{2+}$ in organic media such as toluene-acetonitrile solvent mixture (Fig. 24). The Schiff base has been designed to combine with *α*-lipoic acid, which tends to bind with the AuNPs *via* Au-S linkages, whereas the NOO pocket of the ligand acted as the binding unit. The average particle size of **23** was found to be 10.8 ± 1.1 nm which was further supported by DLS measurement with an average hydrodynamic diameter of 15.45 ± 4.26 nm. A solution color change from red to purple could be observed upon addition of $Fe^{2+}$, $Cu^{2+}$, and $Ni^{2+}$ ions, corresponding to the octahedral complex formation between the cation and the *N*-acylhydrazone moiety. The color change was accompanied by a red shift in the SPR band, and the LoD values were depicted in the range of 1.4 to 11.2 nM. As a proof of concept, **23** was used to detect $Ni^{2+}$ ions in the Kumada coupling reaction. The coupling reaction employed a $Ni^{2+}$-based catalyst for the synthesis of PDE472, which is a drug target for asthma treatment [135]. The toluene waste obtained from the Kumada reaction was added to the solution of **23** and was monitored using UV-Vis analysis. The SPR band showed a redshift of 4 nm after 10 min, and the LoD was determined to be 3.5 µM. This value agreed well with the value calculated by the ICP-MS analysis.

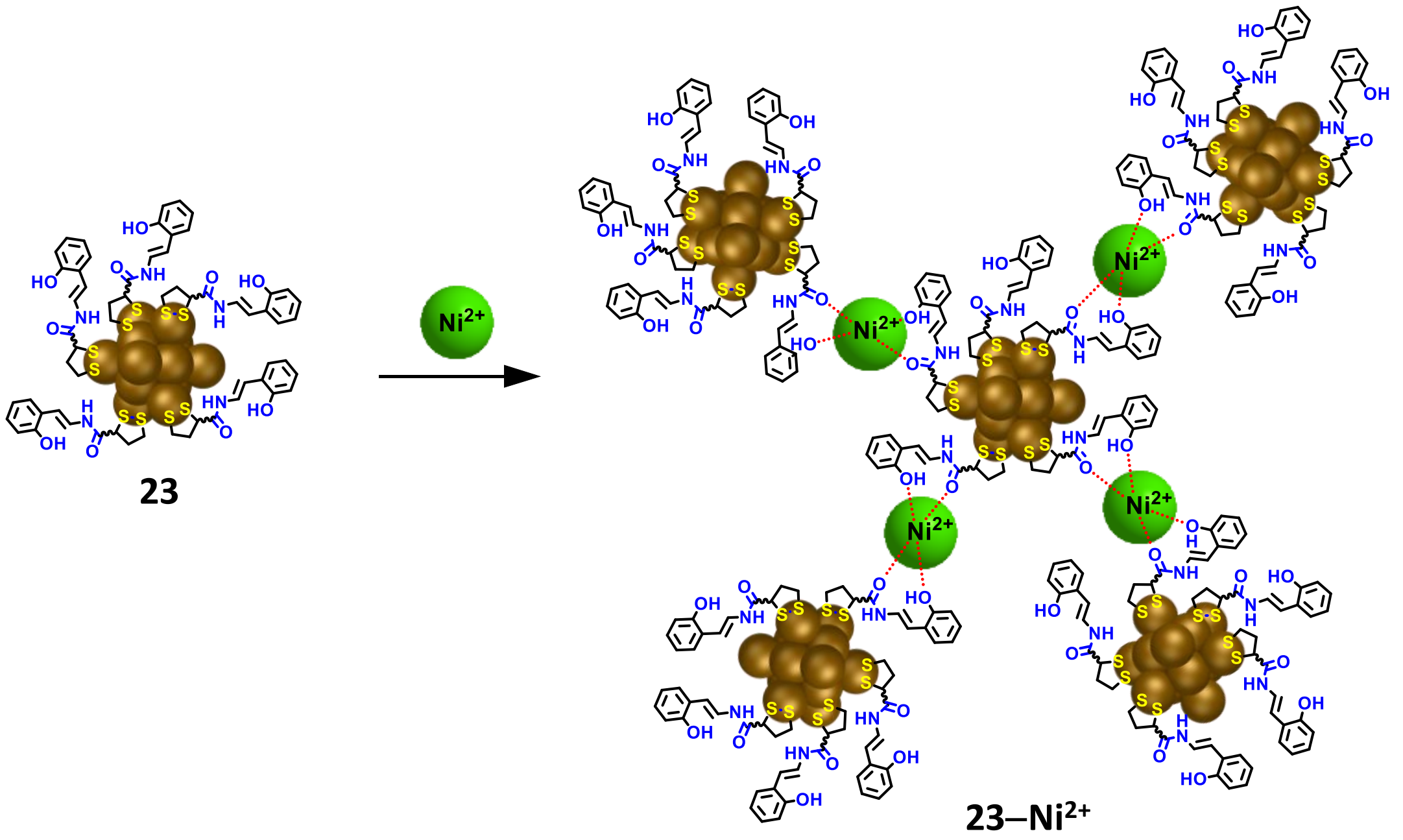

**Fig. 24.** Binding mode of AuNPs based probe **23** with $Ni^{2+}$ ions followed by the NPs aggregation.

**Table 2**. Gold nanoparticles (AuNPs) and other miscellaneous inorganic nanosensors for $Ni^{2+}$ ions.

| Probe | $\lambda_{abs}$ (nm) (SPR) | Particle size (nm) | Stabilizing agent | Target ions | LoD (µM) | Colorimetric effect | Solvent | Practical application | [Ref.] |
|---|---|---|---|---|---|---|---|---|---|
| **14** | 523 | 14<br>45 | L-carnosine | $Cu^{2+}$, **$Ni^{2+}$** | $0.5 \times 10^{3}$ | Raspberry red to Purple | Water | - | [123] |
| **15** | 520 | 20 | Peptide | $Cd^{2+}$, **$Ni^{2+}$**, $Co^{2+}$ | $Cd^{2+} = 0.05$<br>**$Ni^{2+}$** = 0.3<br>$Co^{2+} = 2$ | Red to Purple | Water | River water | [125] |
| **16** | 520 | 20 | Dithiol-surfactant | **$Ni^{2+}$**, $Zn^{2+}$ | - | - | Water | - | [126] |
| **17** | 550 | 8 | *N*-cholyl-l-valine | $Co^{2+}$, **$Ni^{2+}$**, | **$Ni^{2+}$** = $10 \times 10^{-3}$<br>$Co^{2+}$=$10 \times 10^{-3}$ | $Ni^{2+}$=Pink to Purple<br>$Co^{2+}$=Pink to Light-pink | Water | Tap and drinking water | [128] |
| **18** | 525 | 12.5 | Malonate | $Ba^{2+}$, **$Ni^{2+}$** | **$Ni^{2+}$** = 3<br>$Ba^{2+} = 5$ | Pink to Blue | Water | River, tap, and pond water | [129] |
| **19** | 530 | 18 | Ribavirin | $Cr^{3+}$. **$Ni^{2+}$** | **$Ni^{2+}$** = $25.2 \times 10^{-3}$<br>$Co^{2+} = 30.5 \times 10^{-3}$ | Red to Blue | Water | Drinking and tap water | [130] |
| **20** | 522 | 40 | Zwitterionic polypeptide | **$Ni^{2+}$** | $34 \times 10^{-3}$ | Red to Purple | Water | Urine, soil, tap and drinking water | [38] |
| **21** | 530 | - | Nocardiopsis dassonvillei | **$Ni^{2+}$** | $2 \times 10^{3}$ | Ruby-red to Purple | Water | - | [133] |
| **22** | 523 | 25.6 | Phytate Acid | **$Ni^{2+}$** | 0.216 | Red to Blue | Water | Drinking water | [134] |
| **23** | 526 | 15.45 | α-lipoic acid, mono(salicylalehyde)-iminoacetyl-hydrazone | **$Ni^{2+}$**, $Cu^{2+}$, $Fe^{3+}$ | **$Ni^{2+}$** = $1.4 \times 10^{-3}$<br>$Co^{2+} = 9.0 \times 10^{-3}$<br>$Fe^{3+} = 11.2 \times 10^{-3}$ | Brown to Orange | Toluene/ACN | - | [26] |
| **24** | 525 | - | Trisodium citrate GSH | **$Ni^{2+}$** | $5.36 \times 10^{-4}$ | Light red to Grey | Water | Drinking water<br>Tap water | |
| **25** | 470 | 70 | - | **$Ni^{2+}$** | 0.0356 | Yellow to Grey | Water | Tap and river water; Serum | |

### *4.3. Other inorganic colorimetric sensors*

In 2018, Chen's group [136] developed a colorimetric strategy to quantify $Ni^{2+}$ ions at nM scale. This method was based on the usage of $I^-$-responsive Cu-Au-nanoparticles **24** in combination with $Ni^{2+}$-catalysed GSH oxidation reaction. Without $Ni^{2+}$, iodide ions assembled at the surface of **24** by replacing the stabilizers (i.e., citrate groups), leading to the morphology change of NPs. Concomitantly, the color of probe's solution turned red from grey. On the other hand, the addition of $Ni^{2+}$ triggered the catalytic oxidation of GSH to afford bulky GSSG moiety, which prevented the access of $I^-$ to NPs, and therefore, the original grey color of the probe retained. The absorption maximum of **24** assay was observed at 525 nm, which diminished after the addition of $Ni^{2+}$ ions. The detection limit value was found as low as 0.536 nM. Other metal ions such as $Fe^{3+}$, $Co^{2+}$, $Cu^{2+}$, $Sr^{2+}$, $Mn^{2+}$, $Pb^{2+}$, $Cd^{2+}$, $Zn^{2+}$, and $Ag^+$ could not prevent the color change from grey to red when iodide was added in presence of GSH. As a proof of concept, **24** was able to detect $Ni^{2+}$ ions in the spiked tap water in the range of 10 nM to 200 nM. The change in tap water was obvious to monitor with UV-Vis analysis, as the absorbance at 525 nm was reduced with recovery of 95-96%. The LoD value for $Ni^{2+}$ in real sample was found as 0.606 nM. The slight variation in the LoD values has been ascribed to the presence of some organic matter in the tap water sample.

In 2016, Jie *et* al. [137] synthesized hallow nanospheres **25** using zinc silicate ($ZnSiO_3$) integrated with zincon for the selective colorimetric detection of $Ni^{2+}$ ions both in solution as well as onto paper strips. SEM studies showed that the hollow nanoparticles had an average size of 70 nm. The addition of zinc silicate hollow spheres led to a change in the absorption peaks of free zincon from 480 nm to 620 nm, most likely due to $Zn^{2+}$ complexation. The gradual addition of $Ni^{2+}$ to **25** led to a decline in the absorption peak at 620 nm, which splitted into 500 nm and 666 nm upon excessive $Ni^{2+}$ addition. The selectivity study suggested that the presence of other metal ions barely interrupted $Ni^{2+}$ detection. The discerning behavior of **25** could be further improved using $Na_2$-EDTA as a masking agent keeping the optimal volume ratio of probe's solution to $Na_2$-EDTA as 3:2. The **25**-EDTA mixture exhibited maximum absorption at 470 nm, which shifted to 660 nm with increasing $Ni^{2+}$ ion concentration, signifying the formation of the **25-**EDTA-$Ni^{2+}$ complex. The spectral alterations were associated with a change in color from pink to grey, visible to the naked eye. The probe **25** exhibited detection limit in the nanomolar range (36 nM) in solution state. Additionally, it was successfully immobilized on cellulose paper strips to develop a paper-based $Ni^{2+}$ sensor and the detection limit achieved was 83 nM. This probe was also capable of detecting $Ni^{2+}$ in tap water, river water, and serum with a recovery of 89.75% to 108.63%.

### *4.4. Organic colorimetric sensors*

Electrospun nanofibers are gaining wide attention due to their fast response, large surface area to volume ratio, and upscalable fabrication technique [138]. In 2011, Poltue *et* al. [139] developed novel poly(caprolactone) electrospun nanofibrous mats **26** incorporated with DMG. These electrospun fiber mats served as a sensor for detection of $Ni^{2+}$ ions *via* a change in color (colorless to red-pink). The SEM studies confirmed the formation of uniform nanofibers with an average diameter of 275 ± 39 nm. A linear relationship between the reflectance values at 547 nm and the concentration of $Ni^{2+}$ (1-10 ppm) was observed. In the FTIR spectrum, the peak corresponding to -OH of DMG initially present at 3205 $cm^{-1}$ disappeared upon the addition of $Ni^{2+}$ ions. These spectral changes indicated the hydrogen bonding interactions between DMG and nickel ions while the appearance of a new peak at 1572 $cm^{-1}$ ($\nu_{CN}$ stretching) confirmed the formation of $Ni(DMG)_2$ complexes.

Another report on electrospun nanofibers was demonstrated by Sereshti and group [140] in 2016 for highly selective recognition of $Ni^{2+}$ using probe **27**. This group employed polycaprolactam and DMG and incorporated the nanofibrous mat with a polyvinyl alcohol solution to make the probe transparent. SEM images revealed nanofibrous morphology with a uniform diameter ranging from 80 to 160 nm. The fabrication process was carefully optimised by varying PVA/DMG amount, electrospinning time, flow rate, and electrospinning voltage. The addition of $Ni^{2+}$ resulted in a drastic increase in the absorption intensity at 547 nm. Similar to the previous report on **26**, the color of this probe turned red upon increasing the concentration of $Ni^{2+}$, which could be ascribed to the formation of DMG-$Ni^{2+}$ complex. However, in contrast to the previous report, the probe **27** exhibited a higher linear range of 5.0-100 μg/mL. Moreover, successful detection of nickel ions was achieved in tap water, electroplating wastewater, and chemical laboratory wastewater. The selectivity for $Ni^{2+}$ was also tested against various interfering cations ($Sn^{2+}$, $Hg^{2+}$, $OH^{-}$, $Na^{+}$, $Cu^{2+}$, $Cd^{2+}$, $Fe^{2+}$, and $Co^{2+}$); however, none other than $Cu^{2+}$ displayed interference even at very high concentration upto 1000 μg/mL. The detection and the quantification limits for $Ni^{2+}$ were determined to be 2.0 and 5.0 μg/mL, respectively.

### *4.5. Comparative sensing performance of colorimetric sensors*

In case of colorimetric nanosensors, either surface functionalization or target-induced aggregation leads to the variation in optical characteristics of the nanoprobes. For instance, the MNPs acting as colorimetric $Ni^{2+}$ sensors function mainly *via* aggregation-based strategy, often comprised of AuNPs and AgNPs. Such plasmonic sensors utilize simple design and exhibit high sensitivity and selectivity. Hence, an appropriate surface functionalization of MNPs is essential for improving plasmonic activity, colloidal stability, and providing sensitivity and new functionalities [141,142]. Despite their high versatility, $Ni^{2+}$-responsive noble metal-based nanosensors sometimes suffer from major drawbacks: (i) an interference from other *d* block metal ions, especially $Co^{2+}$ and $Cu^{2+}$; (ii) auto-aggregation of NPs due to some external factors (e.g., pH and temperature of the medium) which may lead to a false positive outcome and (iii) a significantly low and inadequate solution color change (sometimes not obvious to the naked eye). Recently, non-aggregation based plasmonic heavy metal responsive nanosensors have seen gradual growth due to their better reliability [143]; however, to the best of our understanding, not a single report is available on non-aggregation-based colorimetric MNPs sensors for recognition of $Ni^{2+}$ ions. Few sensors such as **5**, **8**, and **24** also rely upon a change in morphology for the detection of $Ni^{2+}$ ions. These sensors exhibit very high selectivity/sensitivity and sharp color changes in the presence of $Ni^{2+}$ ions, but their dependence on reaction by $I^{-}$ or $H_2O_2$ for etching further restricts their applied potential in complex biological media.

A direct comparison between various sensing parameters of AgNPs and AuNPs clearly reveals substantial differences between these sensing materials. On comparing the LoDs, which is one of the key parameters for a sensor, it appears that AuNPs tend to be relatively much more sensitive (nM range) than that of AgNPs (μM range) (Tables 1 and 2) even though the latter exhibit a higher SPR effect. The LoD can be further improved by optimizing the LSPR band of MNPs *via* appropriate surface functionalization [71]. The LSPR properties of AuNPs can easily be varied by coating a silver nanoshell on their surface, leading to multicolor changes where the HSAB principle may play an important role in selecting suitable functionality for highly selective detection of $Ni^{2+}$ ions [144]. Indeed, it has been observed that the surface modification with a moiety containing O and N atoms has resulted in sensitive and selective detection of $Ni^{2+}$ ions. Moreover, to increase the selectivity and LoD of the nanosensors, one can select specific DNAzymes, proteins, and aptamers that have high selectivity towards $Ni^{2+}$ ions [38]. So far, more than 20 reports are available on such sensors; however, the solid-state paper strip-based $Ni^{2+}$ sensing is rarely explored with only a single report to our knowledge (i.e., probe **3**).

Although, electrospun nanofibers such as **26** and **27** offer a solid-state detection platform, their sensitivity is relatively low for practical purposes. Interestingly, probe **25** (zincon incorporated $ZnSiO_3$ nanosphere) was able to selectively detect $Ni^{2+}$ onto paper strips with improved sensitivity (nM range) in real water as well as in serum samples. There are many examples of salient detection of $Ni^{2+}$ in real water samples (such as river water, drinking water, and tap water) with moderate to excellent recoveries; nevertheless, $Ni^{2+}$ detection in complex biological media and industrial waste remains underexplored with very few examples (e.g., **20** and **27**). In total, a thorough investigation of various colorimetric nanosenosors is in high demand for the future detection of heavy metal ions including $Ni^{2+}$ ion. The development of improved and innovative optical sensors for nickel is essential for overcoming the drawbacks associated with conventional colorimetric detection techniques.

## 5. Fluorogenic $Ni^{2+}$ nanosensors

The fluorogenic sensors feature certain advantages over colorimetric sensors, such as higher sensitivity and lower background interference [145–147]. More importantly, fluorescence-based sensors can easily be employed in live cell imaging applications in real-life samples. In this context, QDs are gaining wide attention among researchers for heavy metal ion recognition purposes [148,149]. QDs are particles smaller than excitonic diameter/exciton Bohr radius leading to a quantum confinement effect and the formation of discrete energy levels [150]. The photoluminescence of QDs arises from the radiative recombination of exciton (electron-hole pair), where charge carriers recombine either before they are trapped (band gap recombination) or while trapped in shallow traps (near band gap recombination) [49,148]. These well-dispersed semiconducting nanocrystals consist of an inorganic core with atoms typically from IV-VI, III-V, or II-VI group elements, with an outer capping layer of organic ligands. The interaction between these surface ligands and any surrounding analyte or molecule significantly influences the photoluminescence (PL) characteristics of QDs by influencing the radiative electron-hole recombination; as a result, QDs act as one of the most effective NPs for chemosensing/biosensing applications [151]. The unique photophysical properties of QDs that include size- and/or shape-tuneable emission, appreciable quantum yields, broad excitation bands, narrow emission bands, large Stokes shift, and effective resistance towards photobleaching enable QDs to act as excellent optical probes [152,153]. Owing to these inherent features, QDs have found fascinating applications in the fields of light emitting devices, solar cells, photodetectors, luminescent bio-labels, and bio-imaging probes [154–157].

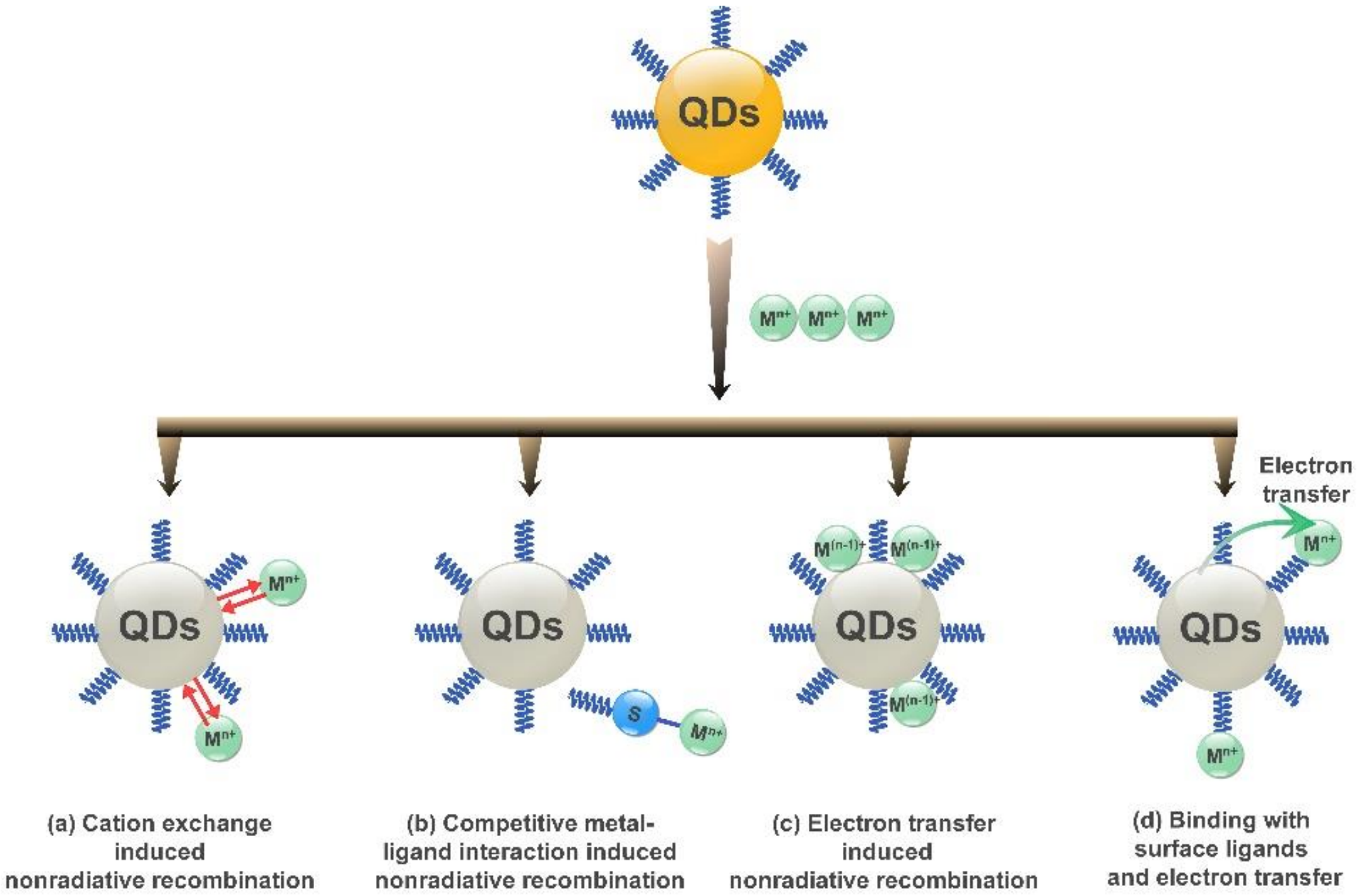

**Fig. 25.** Different mechanistic pathways usually followed by inorganic QDs-based fluorogenic sensors for metal ion detection.

### *5.1. Inorganic quantum dots*

Inspired by the pioneering work by Chen and Rosenzweig [158] based on the usage of CdS-based QDs as luminescent sensors for $Cu^{2+}$ and $Zn^{2+}$ recognition, great progress has been achieved in this field of metal-based QDs, especially for heavy metal ion detection. The instant change in luminescence intensity of QDs that typically occurs from the direct interaction between target ion and the surface/capping ligands has inspired, so far, the development of many QDs-based optical sensors [159]. Inorganic QDs are relatively brighter, and possess good to excellent quantum efficiency and stability features [160]. In particular, chalcogenide QDs display narrow emission bands, high fluorescence yields, and negligible photobleaching [151]. Conceptually, the photoluminescent features of QDs arise from the exciton recombination. Therefore, any change in ligand components and/or the surface state of QDs would significantly affect the recombination process, and subsequently, the luminescence efficacy. In this context, a diverse panel of QDs-based luminescent probes have been developed to detect metal ions by cation exchange induced non-radiative recombination, competition of ligands induced non-radiative combination, electron transfer induced non-radiative recombination, and binding with surface ligands and electron transfer (Fig. 25). Moreover, changes in photoluminescence can occur *via* several mechanisms, such as static/dynamic quenching mechanisms, inner filter effect (IFE), photoinduced electron transfer (PET), and Forster resonance energy transfer (FRET) processes. Notably, the detection of $Ni^{2+}$ ions by QDs is often achieved by binding of the metal ion to the surface ligands followed by static quenching mechanism (Fig. 26). The static quenching mechanism results from the formation of non-fluorogenic ground state complexes between the target metal ion and the QDs. This is then followed by monitoring changes in the absorption spectra, whereas no changes in the fluorescence lifetime ($\tau_0/\tau = 1$) can be observed *via* static quenching mechanism [148,149]. Moreover, the ground state complex should be stabilized with the increase in temperature.

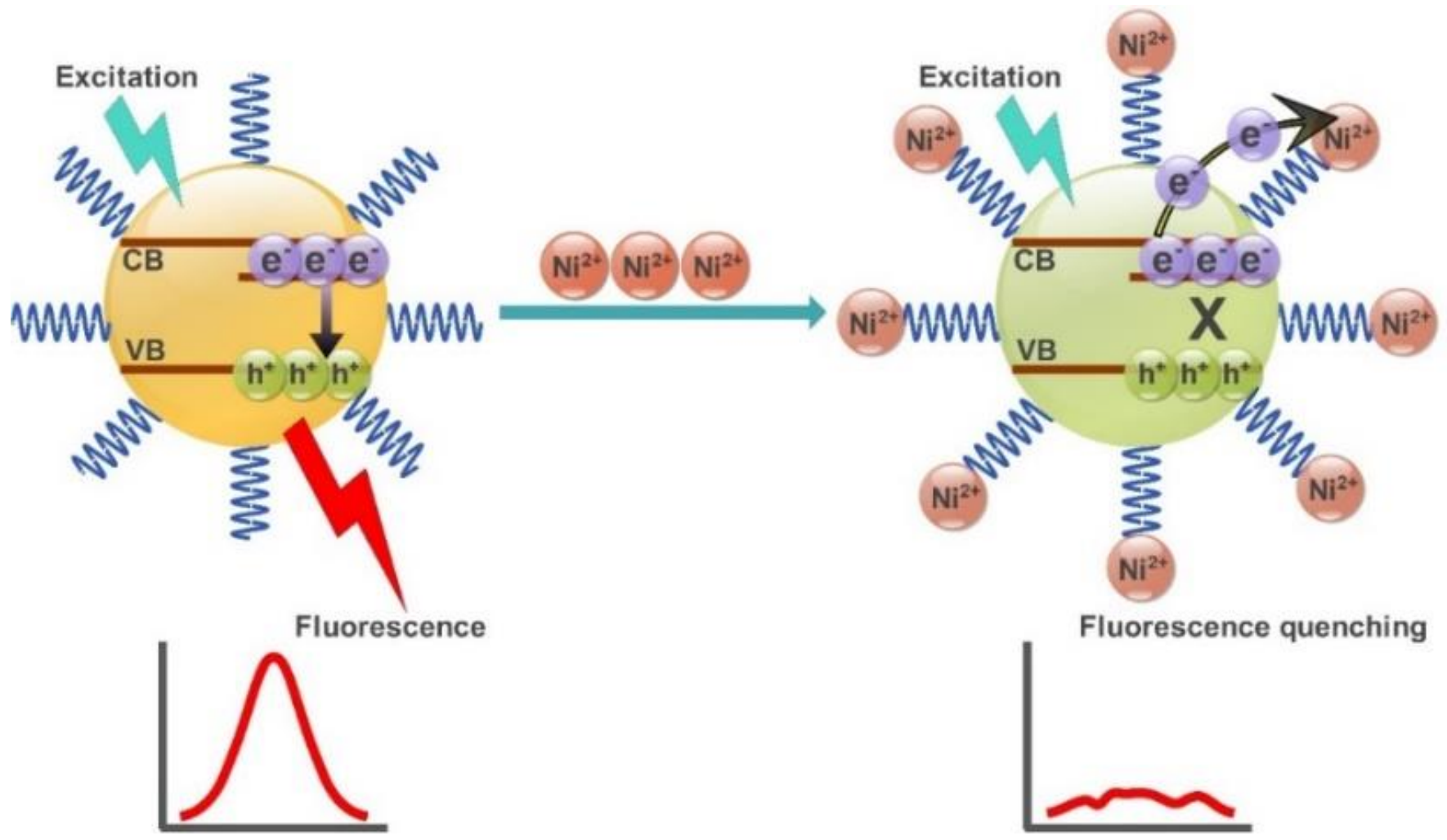

**Fig. 26.** Fluorescence quenching mechanism using QDs sensors for the detection of $Ni^{2+}$ ions.

In year 2014, Mahapatra *et* al. [37] reported one pot synthesis of water soluble CdS-based QDs **28** for the fluorogenic recognition of $Ni^{2+}$ ions (Fig. 27). The 3-mercaptopropionic acid (MPA) has been utilized as a capping agent to stabilize **28**. These QDs exhibited a tuneable emission feature ranging from 510 nm to 650 nm relying on the applied reaction conditions (such as pH, reaction time, and the reactant concentration). The wide range of emission features have been ascribed to the recombination of shallowed-trapped electrons in S (sulphur) vacancy defect states having valance band holes. The sensing ability of MPA-coated QDs having high photoluminescence was examined for fluorogenic detection of various metal ions (such as $Zn^{2+}$, $Cr^{3+}$, $Ni^{2+}$, $Al^{3+}$, $Cd^{2+}$, $Ag^{+}$, $Pb^{2+}$, $Co^{2+}$, $Ca^{2+}$, $Na^{+}$, $K^{+}$, $Hg^{2+}$, $Fe^{3+}$, $La^{3+}$, $Fe^{2+}$, and $Cu^{2+}$). The luminescence of these QDs significantly decreased upon the addition of only $Co^{2+}$ and $Ni^{2+}$ ions; however, other ions displayed negligible emission changes. Besides, UV-Vis spectra clearly distinguished $Ni^{2+}$ from $Co^{2+}$ as the solution color of the QDs become pale yellow solely in presence of $Co^{2+}$ ions (above 5.0 μM concentration). The emission titration experiments revealed a linear correlation range of 50 nM to 1.0 μM for $Ni^{2+}$ with a LoD value of 50 nM. The time-resolved fluorescence studies clearly suggested the non-existence of the electron transfer process, as the original excited-state lifetime of the QDs ($\tau_{avg}$ 34.38 ns) was unaffected upon addition of $Ni^{2+}$ ions. The resulting selective recognition of $Ni^{2+}$/$Co^{2+}$ and the related luminescence quenching has been attributed to the interaction of the target metal ions with the negatively charged carboxylate ions of MPA. The QDs probe was also found reversible with EDTA for $Co^{2+}$ ions; however, no recovery of luminescence was reported for $Ni^{2+}$ even after 30 min. These data further corroborate the distinguished selective behaviors of the probe for $Co^{2+}$ and $Ni^{2+}$ ions.

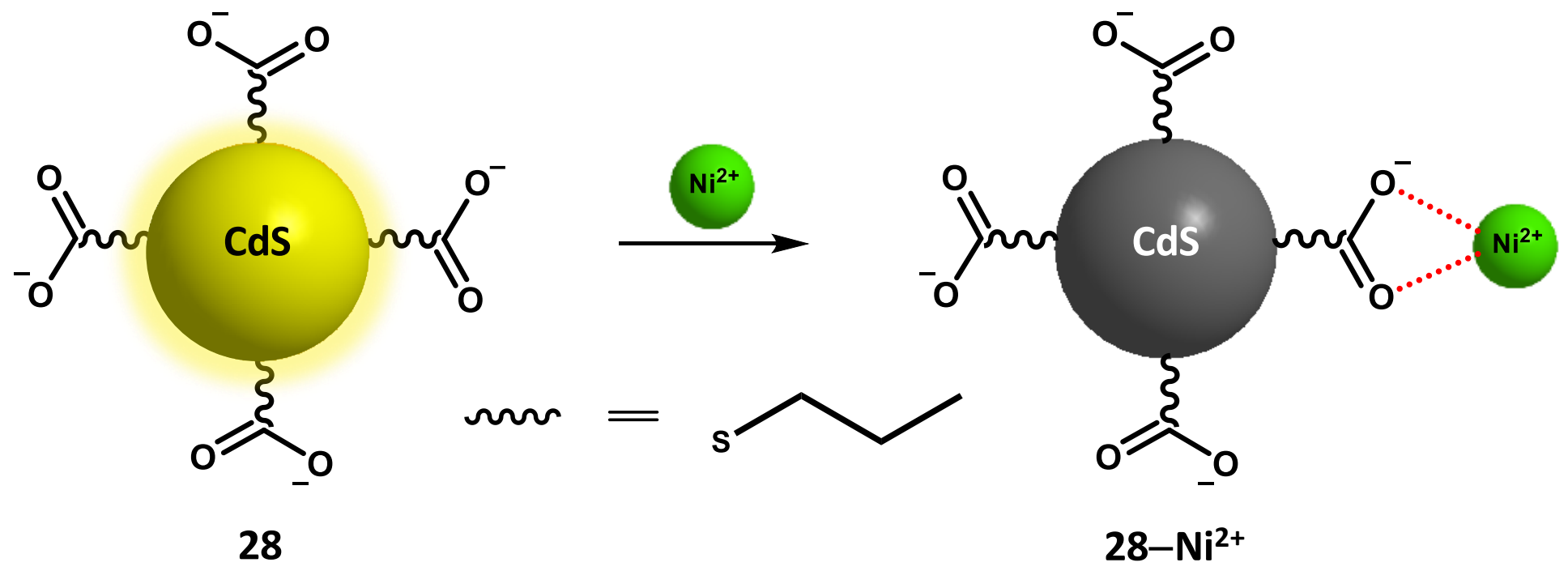

**Fig. 27.** Binding mode of QDs based probe **28** with $Ni^{2+}$ ions.

Three years later in 2017, Zare *et* al. [161] demonstrated thioglycolic-capped CdTe QDs (probe **29)** for a highly sensitive and selective detection of $Ni^{2+}$ in ammonia buffer solution (pH 9.0) (Fig. 28). The probe exhibited significant luminescence at 531 nm upon excitation at 360 nm. The addition of $Ni^{2+}$ ions (~10 μM) to **29** resulted in rapid quenching of the luminescence intensity. The addition of 30 μM dimethylglyoxime (DMG) to QDs-$Ni^{2+}$ adduct restored the original luminescence of the probe. DMG itself, even in excess, did not disturb the original luminescence of the probe without $Ni^{2+}$ ions. The luminescence quenching in **29** has been attributed to the binding of $Ni^{2+}$ with deprotonated carboxylic (-COOH) groups onto the surface of QDs. The probe displayed a low LoD of 7.0 nM with a linear correlation in the range of 0.01-10 μM. It is worth mentioning that some other metal ions such as $Hg^{2+}$, $Ag^{+}$, $Cr^{3+}$, $Cu^{2+}$, and $Co^{2+}$ also exhibited a quenching effect; however, the luminescence could not be revived upon DMG addition in these cases. This QDs-based probe has been successfully employed for $Ni^{2+}$ detection in real water samples such as tap water and wastewater with a recovery of 94.3-108%, and the obtained results were in complete agreement with the samples analysed by ICP-AES studies.

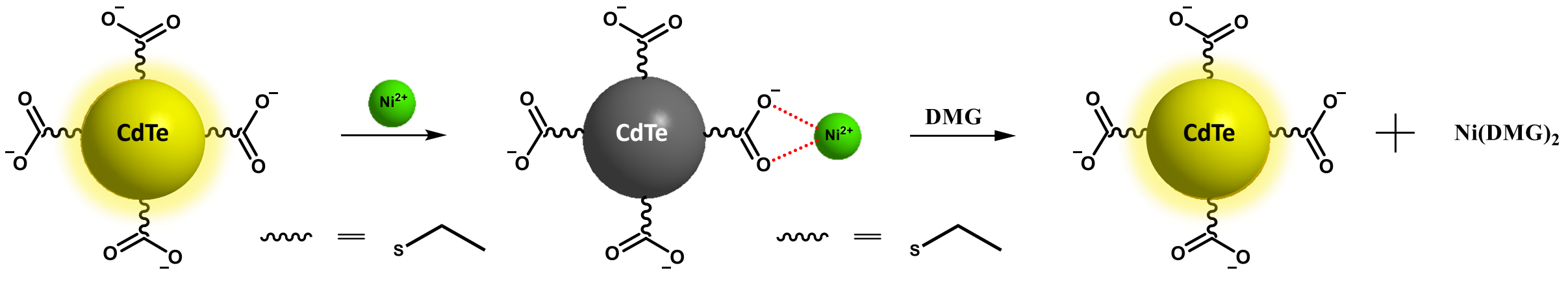

**Fig. 28.** Binding mode of QDs based probe **29** with $Ni^{2+}$ ions.

In 2019, Liu and coworkers [162] designed $SnO_2$-based QDs **30** acting as a luminescent probe for recognition of $Ni^{2+}$ ions. This probe emitted at 300 nm with a small Stokes shift when the excitation wavelength was 280 nm. The addition of $Ni^{2+}$ ions to **30** resulted in a significant quenching of the emission band, which could be attributed to the similar ionic radii of $Sn^{4+}$ and $Ni^{2+}$ ions acting as surface states for QDs. The probe had a LoD of 0.01 ppm for nickel ions. It was also reported that 71% of the vacancies on the surface of the QDs were unoccupied, whereas only 29% of the sites were filled by $Ni^{2+}$ ions. These results indicated that the vacancies or the surface states are vital for the detection of heavy metal ions by $SnO_2$ QDs.

In recent years, core-shell QDs have gained wide attention because of their tuneable transition energy, which can lead to the potentially modified emission/absorption features [163]. Fundamentally, the shell encapsulating the QDs increases the confinement of the electron-hole pair in the core and passivates the dangling bonds present onto the surface of the QDs [164,165]. Consequently, the lowered non-radiative exciton recombination leads to the improved luminescence properties. The thickness of a shell also plays an important role in fine tuning the photophysical properties. With an increasing thickness, significant changes can be observed in CSQDs, e.g., refractive index changes, enhanced non-linear absorption coefficient, and noticeable red-shift in the threshold energy [166]. Moreover, CSQDs are much more biocompatible than their counterpart QDs. Relatively, the core and the shell in CSQDs are primarily comprised of the following configurations: (i) CdS-ZnS; (ii) CdSe-CdS; (iii) CdSe-Zns and (InAs)CdSe. Notably, these CSQDs have found applications in various fields such as catalysis, electronics, medicines, and sensing.

In 2013, Sui and co-workers [167] synthesized *L*-cysteine capped CdTe/ZnS QDs based probe **31** with CdTe as a core template (Fig. 29). The ZnS shell assisted in removing the dangling bond and minimizing the surface defects to improve the photostability. The emission maximum for the CdTe QDs was observed at 550 nm, whereas probe **31** (CdTe/ZnS core shell) emitted at 610 nm along with significant resistance to photobleaching. The addition of metal ions such as $K^{+}$, $Na^{+}$, $Ca^{2+}$, $Zn^{2+}$, $Mg^{2+}$, $Al^{3+}$, $Fe^{2+}$, and $Mn^{2+}$ displayed negligible changes in the emission spectra of **31**; however, an immediate slightly red-shifted quenching could be noticed upon $Ni^{2+}$ addition. This emission quenching was attributed to the binding of $Ni^{2+}$ to *L*-cysteine of the probe **31**. UV-Vis analysis also supported the formation of a complex between $Ni^{2+}$ and **31** as a new slightly red-shifted band appeared in the presence of the target ion. Interestingly, no significant change could be noticed in the lifetime of the probe upon $Ni^{2+}$ addition ions, most likely due to an electron transfer from the excited state of the QDs to $Ni^{2+}$ ions. The incremental addition of $Ni^{2+}$ resulted in remarkable fluorescence quenching, and a linear relation was established from $4 \times 10^{-9}$ to $5 \times 10^{-7}$ mol/L. The detection limit was depicted to be $5.9 \times 10^{-10}$ mol/L with a high $K_{sv}$ value of $3.09 \times 10^{6}$ mol $L^{-1}$. The probe **31** was properly tested to trace $Ni^{2+}$ ions in real drinking water and river water samples with a good recovery in the range of 87.6-117.1%.

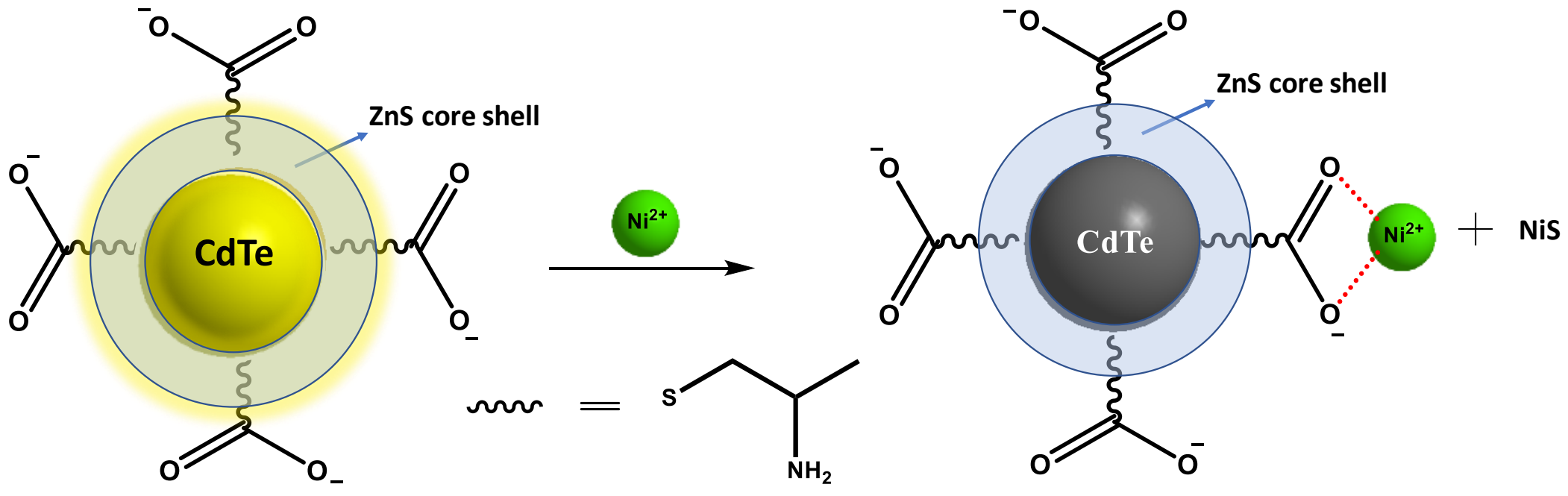

**31** **31–Ni2+**

**Fig. 29.** Binding mode of CSQDs based probe **31** with $Ni^{2+}$ ions.

The CdTe/CdS, CdSe/ZnS, and CdTe/CdSe QDs were reported to have better selectivity and sensitivity towards various metal ions in comparison to single component QDs [168,169]. Hence, in 2021, Dutta and Das synthesized CdTe/CdS QDs **32** for the selective detection of metal ions with glutathione utilized as a stabilizing agent [169]. The excitation of the probe at 365 nm resulted in a bright orange luminescence. The average particle size of **32** was depicted as 3.3 nm using HR-TEM studies. Unlike the previous reports, pH of 7.4 was ideal for the detection of the metal ions. The aqueous solution of the probe exhibited no quenching in the presence of $Na^{+}$, $K^{+}$, $Ca^{2+}$, $Mg^{2+}$, $Mn^{2+}$, $Cr^{3+}$, $Co^{2+}$, $Fe^{2+}$, $Cd^{2+}$, $Zn^{2+}$, $Pb^{2+}$, and $Hg^{2+}$; however, a remarkable quenching was observed in the case of $Cu^{2+}$ and $Ni^{2+}$ ions. The increment in temperature prompted a higher quenching efficiency, which indicated the existence of a collision dynamic quenching. The lifetime studies revealed that the average lifetime decay declined with the increase in the concentration of $Ni^{2+}$ ions due to the electron transfer process. The LoD was calculated with the help of lifetime studies to be 23 μg/L.

Due to their low toxicity, biocompatibility, chemical stability, and good photoluminescence stability, Yao *et* al. in 2017 [170] reported boron nitride based QDs (BNQDs) **33** containing amino and carboxyl groups for the detection of $Ni^{2+}$ ions (Fig. 30). The absorption profile of **33** displayed a band in the high energy region (below 250 nm), followed by a weak shoulder near 300 nm. The BNQDs were uniformly dispersed with an average particle size of 4 nm. The probe **33** exhibited an emission band at 400 nm when excited at 305 nm in phosphate buffer. The maximum fluorescence was observed at pH 5.0 that was selected as an optimum pH for the evaluation of sensing performance. A quantum yield of 32.3% was reported using quinine sulfate as a reference. The detection limit calculated was 0.1 μmol/L. Upon the addition of $Ni^{2+}$, probe **33** displayed a rapid fluorescence quenching response (<5 mins), which was attributed to the PeT between **33** and $Ni^{2+}$ ions. The probe could also detect $Ni^{2+}$ ions in lake water and river water with good recoveries in the range of 102−109%.

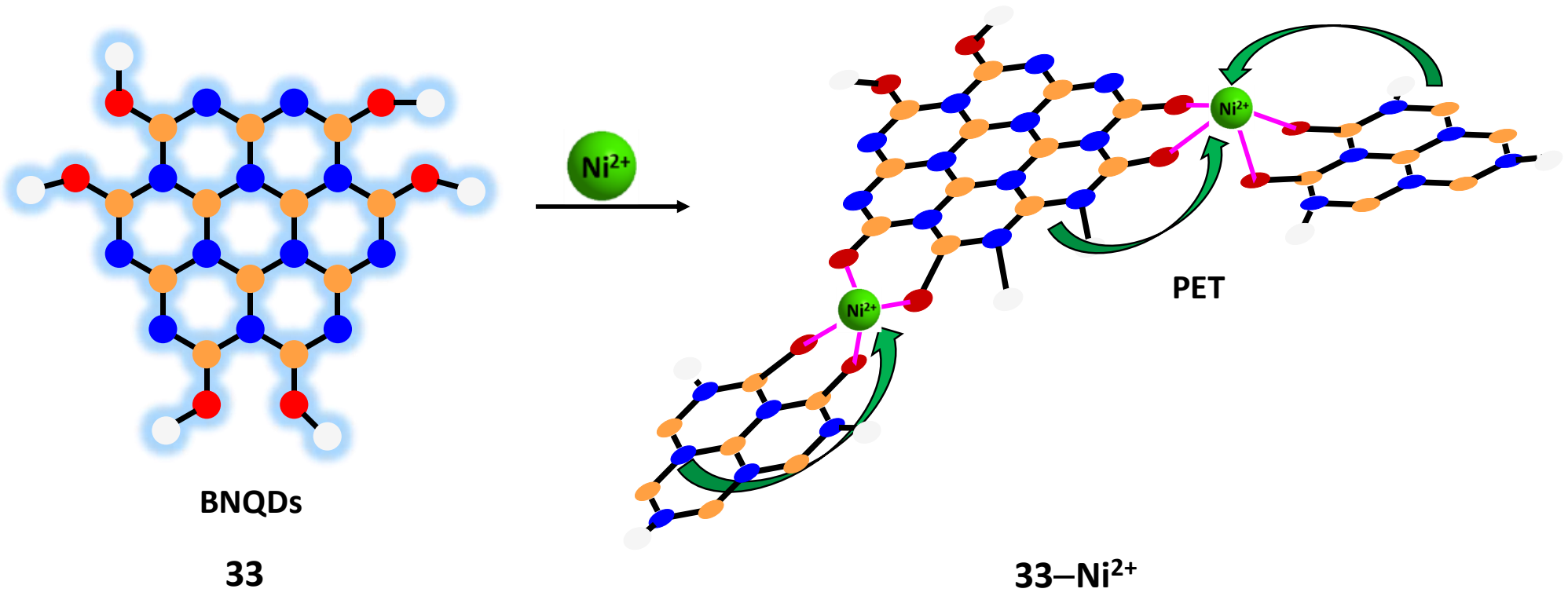

**Fig. 30.** Binding mode of BNQDs based probe **33** with $Ni^{2+}$ ions.

### *5.2. Other inorganic nanomaterials*

Majority of the AuNPs based $Ni^{2+}$ probes reported were colorimetric and relied on aggregation followed by changes in the SPR band (Table 2). However, recently (in 2024), Tiwari *et* al. [171] synthesized aggregation-resistant vancomycin conjugated polyethyleneimine (PEI) stabilized fluorogenic AuNPs **34** for highly selective detection of $Ni^{2+}$ ions. The probe **34** exhibited characteristic SPR absorption band at 508 nm. An intense peak was also observed at 384 nm indicating the spectral overlap between vancomycin (COO-) and PEI (NH-) groups The incremental addition of $Ni^{2+}$ (51.2 μM) to **34** showed negligible impact on the SPR band, and no noticeable color change was observed even after 4 hours. The excitation of vancomycin in the range of 280-310 nm revealed an emission band at 350 nm. However, a bathochromic shift to 420 nm was observed in the emission spectrum upon conjugation with gold nanoparticles, and the emission was quenched upon $Ni^{2+}$ addition. The emission lifetime of **34** changed from 3.2 ns to 2.5 ns after the introduction of $Ni^{2+}$ ions in solution, suggesting electron transfer from vancomycin to $Ni^{2+}$ ions. The selectivity of probe **34** was further tested against various metal ions revealing that only $Hg^{2+}$ could moderately interfere in the detection of $Ni^{2+}$ ions. Probe **34** exhibited a LoD of 90.5 nM with a linear range of 0.05-6.4 μM.

In 2018, Shah *et* al. [172] synthesized $Fe_3O_4$-based nanoparticles probe **35** with the fluorescein dye attached to an amino silica moiety with a metallic core. The average size of the nanoparticles was depicted to be 15.0 nm using TEM analysis. The absorption and emission maxima of **35** were observed near 500 nm and 515 nm, respectively, and the luminescence intensity of **35** was drastically quenched after the addition of $Ni^{2+}$. The resultant non-linearity in the Stern-Volmer plot suggested the possibility of both static as well as dynamic quenching mechanisms. The static quenching could be attributed to the binding between $Ni^{2+}$ and the fluorescein group of **35** whereas the dynamic mechanism occurred due to the collision between **35** and $Ni^{2+}$ ions. No remarkable emission changes have been observed in the presence of other assorted metal ions such as $K^+$, $Na^+$, $Ca^{2+}$, $Mn^{2+}$, $Zn^{2+}$, $Co^{2+}$, $Pb^{2+}$, $Cd^{2+}$, $Fe^{2+}$, $Fe^{3+}$ and $Ag^+$, ions. The binding constant for $Ni^{2+}$ was depicted as $3.2 \times 10^4$ $M^{-1}$ with a 1:1 binding stoichiometry. The authors claimed that the response time of **35** for $Ni^{2+}$ was almost instantaneous at pH 7.0, with the LoD value as $8.3 \times 10^{-10}$ M. This probe could also be applied for $Ni^{2+}$ detection in real models such as spiked tap water with a recovery rate of 88−105%.

The Eu-Tb based nanocomposite **36** with pyridine-2,6-dicarboxylic acid immobilized on Laponite was reported by Yao *et* al. [173]. Probe **36** acted as a pH sensor and was tested against different metal ions in water. The excitation of **36** at 282 nm resulted in the sharp characteristic emission bands of $Tb^{3+}$ at 490, 544, 584 and 620 nm (attributed to $^5D_4 \longrightarrow {}^7F_J$ transitions; J = 0-3) whereas the typical $Eu^{3+}$-centred bands were observed at 579, 593, 612, 650, and 698 nm (attributed to $^5D_0 \longrightarrow {}^7F_J$ transitions; J = 0-4). The dispersion of **36** in 0.01 M aqueous solution of metal ions resulted in a colors array ranging from orange-yellow to green. The detected metal ions such as $Cu^{2+}$, $Al^{3+}$. $Bi^{2+}$, $Ni^{2+}$, $Fe^{3+}$, $Ce^{3+}$ and $Co^{2+}$ exhibited a unique ratio of emission intensity of $I_{Eu}/I_{Tb}$. The probe was found most sensitive to $Cu^{2+}$ with a LoD value of 0.05 mM.

In 2020, Wen *et* al. [174] synthesized silicon based carbon dots **37** by utilizing (3-aminopropyl)trimethoxysilane (APS) as a stabilizer (Fig. 31). The absorption spectrum of **37** showed a peak around 279 nm and a weak tail near 350 nm. The probe exhibited a red shifted emissions ranging from 475-545 nm when the excitation wavelength was varied from 350-460 nm in phosphate buffer. The successive red shifted emissions were ascribed to the quantum size effect or surface states. The maximum emission (530 nm) intensity was observed upon excitation at 430 nm. The presence of amino, hydroxyl, and carboxyl groups on the surface of the probe was confirmed by FTIR studies. The sensing performance of **37** was assessed using a range of cations such as $Na^{+}$, $K^{+}$, $Ag^{+}$, $Mg^{2+}$, $Mn^{2+}$, $Ni^{2+}$, $Cu^{2+}$, $Cd^{2+}$, $Fe^{2+}$, $Hg^{2+}$, $Sr^{2+}$, and $Fe^{3+}$ where a turn-off response was observed only in the presence of $Ni^{2+}$ ions due to strong binding of $Ni^{2+}$ with the surface hydroxyl and amino groups. The average particle size was reported to be 2.5 nm as depicted by TEM analysis, and the LoD determined was 1.73 µM.

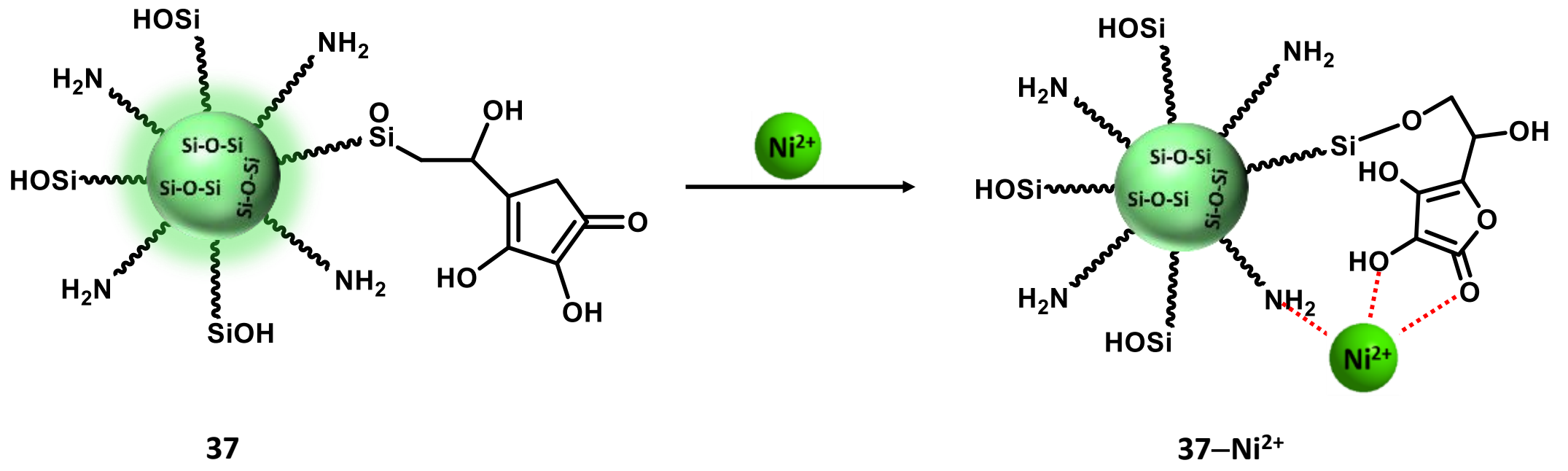

**Fig. 31.** Binding mode of SiNPs based probe **37** with $Ni^{2+}$ ions.

**Table 3.** Inorganic fluorogenic nanosensors used as optical probes for $Ni^{2+}$ ions.

| Probe | $\lambda em$ (nm) (SPR) | Particle size (nm) | Stabilizing agent | Target ions | LoD (μM) | Quenching/ Enhancement | Solvent | Practical application | [Ref.] |
|---|---|---|---|---|---|---|---|---|---|
| **28** | 510-650 | 3.77 | 3-mercaptopropionic acid | **$Ni^{2+}$**, $Co^{2+}$ | **$Ni^{2+}$ = 50 × $10^{-3}$** $Co^{2+}$ = 20 × $10^{-3}$ | Quenching | Water | - | [37] |
| **29** | 531 | 2.96 | Thioglycolic acid | **$Ni^{2+}$** | **7 × $10^{-3}$** | Quenching | Ammonia Buffer | Tap water and waste water | [161] |
| **30** | 300 | 2.23 | Ammonium thiocyante | **$Ni^{2+}$**, $Fe^{3+}$ | 0.01 | Quenching | Water | Reclaimed water and sea water | [162] |
| **31** | 610 | | L-cysteine | **$Ni^{2+}$**, $IO_4^-$ | $Ni^{2+}$ = 4.0 ×$10^{-3}$ $IO_4^-$ = 3.6 ×$10^{-3}$ | Quenching | Alkaline Medium | Drinking and river water | [167] |
| **32** | 580 | 3.3 | Glutathione | **$Ni^{2+}$**, $Cu^{2+}$ | 23 | Quenching | Tris-HCl Buffer | - | [169] |
| **33** | 400 | 4.0 | | **$Ni^{2+}$** | 0.1 | Quenching | Phosphate Wuffer | Lake water | |
| **34** | 420 | 7.5 | Vancomycin conjugated polyethyleneimine | **$Ni^{2+}$** | $Ni^{2+}$ = 0.2059 | Quenching | Water | River water | |
| **35** | 515 | 15 | Fluorescein | **$Ni^{2+}$** | 0.83×$10^{-3}$ | Quenching | Water | Tap water | [172] |
| **36** | 612, 542 | | Pyridine-2, 6-dicarboxylic acid | **$Ni^{2+}$**, $Cu^{2+}$, | - | Quenching | Water | - | [173] |
| **37** | 542 | 2.5 | (3-aminopropyl)trimethoxysilane | **$Ni^{2+}$** | 1.73 | Quenching | PBS | - | [174] |

### *5.3. Organic quantum dots*

Although a diverse panel of fluorogenic organic nanosensors (such as organic nanoparticles, nanofibers, nanotubes and carbon-based QDs) have been reported for heavy metal ion detection over the preceding years [50,149,175], $Ni^{2+}$ sensing is majorly achieved by carbonaceous QDs (CQDs). Notably, these sensors display some significant advantageous features over their inorganic counterparts. Inorganic QDs, primarily composed of cadmium-based materials, exhibit remarkable optical properties and stability; however, their applications in environmental and biological settings are constrained due to their low biocompatibility, high cost, and leaching of heavy toxic metal ions [151]. Carbon-based QDs have emerged as potential alternatives to semiconductor QDs due to their excellent structural diversity, good water solubility and improved biocompatibility [176,177]. Suitable surface functionalization of organic QDs using polar entities (such as -OH, -COOH, $-NH_2$, $-NO_2$, etc.) not only offers binding sites to the target analyte but also imparts water solubility. Indeed, several carbonaceous QDs have displayed excellent permeability and are safe for plants [178,179]. CQDs are zero dimensional NPs exhibiting discrete quantized energy levels, crystal lattice, and variable density of states [79,178]. For spherical CQDs, the optical quantum confinement effect becomes prominent when size lies in the range 2-10 nm [180], approaching the Bohr exciton radius. The optical properties of CQDs directly rely on their intrinsic quantum confinement effect, band edge position, band gap, and dimension of the NPs. The reduced size of CQDs is accompanied by a change in their transition from band to discrete energy levels [181]. Further decrease in particle size confines the exciton, causing the energy band gap structure resembling to that of a molecular electronic structure. The fluorescence properties of CQDs are influenced due to the variations in their HOMO-LUMO gap with the varying particle size. Upon excitation of CQDs, the non-radiative process lowers the energy of the first excited singlet state ($S_1$) mediated by defect states, resulting in an emitted photon with lower energy than that of initial excitation energy [181,182].

Furthermore, due to size dependence nature of the emissive properties of CQDs, their photophysical properties can be fine-tuned by adjusting the HOMO-LUMO gap by employing different surface functional groups [183]. For instance, electron-donating groups raise the HOMO to higher energy with a concomitant reduction in the energy gap. Overall, the excitation and emission features rely on both the size of NPs and the surface functionalization. It is important to mention that the exact origin of fluorescence in CQDs is still a topic for debate [184]. Recently, several review articles have been documented covering the detection of heavy metal ions employing fluorogenic organic nanoparticles; however, to the best of our knowledge, $Ni^{2+}$ detection has never been comprehensively reviewed. The ensuing sections will focus on various carbonaceous QDs employed for $Ni^{2+}$ detection.

In an elegant study (in 2018), Su *et* al. [185] developed a multifunctional carbon dots based platform **38** *via* a facile one-step hydrothermal method to recognize $Ni^{2+}$ ions in aqueous medium (Fig. 32). Probe **38** was also utilized to detect doxycycline via a turn-on fluorescence response. The surface of the probe was functionalized by carboxyl and amino groups. The absorption profile of carbon dots in water displayed $\pi$- $\pi^*$ (>C=N- and >C=C<) transitions at 220 nm and 227 nm and a peak at 290 nm, recognized as n-$\pi^*$ transition (>C=O). The particles were well dispersed, with size ranging from 1.12-4.19 nm as observed in HR-TEM and DLS studies. Probe **38** emitted at 430 nm when the excitation wavelength was 310 nm. Various other metal ions including $Na^+$, $K^+$, $Mn^{2+}$, $Co^{2+}$, $Pb^{2+}$, $Zn^{2+}$, $Ca^{2+}$, $Ba^{2+}$, $Sr^{2+}$, $Hg^{2+}$, $Cu^{2+}$, $Ni^{2+}$, and $Al^{3+}$, have been assessed; however, only $Ni^{2+}$ exhibited a remarkable fluorescence quenching response. The detection limit of **38** was depicted to be 16.57 µM, which was poor in comparison to the permissible limit prescribed by WHO [16], which limits its application in real world scenarios. The selective detection of $Ni^{2+}$ was attributed to the presence of oxygen-based −COOH and −OH groups as well as $-NH_2$ present on the surface of carbon dots. These groups acted as receptor for $Ni^{2+}$ ions which inhibited the existing charge transfer process, leading to fluorescence quenching event.

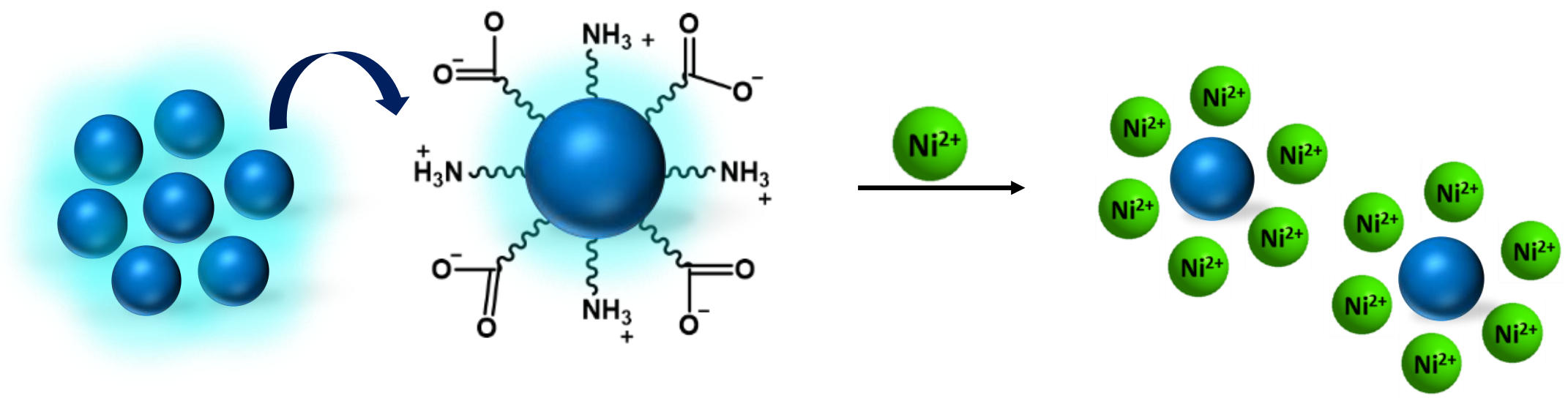

**38** **38–Ni$^{2+}$**

**Fig. 32**. Binding mode of $Ni^{2+}$ ions with carbon dots **38**.

To date, numerous imidazole-based sensors have been reported that can selectively detect anions and cations including $Ni^{2+}$ ions. Imidazole and its derivatives are widely known for its function as a bioagent, binding with metal ions and exhibiting hydrogen bonding with proteins and drugs [186]. In 2019, Gong and Liang [187] developed a carbon dots-based sensor **39** functionalized with imidazole carboxylic acid (Fig. 33). The average size of the carbon dots was determined to be 5 nm by TEM studies. UV-Vis spectrum of **39** revealed two peaks at 223 nm and 275 nm together with a broad band ranging from 300-600 nm. Upon excitation at 360 nm, the probe **39** emitted at 456 nm. A significant quenching response could be noticed solely with $Ni^{2+}$ addition when sensing ability of **39** was examined for a pool of metal ions (e.g., $K^+$, $Na^+$, $Li^+$, $Cu^{2+}$, $Zn^{2+}$, $Mn^{2+}$, $Mg^{2+}$, $Ca^{2+}$, $Cd^{2+}$, $Co^{2+}$, $Ni^{2+}$, $Hg^{2+}$, $Al^{3+}$, and $Fe^{3+}$). The emission quenching was attributed to the strong binding ability of imidazole derivative with $Ni^{2+}$, and the sensing mechanism underwent a dynamic PET process. This probe could be successfully applied in real water samples such as tap water with good recovery values (85-102%).

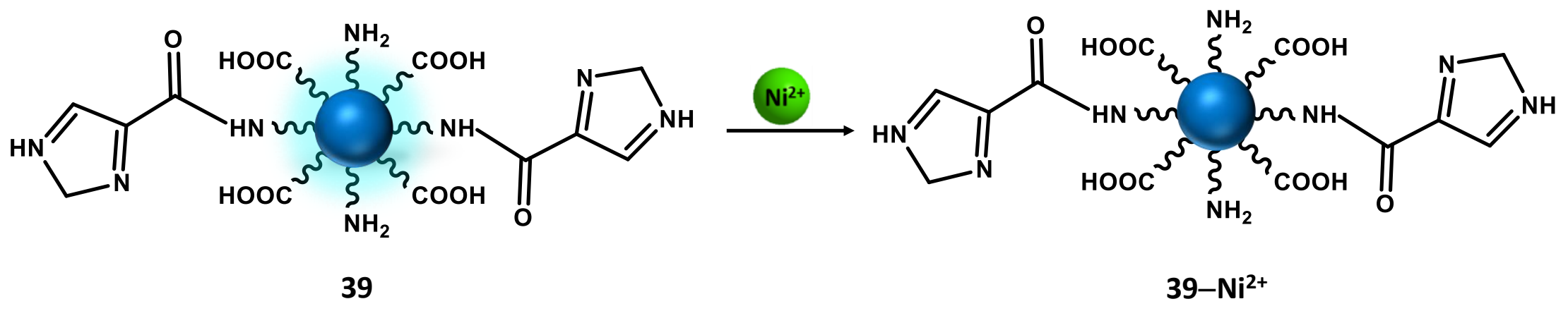

**Fig. 33**. Interaction of $Ni^{2+}$ ions with carbon dots **39**.

The obtained detection limit of probe **39** was 0.93 mM, which is higher than the permissible limit set by WHO. Moreover, the development of this probe involved several synthetic steps. Therefore, the implementation of simple and facile synthetic routes leading to highly sensitive carbon dots for $Ni^{2+}$ is considerably desirable.

The use of natural biomolecules as the carbon source for the synthesis of CQDs is vital in terms of green and sustainable chemistry [188]. In an effort, Wei and co-workers [189] designed tryptone and yeast extracts derived carbon dots **40** by a hydrothermal process. The surface of the carbon dots was functionalized with amino groups to sense various metal ions. TEM studies revealed an average diameter of 6 nm for **40** and the particles were monodispersed and spherical. The absorption and emission maxima of **40** were measured at 266 nm and 422 nm ($\lambda_{ex}$ = 344 nm), respectively, in aqueous medium. This probe displayed high selectivity towards $Ni^{2+}$ ions among the broad range of investigated cations ($Ca^{2+}$, $Cd^{2+}$, $Mn^{2+}$, $Cu^{2+}$, $Pb^{2+}$, $Hg^{2+}$, $Mg^{2+}$, $Ba^{2+}$, $Zn^{2+}$, $Al^{3+}$, and $Fe^{3+}$). The incremental addition of $Ni^{2+}$ (0.1-32.0 μM) to **40** resulted in a dramatic fluorescence quenching response. Interestingly, probe **40** could be applied over a wide range of pH from 2.0-8.0 with a $Ni^{2+}$ detection limit of 46 nM. The amino groups present on the surface of **40** bind with $Ni^{2+}$ to form a stable complex. UV-Vis spectra showed negligible changes upon $Ni^{2+}$ addition indicating the absence of any energy transfer process. The Stern-Volmer plot revealed that fluorescence quenching followed dynamic electron transfer process. This probe exhibited improved LoD in the nM range along with excellent biocompatibility, and thus, can be potentially utilized for real practical applications. The probe **40** also exhibited low cytotoxicity and successfully employed for live cell imaging of $Ni^{2+}$ in HeLa cells, E. coli, and S. cerevisiae.

The modulation of synthetic conditions can significantly impact the sensing behaviour and photophysical properties of carbon dots. In 2019, Paoprasert and co-workers detected $Ag^{+}$ ions using carbon dots prepared from polyurethane *via* the pyrolysis method [190]. In the subsequent year, the same group synthesized carbon-based QDs **41** with the help of polyurethane for selective detection of $Ni^{2+}$ ions by varying the reaction conditions [191] (Fig. 34). Polyurethane was chosen due to its nitrogen rich content which improves the chelating ability as well as the quantum efficiency of the sensory probe. The absorption spectrum of **41** in aqueous medium displayed peaks at 250 nm and 300 nm ascribed to the π- π* and n- π* transitions, respectively. This probe exhibited an emission maximum at 492 nm ($\lambda_{ex}$ = 380 nm) in aqueous solution with an emission quantum yield of 24%. The substantial quenching of fluorescence intensity (~70%) was observed after the addition of nickel ions within 50 seconds. The fluorescence titration experiment was performed and the LoD was estimated to be 3.14 μM with a linear range of 0-150 μM. Other metal ions ($K^+$, $Ag^+$, $Sn^{2+}$, $Mg^{2+}$, $Ca^{2+}$, $Zn^{2+}$, $Co^{2+}$, $Cr^{6+}$, $Cd^{2+}$, $Fe^{3+}$, and $Cu^{2+}$) failed to interfere in the detection process of $Ni^{2+}$ ions. The rationale behind quenching was due to the static complex formation between the **41** and $Ni^{2+}$ ions. It was observed that a single carbon dot can adsorb and/or accommodate approximately 182 nickel ions (234.8 mg/g). UV-Vis spectra were utilized to further investigate the sensing mechanism. It was observed that the absorption profile of $Ni^{2+}$ partially overlapped with the emission spectrum of **41**. Additionally, the excitation spectrum of **41** exhibited substantial overlap with the absorption curve of $Ni^{2+}$ ions, suggesting that the inner filter effect played a prominent role in the fluorescence quenching. The optical band gap of **41** (5.3 eV) was also larger than that of $Ni^{2+}$ (3.8 eV), which implied that the emission energy might be absorbed by $Ni^{2+}$ ions. It was concluded that $Ni^{2+}$ interacted with the carbonyl, hydroxyl, and carboxylate functional groups of the carbon dots resulting in the formation of a non-fluorogenic complex. The synergistic effect of this interaction together with the inner filter effect resulted in the observed quenched emission. Probe **41** was further tested to detect $Ni^{2+}$ ions in spiked bottled water samples with recovery values in the range of 95-99%. As mentioned earlier, test paper strips are vital in terms of practicality and affordability, therefore, paper-based sensors were also developed as a portable and convenient kit for on-site $Ni^{2+}$ testing. The calculated RGB values decreased linearly with the increasing $Ni^{2+}$ concentration, and a solid-state detection limit of 43.3 μM has been determined. The RGB values are crucial to incorporate these sensor technologies into smart portable/smartphone devices for accurate on-site detection.

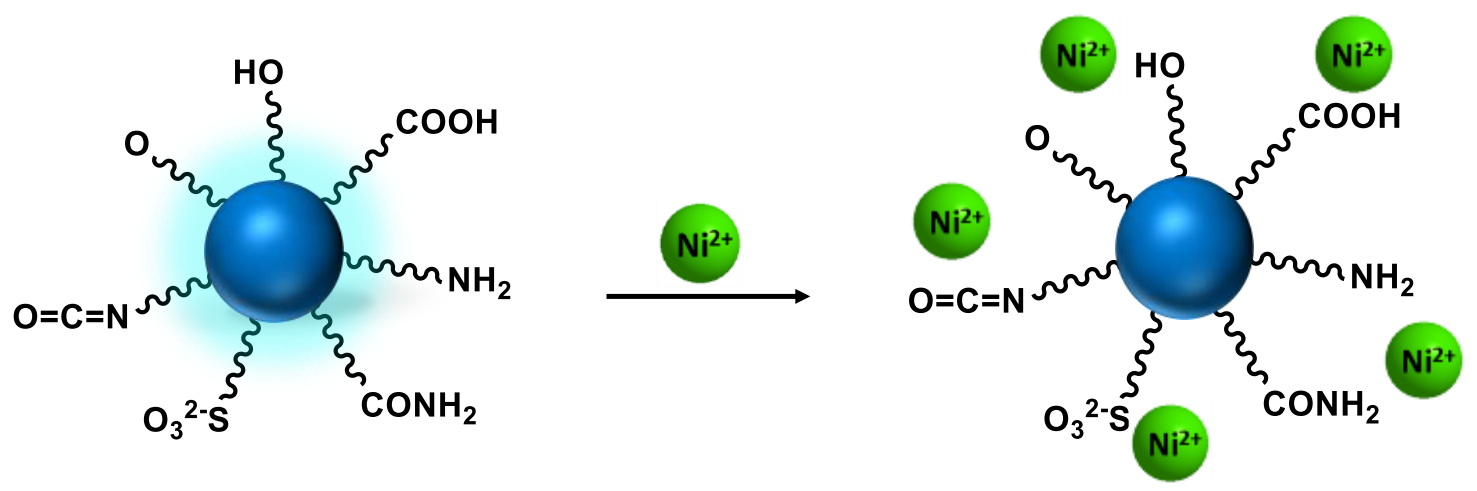

**Fig. 34.** Binding mode of $Ni^{2+}$ ions with carbon dots **41**.

Recently, efforts have been made to employ new synthetic approaches for organic nanosensors development using green solvents as organic solvents can possess detrimental effects on humans and environment. Indeed, solvent-free methods are gaining increasing attention of researchers in the context of economy and the prevention of pollution [192]. In 2022, Lin's group [193] synthesised carbon dots **42** by adopting a green, solvent free pyrolysis method and the natural polyphenol (i.e., tannic acid) (Fig. 35). The particle size of the carbon dots averaged around 3.4 nm and the surface sulfhydryl and hydroxyl groups of **42** acted as metal binding sites. UV-Vis studies revealed a high energy absorption appearing near 220 nm in ethanol, corresponding to the π-π* transitions. The probe exhibited emission maxima at 450 nm when excited at 370 nm with a quantum yield of 35.4%. Preferential coordination of $Ni^{2+}$ with the surface hydroxyl groups resulted in a turn-off fluorescence response over other investigated metal ions (such as $Li^{+}$, $Na^{+}$, $Ca^{2+}$, $Mn^{2+}$, $Mg^{2+}$, $Zn^{2+}$, $Cu^{2+}$, $Fe^{3+}$, $Ir^{3+}$, and $Al^{3+}$). A similar set of sensing experiments was also conducted with carbon dots **43** synthesized using tannic acids *via* hydrothermal process. The probe **43** also exhibited quenching in the presence of $Ni^{2+}$ ions; however, the LoD was much higher (100 μM) when compared to **42** (20 μM). This report clearly demonstrated that the appropriate reaction conditions can drastically modify the sensitivity of the probe.

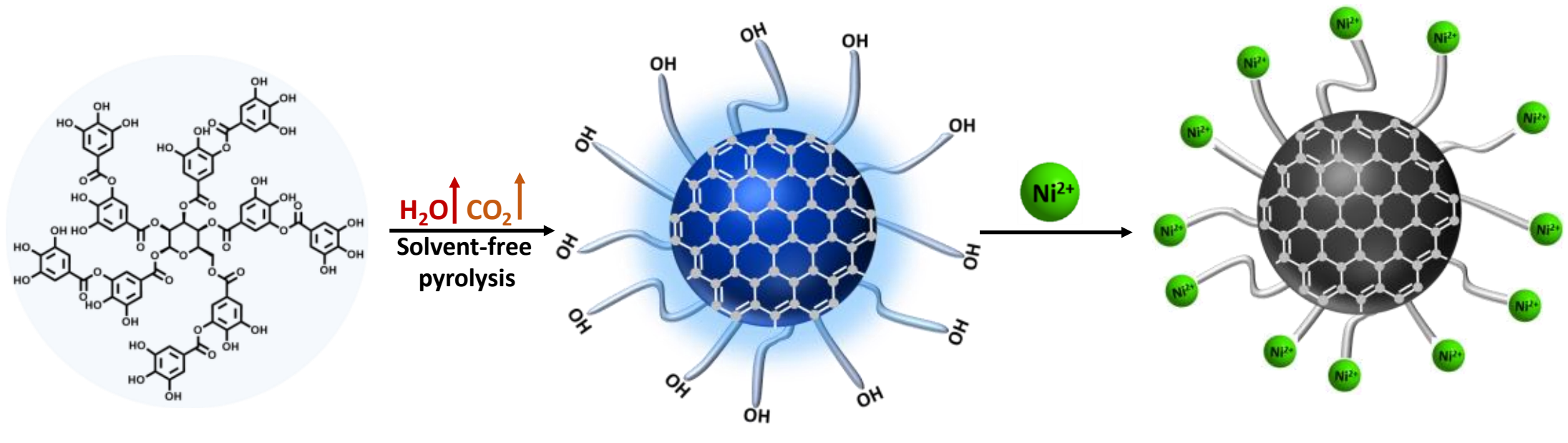

**Fig. 35.** Binding mode of $Ni^{2+}$ ions with carbon dots based probe **42**.

In 2022, Wang and coworkers demonstrated the use of sulfur incorporated CQDs **44** for selective detection of $Ni^{2+}$ ions [194]. The diameter of these $Na_2S_2O_3$ derived QDs was 6.54 nm as evidenced by TEM analysis. UV-Vis spectrum of **44** displayed two absorption bands near 336 and 363 nm, which could be ascribed to n-π* transition of >C=O functionality and the surficial polysulfide ions, respectively. When excited at 375 nm, this probe emitted near 488 nm in aqueous medium. TRF analysis revealed that **44** exhibited bi-exponential decay behavior. The short lifetime component (1.35 ns) represented the recombination of the generated exciton whereas the long component (3.42 ns) was attributed to the surface state recombination. Probe **44** selectively recognized $Ni^{2+}$ ions among a range of assorted cations ($Na^{+}$, $Ag^{+}$, $K^{+}$, $Cu^{2+}$, $Zn^{2+}$, $Fe^{2+}$, $Cd^{2+}$, $Ba^{2+}$, $Mg^{2+}$, $Pb^{2+}$, $Hg^{2+}$, $Al^{3+}$, and $Fe^{3+}$) and anions ($F^{-}$, $I^{-}$, $Br^{-}$, $Cl^{-}$, $NO^{-}$, $NO_3^{-}$, $C_2O_3^{2-}$, $CO_3^{2-}$, $S_2O_3^{2-}$, $SO_4^{2-}$, and $PO_4^{3-}$). Both UV-Vis and emission studies were employed to examine the sensing ability of this probe toward $Ni^{2+}$, as its spectral signatures significantly altered upon $Ni^{2+}$addition. The average emission lifetime of **44** was barely affected upon $Ni^{2+}$ addition, most likely due to a static complex formation. This migrates the electron from the metal ion to the ground state of the surface functional groups (such as -OH and -COOH) prohibiting the exciton recombination. The fluorescence titration experiment revealed that probe **44** exhibited good linearity in the range of 0-400 μM, with a LoD value of 0.18 μM. The practical utility of this probe was demonstrated *via* trace amount detection of $Ni^{2+}$ in tap water and drinking water samples having recovery values in the range of 96-104%.

In recent years, the integration of machine learning (ML) into sensor technology has inevitably played a crucial role in removing false positives and achieving accurate detection of metal ions. In this context, Xu *et* al. [195] designed a sensor array **45** based on carbon dots and $Tb^{3+}$-EDTA lanthanide complex for the detection of $Cr^{6+}$, $Fe^{3+}$, $Cu^{2+}$, $Ni^{2+}$, $Mn^{2+}$, $Co^{2+}$, and $Fe^{2+}$. The sensor array emitted at 400 nm after being excited at 320 nm, which was attributed to the carbon dots. Moreover, $Tb^{3+}$ was sensitized by EDTA (antenna effect) to emit at 489 and 545 nm. Among the various metal ions, only $Ni^{2+}$ quenched all the characteristic $Tb^{3+}$-based emission bands while the emission of CDs remained intact. The quenching effect could be attributed to the formation of complex between nickel ions and EDTA. The authors utilized TPOT (tree-based pipeline optimization technique), under which classification and concentration models were integrated. A SX model was prepared for machine learning to selectively differentiate the metal ions. The above-mentioned metal ions can be detected upto 0.05-50 μM with a 95.6% accuracy. Probe **45** could detect a range of metal ions in soil samples and lake water with recovery values of 93.3-100%. Moreover, the sensor array could accurately differentiate between the binary mixture of $Fe^{2+}/Cr^{6+}$, $Fe^{2+}/Fe^{3+}$, and $Cu^{2+}/Cr^{6+}$ with 100% accuracy. These sensor array with multiple emissions incorporating ML and AI will eventually pave the way to successfully commercialize a highly selective and sensitive sensor.

The fluorogenic and sensing properties of QDs can also be fine-tuned by altering their organic precursor. Recently, in 2023, Jia and co-workers [196] synthesized nitrogen doped carbon nanodots **46** by utilizing phthalonitrile-based compounds (melamine and benzoxazine) as precursors (Fig. 36). The probe **46** displayed an emission maximum centred at 490 nm when excited at 365 nm, and the variations in the emission maxima could be experienced with different excitation wavelengths between 325 nm to 405 nm. The variations in the emission spectrum depending on the excitation wavelength indicated towards the non-uniform size of **46**. A large stokes shift of 125 nm not only rules out the background fluorescence interference but also circumvents any self-quenching event, ultimately leading to improved sensitivity and reliability of the sensory system. This probe exhibited good photostability, as evidenced by its unaltered fluorescence intensity upto 90 min of UV light illumination. The particle size of the **46** determined from TEM analysis was *ca.* 40 nm, possibly due to the self-cross linking reaction. The fluorescence sensing experiments of **46** towards various metal ions revealed the selective nature of the probe towards $Ni^{2+}$ ions. The fluorescence titration further displayed a good linear relationship for $Ni^{2+}$ (1.0-5.0 μM) ions and a very low detection limit of 1.58 nM was calculated. The absorption spectrum of the probe displayed a peak near 197 nm due to the $\pi$- $\pi^*$ transition of surface nitro group, which significantly decreased upon addition of the $Ni^{2+}$ ions. The modulations in the UV-Vis and fluorescence spectra were attributed to the coordination of $Ni^{2+}$ with the O atom of the $-NO_2$ group present on the surface. These results were finally supported by XPS and FTIR studies. The LoD value reported is one of the lowest values for fluorogenic $Ni^{2+}$ based probes.

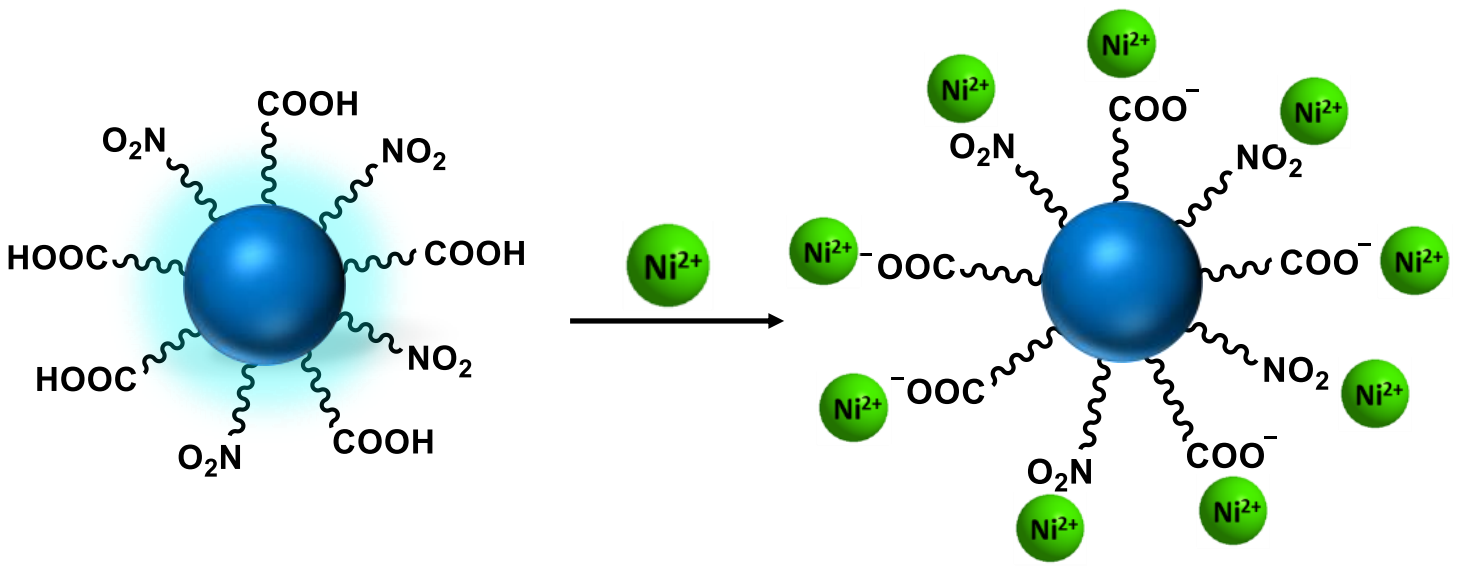

46

46–Ni2+

**Fig. 36.** Binding mode of $Ni^{2+}$ ions with CQDs based probe **46**.

It can be clearly noticed that the fluorescence emission of most of the previously discussed carbon dots is limited to short wavelengths. However, the development of fluorogenic probes emitting at longer wavelengths is relatively more desirable in terms of practical applications. This is due to their related advantageous features such as deeper tissue penetration, minimal biological tissue damage, higher sensitivity, and easy visualization [197,198]. In a recent report (2024), Qian's group [199] developed orange luminescent carbon dots **47** for the detection of $Ni^{2+}$ in aqueous media (Fig. 37). These carbon dots having uniform and homogenous dispersion exhibited an average diameter of 3.15 nm. The surface of **47** was judiciously functionalized with carbonyl and amine receptors for metal ion binding. Two distinct peaks at 199 nm and 400 nm were observed in the absorption spectra of **47**, which were attributed to the π- π* (>C=C<) and the n-π* (>C=O or >C-N-) transitions, respectively. The absorption band present at 566 nm corresponded to the resazurin moiety. The emission maximum for this probe was measured at 592 nm ($\lambda_{ex}$ = 575 nm) which was significantly quenched in the presence of $Ni^{2+}$ ions over a variety of other competing ions. Interestingly, the fluorescence intensity revived after the addition of morin in the **47**-$Ni^{2+}$ solution. The sensing mechanism was established by monitoring the changes in the zeta potential value of the probe in presence of $Ni^{2+}$ ions. $Ni^{2+}$ addition induced a notable increase in the zeta potential of probe from +3.03 mV to +3.50 mV; however, no considerable change was noticed in its emission lifetime. These results suggested a static quenching mechanism as a consequence of stable ground-state complexation between $Ni^{2+}$ and **47**. The probe exhibited a low LoD of 42.2 nm and was successfully employed to detect $Ni^{2+}$ in food samples such as lettuce and chocolate.

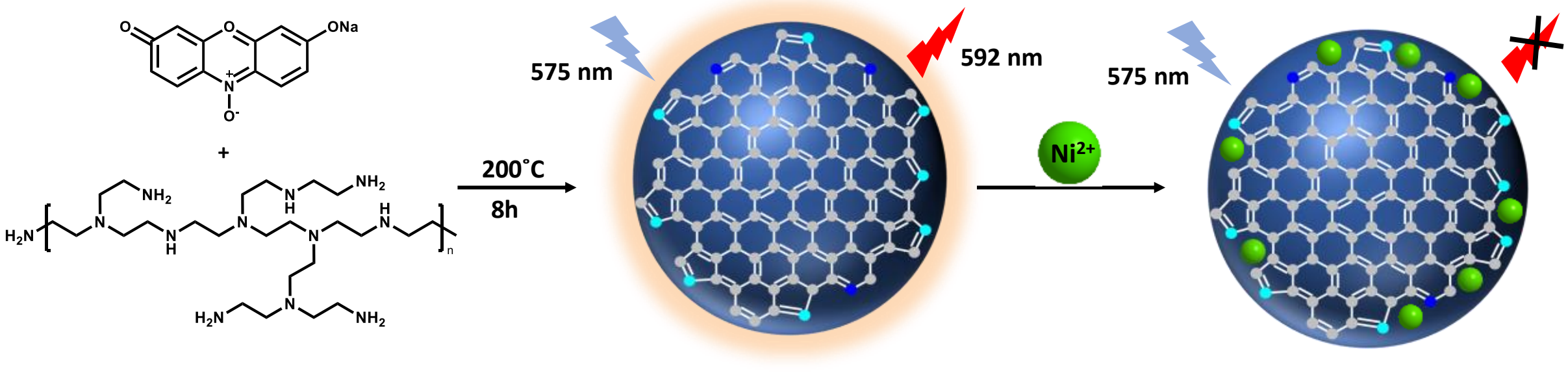

47 47–$Ni^{2+}$

**Fig. 37.** Binding mode of $Ni^{2+}$ ions with carbon dots **47**.

Graphene QDs are 0D nanoparticles with particle size between 1-20 nm having $sp^2$ hybridized carbon atoms. The ease of post synthetic functionalization, compatibility with an aqueous system, and excellent photostability make GQDs ideal platform for metal ion sensing [175,200]. The small size results in quantum confinement effect and causes some band gap energy which can be modified by varying functional groups, particle size, and by doping of heteroatoms. However, to the best of our knowledge, only a single report is available for their use in fluorogenic detection of $Ni^{2+}$ ions.

In 2018, Xu *et* al. [201] synthesized **48**, an ethylenediamine functionalized GQDs, for selective detection of $Ni^{2+}$ in aqueous medium. TEM analysis of **48** revealed an average particle size of 4.2 nm, indicating the successful synthesis of QDs. UV-Vis and Raman spectra confirmed the formation of GQDs **48**. Excitation of **48** at 505 nm led to an intense fluorescence emission at 590 nm due to the inter-band transition. The probe demonstrated excellent photostability under UV (320 nm) for 2 days as well as visible (450-850 nm) light illumination for up to 60 days. The sensing performance was evaluated by a range of cations, anions, and various amino acids. The addition of $Ni^{2+}$ ions resulted in substantial quenching of the fluorescence intensity. The photoluminescence titration with $Ni^{2+}$ revealed that the probe **48** exhibited a detection limit of $3 \times 10^{-8}$ M and a linear concentration range of $1 \times 10^{-7}$ M to $5 \times 10^{-5}$ M. The reversible binding capability of probe **48** was evaluated using EDTA. Notably, the lifetime of the probe exhibited minimum change (~ 4.2 ns) after the addition of $Ni^{2+}$ ions, suggesting a static quenching process due to strong interaction between ethylenediamine in **48** and $Ni^{2+}$ ions. The practical utility of probe **48** was demonstrated for intracellular detection of $Ni^{2+}$ ions in the rADSC cell lines. The probe possessed low toxicity and maintained good viability even after 24 hours of incubation. However, the bright yellow luminescence of rADSC cells incubated with **48** faded when $Ni^{2+}$ ions were introduced.

### *5.4 Other organic nanomaterials*

Graphite materials such as graphene oxide (GO) and reduced graphene oxide (rGO) have been extensively utilized in various research areas including catalysis, energy storage, sensors, drug delivery, and environmental remediation [202–204]. In particular, rGO is gaining growing attention among researchers to develop novel sensors because of its large active surface area, non-toxicity, robust mechanical properties, and ease of post-functionalization. Ramesh *et* al. [205] successfully synthesized a GO-based probe **49** for selective turn-on detection of $Ni^{2+}$ ions in dispersed aqueous medium (Fig. 38). A bis-(phthalimidoethyl)-amine was used as a functionalizing agent, and TEM analysis clearly validated the formation of GO sheets. This probe was weakly fluorogenic in aqueous medium but displayed an intense emission at 354 nm in presence of only $Ni^{2+}$ among other ions. pH effect study (2.0 to 9.0) revealed that the probe's fluorescence enhanced in acidic conditions due to the protonation of nitrogen atoms on the basal plane of the GO sheet, resulting in a stronger PET response. The ability to sense $Ni^{2+}$ ions in the aqueous medium makes probe **49** viable for environmental and biological sensing applications at physiological pH.

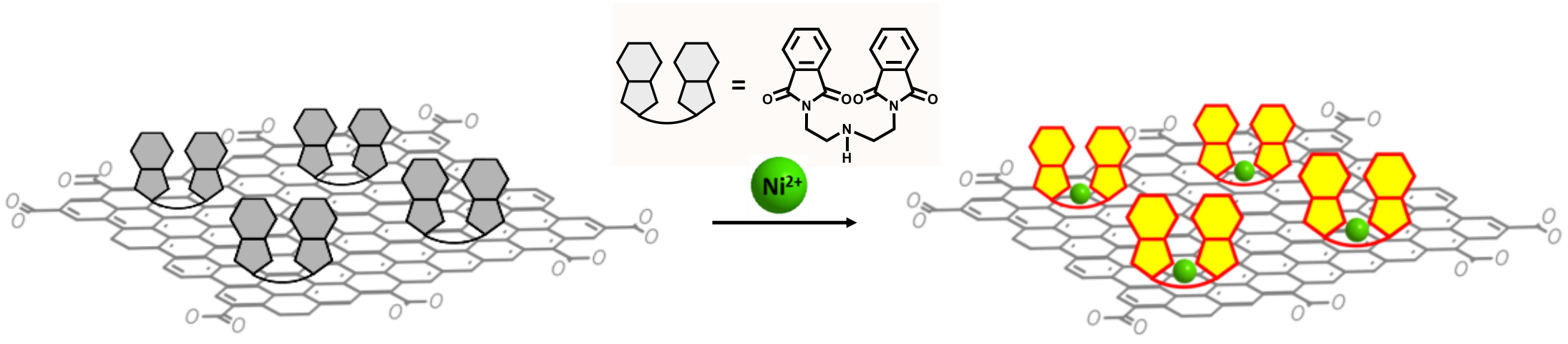

**49** **49–Ni²⁺**

**Fig. 38.** Binding mode of $Ni^{2+}$ ions with QDs based probe **49**.

In 2012, Torto and co-workers [206] fabricated a fluorogenic nanofiber **50** comprising pyridylazo-2-naphthol and poly(acrylic acid) for selective detection of $Ni^{2+}$ ions. These nanofibers were uniformly dispersed with an average size of 230-800 nm. The probe exhibited emission at 557 nm when excited at 479 nm. The nanofibers were coated on a glass slide and exhibited a turn-off response upon addition of $Ni^{2+}$, which could be attributed to the formation of **50-Ni²⁺** complex promoting the intersystem crossing event. Interestingly, the fluorescence response restored by rinsing the slides in an acidic solution which destabilized the formed complex by protonating the receptor nitrogen atoms. This reversibility was demonstrated up to 4 cycles; however, the fluorescence intensity was slightly reduced after each cycle which could be attributed to the change in morphology of the probe.

**Table 4.** Organic fluorogenic nanosensors employed as optical probes for $Ni^{2+}$ ions.

| Probe | *λem* (nm) (SPR) | Particle size (nm) | Stabilizing agent | Target ions | LoD (µM) | Quenching/ Enhancement | Solvent | Practical application | [Ref.] |
|---|---|---|---|---|---|---|---|---|---|
| **38** | 430 | 1.12-4.19 | - | $Ni^{2+}$ | 16.57 | Quenching | Water | - | [185] |
| **39** | 456 | 5 | Urea | $Ni^{2+}$ | 930 | Quenching | Water | Tap water | [187] |
| **40** | 422 | 6 | - | $Ni^{2+}$, p-NP | $Ni^{2+}$ = 0.046<br>p-NP = 0.012 | Quenching | Water | Bio-imaging | [189] |
| **41** | 492 | 4 | Polyurethane | **$Ni^{2+}$** | 3.14 | Quenching | - | Paper based sensor | [191] |
| **42** | 450 | 3.4 | Tannic acid | **$Ni^{2+}$** | 20 | Quenching | Ethanol | - | [193] |
| **43** | - | 3.3 | Tannic acid | **$Ni^{2+}$** | 100 | Quenching | Ethanol | - | [193] |
| **44** | 488 | 50 | Poly(ethyleneglycol) | **$Ni^{2+}$** | 0.18 | Quenching | Water | Tap and drinking water | [194] |
| **45** | - | - | - | **$Ni^{2+}$** | 0.05 | Quenching | Water | Smartphone based detection | [195] |
| **46** | 490 | 40 | - | **$Ni^{2+}$** | 0.00158 | Quenching | Water | - | [196] |
| **47** | 592 | 3.15 | - | **$Ni^{2+}$** | 0.0422 | Quenching | PBS | Food samples | [199] |
| **48** | 590 | 4.2 | EDTA | **$Ni^{2+}$** | 0.03 | Quenching | Water | Bio-imaging | [201] |
| **49** | 354 | - | Bis-(pththalimidoethyl)-amine | **$Ni^{2+}$** | - | Enhancement | Water | - | [205] |
| **50** | 557 | | Pyridylazo-2-naphthol, Poly(acrylic acid) | **$Ni^{2+}$** | 1.19 | Quenching | Water | - | [206] |

### *5.5. Comparative sensing performance of fluorogenic sensors*

The resultant sensing parameters presented in Tables 1-4 clearly demonstrate that the fluorogenic nanosensors offer superior $Ni^{2+}$ sensitivity in comparison to their colorimetric counterparts. In case of inorganic fluorogenic sensors, mostly chalcogenide-based QDs have been employed due to their unique optical properties. These QDs typically rely on a static quenching mechanism *via* stable ground-state complex formation with $Ni^{2+}$ ions. The majority of such sensors display pH sensitivity, and their LoDs generally fall in the micromolar range. However, the related high toxicity of inorganic QDs limits their practical applicability in many biological and environmental settings. To the best of the author's knowledge, there are only few reports based on non-chalcogenide inorganic nanomaterials acting as fluorogenic $Ni^{2+}$ probes (i.e., **33**, **35** and **37**). While these probes exhibited good sensitivity and selectivity in water, their utility in biological media still remains understudied.

Among organic fluorogenic nanosensors, CQDs dominate the detection of nickel ions. The majority of the reported probes exhibited a 'turn-off' fluorescence response upon interaction with nickel, with an exception in case of probe **49** showing fluorescence enhancement. The focus must be given to develop ‘turn-on’ $Ni^{2+}$ sensors that offer various advantageous features such as low background interference, wide detection range, and the minimization of false positive signals. In contrast to their inorganic counterparts, organic nanosensors have successfully been employed to image $Ni^{2+}$ in biological/environmental samples (such as **40** and **47**). *In fine*, carbonaceous QDs are found to be highly promising in developing nanosensors with excellent stability, sensitivity, detection limit, and biocompatibility. Moreover, artificial intelligence and machine learning (AIML) tools, known for their potential ability to assess complex datasets and improve pattern recognition, have been investigated in the design of inorganic-organic hybrid nanosensors (such as probe **44**). Such nanosensors could be tailored for the selective detection of various heavy metal ions, including $Ni^{2+}$, enhancing recognition accuracy and related efficiency. AIML algorithms enable real-time processing and predictive analysis, making them a valuable addition in refining sensor performance for multi-ion detection.

## 6. Conclusions and future perspectives

There has been increasing ecological and global public health concerns due to environmental contamination caused by $Ni^{2+}$ ions. Human exposure to $Ni^{2+}$ has increased significantly as a result of its widespread usage in industries such as agriculture, metallurgy, technology, and even domestic applications. This growing health risk has prompted researchers to develop more accurate, facile, user-friendly, and highly sensitive methods for detecting $Ni^{2+}$ contamination. Although the conventional analytical techniques are efficient, they are typically restricted to laboratory based $Ni^{2+}$ detection and are therefore hardly suitable to be deployed for on-site field applications. Hence, the development of new sensors for selective and sensitive recognition of $Ni^{2+}$ in real-life samples is evolving as a socio-economic challenge in the 21$^{st}$ century. In this context, optical sensors (colorimetric/fluorogenic) have emerged as gold-standard due to their ease of operation, technical simplicity, high sensitivity/selectivity, and remarkably low background effects.

During the past decade, nanomaterials have dramatically replaced the usage of common organic dyes in optical sensors due to their unique characteristic features. The present study highlights the optical detection of $Ni^{2+}$ using nanomaterials, especially MNPs, inorganic and organic QDs. In case of noble MNPs (e.g., AgNPs and AuNPs), the detection is usually achieved by a large shift in their SPR band leading to a color change of the colloidal probe’s solution. The modification of SPR absorption could be attributed to the variation in the nanoparticle’s size, size distribution and the spacing between them. The sensing mechanism mainly relies then on the aggregation of the NPs in the presence of $Ni^{2+}$ ions. The majority of reported MNPs are typically based on AgNPs and AuNPs functionalized with metal ion chelating agents. The presence of N and O containing groups is found crucial for the detection of $Ni^{2+}$ ions; however, $Co^{2+}$ and $Cu^{2+}$ can act as major interfering species due to their competitive ionic radii and borderline acidic character. These observations further cemented the vital aspect of a judicious choice of donor groups in receptors to achieve highly selective sensing of a specific metal ion. Furthermore, to viably develop a cheap and commercially available nanosensor, the Ag and Au might eventually be replaced by more sustainable alternatives, e.g., earth-abundant metals. More generally, for the syntheses, stabilization and/or functionalization of the nanomaterials, focus could be shifted to biogenic resources, for example involving plant extracts and the microbe-mediated synthetic processes.

QDs stand out among fluorogenic nanosensors for $Ni^{2+}$ detection, surpassing other nanomaterials in sensing performance. These QDs are mainly composed of metal chalcogenides or carbon-based materials. While the emission quantum yield of organic QDs is typically lower than that of their inorganic counterparts, it can be improved by utilizing appropriate capping agents. As highlighted by Paoprasert and group [190,191], the synthesis conditions and doping of QDs with sulfur or nitrogen are critical factors in fine-tuning their selectivity and sensitivity. However, doping the carbon core remains a significant challenge, with its effects still not fully understood. To address this, comprehensive ultrafast photophysics studies are urgently needed to clarify optical properties. The relative complexity of some sensing nanomaterials coupled with unsatisfactory synthesis (e.g., low yields, tedious syntheses or purification steps, high costs) may also harm the environment as well as humans or living beings in general. Therefore, with the support of green chemistry, the eco-friendly synthetic strategy is urgently required to be adopted for future sensing materials. An alternative way is to design and develop recyclable and degradable sensors that may help in reducing the environmental pollution at large scale.

Though there have been significant advancements in the development of optical $Ni^{2+}$ sensors, the emphasis must be laid down to their stability of usage in the complex matrices or relevance. The selectivity/sensitivity of such nanomaterials can be further improved by using various recognition units (such as aptamers, DNAzymes, etc.) that are quite stable under variable environmental circumstances with typically no requirement of any specific storage conditions. Moreover, the tuneable emission features of some nanomaterials (e.g., upconverting nanoparticles, carbon dots, etc.) may help in the expansion of multi-detection sensory systems to further improve their practical employment in environmental and biological settings. Due to the predictable rising demand for on-site recognition devices for testing of samples, recent evolution in technology may contribute to develop adequate portable devices and related software tools i.e. new digitalized mobile apps and/or software that can be utilized to monitor instant colorimetric/fluorogenic changes in presence of a specific target analyte. Machine learning and artificial intelligence may also serve as an important tool for effective simultaneous detection of not one but multiple competing cations. Hence, attention must be given to integrate such optical nanosensors into easy-to-assess solid platforms (e.g., paper-based strips, microfluidic chips, etc.) and easily transportable devices for an efficient on-site detection application.

Ultimately, the authors anticipate that such breakthrough in the sensory systems must lead to the future development of user-friendly, robust, highly sensitive/selective, and commercially available detection kits for heavy metal ions.

**Symbols/Abbreviations:**

| | |
|---|---|
| $Ni^{2+}$ | nickel(II) |
| NPs | nanoparticles |
| MNPs | metal nanoparticles |
| AuNPs | gold-nanoparticles |
| AgNPs | silver-nanoparticles |
| SPR | surface plasmon resonance |
| SERS | surface enhanced Raman scattering |
| QDs | quantum dots |
| CSQDs | core-shell quantum dots |
| PL | photoluminescence |
| LoD | limit of detection |
| $K_b$ | binding or association constant |
| $K$sv | Stern-Volmer constant |
| $\Phi_{em}$ | emission quantum yield |
| nM | nanomolar |
| mM | millimolar |
| μM | micromolar |
| UV | ultraviolet |
| NIR | near-infrared |
| FE-SEM | field emission scanning electron microscopy |
| HR-TEM | high-resolution transmission electron microscopy |
| SAR | structure-activity relationship |
| HSAB | hard and soft acids and bases |

**Acknowledgements**

SK thanks Department of Science & Technology, New Delhi, India for INSPIRE Research Grant (IFA15/CH-213). The authors acknowledge the support received from CIC and Applied Science Cluster, UPES, Dehradun India.

**Declaration of Competing Interest**

The authors declare that they have no known competing financial interests or personal relationships that could have appeared to influence the work reported in this paper.